\documentclass[11pt,english,american,nofonttune, onecolumn]{IEEEtran}
\usepackage[T1]{fontenc}
\usepackage[latin9]{inputenc}
\usepackage[a4paper]{geometry}
\usepackage{amsmath}
\usepackage{amssymb}
\usepackage{graphicx}
\usepackage{esint}

\makeatletter

\newcommand{\lyxaddress}[1]{
\par {\raggedright #1
\vspace{1.4em}
\noindent\par}
}

\usepackage{cite}

\newcommand{\PreserveBackslash}[1]{\let\temp=\\#1\let\\=\temp}

\allowdisplaybreaks

\usepackage{babel}

\makeatother

\usepackage{babel}
\begin{document}

\title{Semi-analytic single-channel and cross-channel nonlinear interference
spectra in highly-dispersed WDM coherent optical links with rectangular
signal spectra}

\author{Alberto Bononi$^{a}$ and Ottmar Beucher$^{b}$ }

\maketitle

\lyxaddress{{\vspace*{-1cm} \small$^{a}$Dip. Ingegneria Informazione, Università
degli Studi di Parma, Parma, Italy\foreignlanguage{english}{. $^{b}$Fakultaet
Maschinenbau und Mechatronik, Hochschule Karlsruhe, Technik und Wirtschaft,
Karlsruhe, Germany.} }}
\begin{abstract}
We provide new single-integral formulas of the power spectral density
of single-channel and cross-channel nonlinear interference in highly-dispersed
coherent optical links for which the Gaussian Noise model \cite{Carena_JLT,poggio_solo}
applies. \end{abstract}
\selectlanguage{english}%

\selectlanguage{american}%

\section{Introduction}

The Gaussian Noise (GN) model has recently been shown to effectively
predict the system performance of highly-dispersed wavelength division
multiplexed (WDM) coherent optical transmission systems, such as high
baud-rate dispersion-uncompensated (DU) systems \cite{Carena_JLT,poggio_solo}.
In such a model, the GN reference formula (GNRF) provides a formally
elegant and compact expression of the power spectral density (PSD)
of the received nonlinear interference (NLI). However, the GNRF involves
a double frequency integral which poses non-trivial numerical problems
for multi-span wavelength division multiplexed (WDM) systems. Many
of the numerical integration issues have been already addressed in
\cite{poggio_solo}. Given the practical importance of developing
an accurate GNRF numerical evaluator, however, for debugging purposes
it proves quite useful to have exact expressions of the NLI PSD in
special realistic cases. The case of rectangular per-channel input
spectra has already served in \cite{poggio_solo} as a basic example
to clarify the integration regions, and in \cite{savory_PTL} to obtain
novel explicit expressions of both NLI PSD and total received NLI
power in the single-channel case, or equivalently in the Nyquist WDM
case where the whole WDM spectrum is rectangular. 

In this paper, we derive exact single-integral semi-analytic expressions
of the NLI PSD in the GNRF for both Nyquist and non-Nyquist WDM systems
with input rectangular per-channel spectra. We provide explicit PSD
formulas for both the single-channel interference (SCI) and the cross-channel
interference (XCI) \cite{poggio_solo}. We formulate the GNRF in a
generalized form that applies to any link configuration, be it with
concentrated or distributed amplification, with or without in-line
compensation, and with possibly different spans: the whole link complexity
is summarized within the \emph{kernel} frequency function \cite{curri_OE_raman,pontus_JLT,noi_ARXIV}.

\section{\label{sec:The-GN-reference}The GN reference formula }

In dual-polarization transmission, assuming uncorrelated signals with
identical spectra on the two polarizations, the GN reference formula
(GNRF) yields the power spectral density (PSD) of the nonlinear interference
(NLI) as \cite{Carena_JLT,poggio_solo,pontus_JLT,noi_ARXIV}:
\begin{equation}
\begin{array}{rcl}
G_{NLI}(f) & = & \frac{16}{27}I(f)\\
I(f) & := & \iintop_{-\infty}^{\infty}|\mathcal{K}(f_{1}f_{2})|^{2}G(f+f_{1})G(f+f_{2})G(f+f_{1}+f_{2})\mbox{d}f_{1}\mbox{d}f_{2}
\end{array}\label{eq:Glausgangsbasis}
\end{equation}
where $G(f)$ is the input PSD (i.e., that of the propagated channel
in single-channel transmission, or the whole wavelength division multiplexed
(WDM) spectrum in multi-channel transmission), and the scalar \emph{frequency-kernel}
when higher-order dispersion is neglected is \cite{noi_ARXIV,pontus_JLT}:
\begin{equation}
\mathcal{K}(v):=\int_{0}^{L}\gamma(s)\mathcal{G}(s)\mbox{e}^{-j(2\pi)^{2}C(s)v}\mbox{d}s\label{eq:frequency-kernel}
\end{equation}
where $L$ is total system length, $\gamma(z)$ is the fiber nonlinear
coefficient, $\mathcal{G}(s)$ is the power gain from $0$ to $s$,
and $C(s)\triangleq C_{0}-\int_{0}^{s}\beta_{2}(s')\mbox{d}s'$ is the\emph{
}cumulated dispersion from $0$ to $s$. $C_{0}$ is the (possibly
present) pre-compensation, and $C$ has here the sign of the dispersion
coefficient. Note that the system function $\mathcal{K}$ depends
only on the product $v\equiv f_{1}f_{2}$. A generalization including
third-order dispersion is provided in \cite{pontus_JLT}.

Whenever the input PSD $G(f)$ is symmetric in $f$, then also $G_{NLI}(f)$
is symmetric. In fact, for $f\geq0$ we have: 
\begin{equation}
\begin{split}I(f)=\iint_{-\infty}^{\infty} & \left|\mathcal{K}(f_{1}f_{2})\right|^{2}G(-f+f_{1})G(-f+f_{2})G(-f+f_{1}+f_{2})\mbox{d}f_{1}\mbox{d}f_{2}\\
 & =\iint_{-\infty}^{\infty}\left|\mathcal{K}(f_{1}f_{2})\right|^{2}G(f-f_{1})G(f-f_{2})G(f-f_{1}-f_{2})\mbox{d}f_{1}\mbox{d}f_{2}
\end{split}
\label{Glausgangsymmetrie-1-1}
\end{equation}
because of the symmetry of $G(\cdot)$. By substituting $f_{1}$ by
$-f_{1}$ and $f_{2}$ by $-f_{2}$ we get $I(f)$ again. Hence with
symmetric input PSDs the $G_{NLI}(f)$ needs to be calculated only
at positive frequencies.

The trouble with the analytic formula (\ref{eq:Glausgangsbasis})
is that it involves a double frequency integration where the squared
kernel $|\mathcal{K}(f_{1}f_{2})|^{2}$ is oscillating in frequency
faster and faster as the number of spans increases and poses non-trivial
integration convergence problems \cite{poggio_solo}. A first step
towards easing the double integration comes from a suitable change
of integration variables. In \cite{poggio_solo} the change to hyperbolic
coordinates $u=-\frac{1}{2}\ln(f_{2}/f_{1})$, $v=\sqrt{f_{1}f_{2}}$
was proposed. The rationale was that the squared kernel is a function
of $v$ only, hence at fixed $v$, integration in the $(f_{1},f_{2})$
plane follows the constant contour levels of $|\mathcal{K}(f_{1}f_{2})|^{2}$.

With a similar rationale, we use here the alternative change $u=f_{1}$,
$v=f_{1}f_{2}$, whose Jacobian is $J=|u|$ and whose inverse is $f_{1}=u$,
$f_{2}=v/u$. With such a change, the double integral in (\ref{eq:Glausgangsbasis})
becomes
\begin{eqnarray}
I(f)= & \int_{0}^{\infty}|\mathcal{K}(v)|^{2} & \left[\int_{0}^{\infty}\frac{1}{u}G(f+u)G(f+\frac{v}{u})G(f+u+\frac{v}{u})\mbox{d}u\right.\label{equno}\\
 &  & +\int_{0}^{\infty}\frac{1}{u}G(f-u)G(f+\frac{v}{u})G(f-u+\frac{v}{u})\mbox{d}u\label{eqdue}\\
 &  & +\int_{0}^{\infty}\frac{1}{u}G(f-u)G(f-\frac{v}{u})G(f-u-\frac{v}{u})\mbox{d}u\label{eq:tre}\\
 &  & \left.+\int_{0}^{\infty}\frac{1}{u}G(f+u)G(f-\frac{v}{u})G(f+u-\frac{v}{u})\mbox{d}u\right]\mbox{d}v\label{eq:quattro}
\end{eqnarray}
where $\mathcal{K}(v)$ is given in (\ref{eq:frequency-kernel}),
and the four lines correspond to integration over the four quadrants
of the $(f_{1},f_{2})$ plane. The pole at $u=0$ in the inner integral
does not pose convergence problems for any finite-power spectrum,
since $\lim_{f\to\pm\infty}G(f)=0$ and thus all triple products $G(.)G(.)G(.)$
in the integrand go to zero sufficiently fast as $u\to0$.

When the input WDM signals have rectangular spectra, the inner integral
(\ref{equno})-(\ref{eq:quattro}) can be solved exactly, and in the
next sections we will present numerically stable single-integral formulas
of the NLI PSD in such a case. The usefulness of these single-integral
formulas is that they provide a case against which numerical double-integration
routines of (\ref{eq:Glausgangsbasis}) can be checked for debugging.

\section{Single-channel / Nyquist-WDM systems }

\begin{figure}
\centering \foreignlanguage{english}{\includegraphics[width=0.23\linewidth]{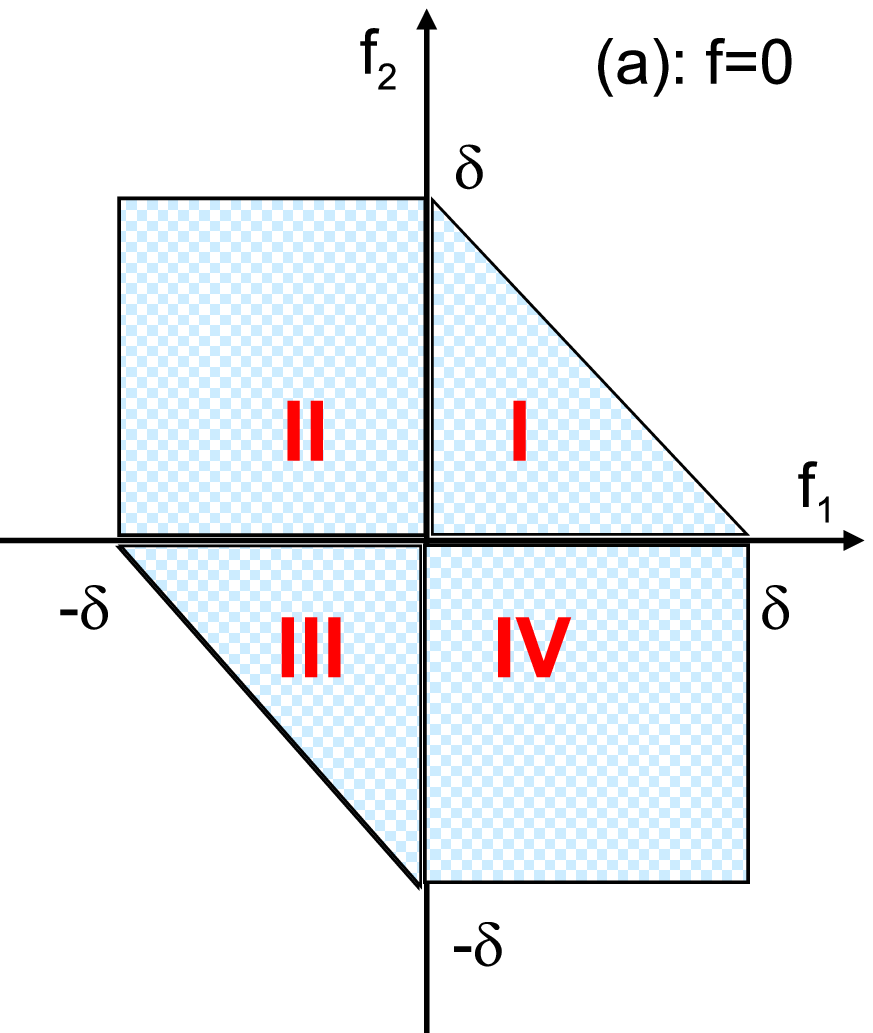}~~\includegraphics[width=0.33\linewidth]{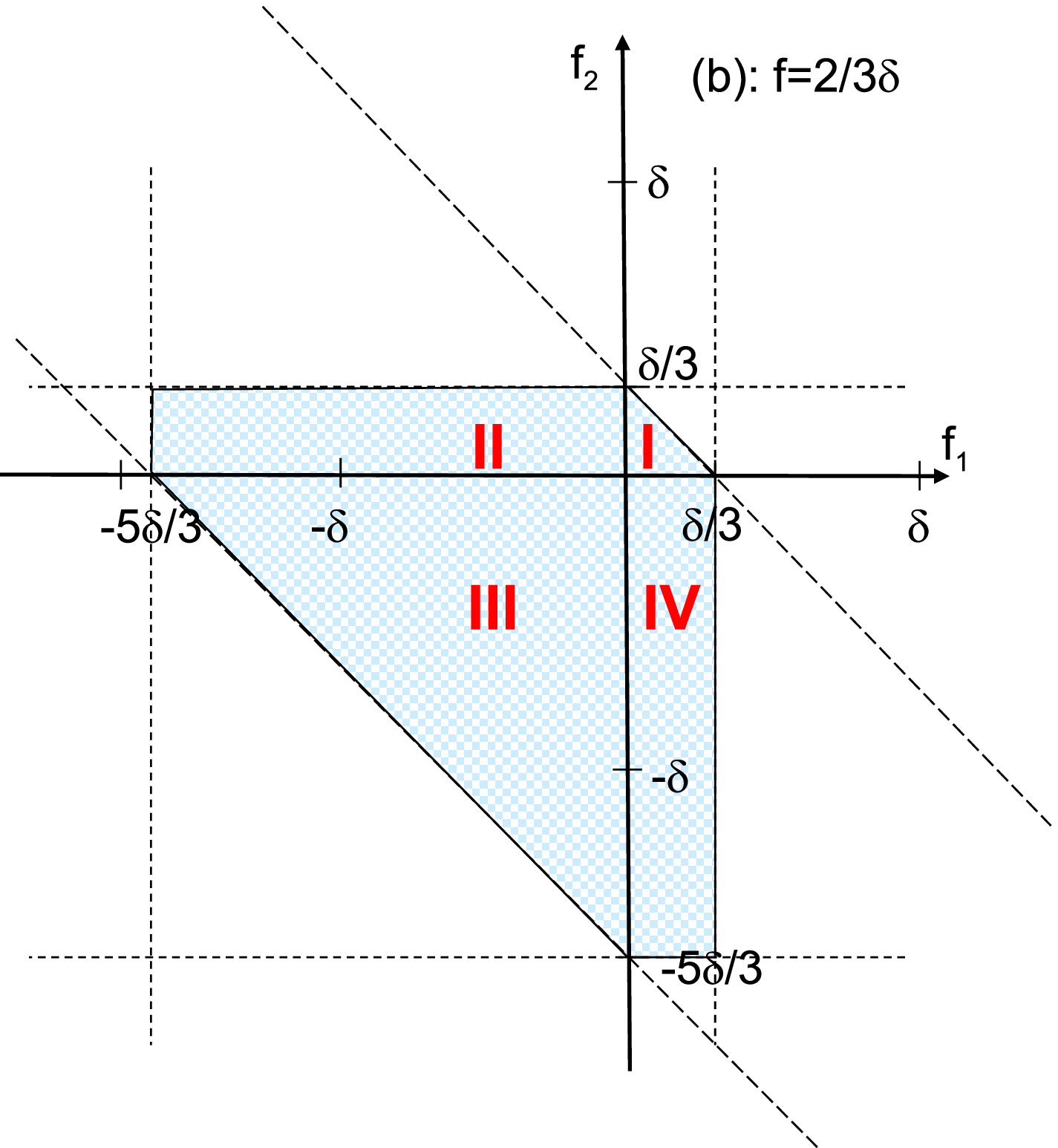}~~\includegraphics[width=0.35\linewidth]{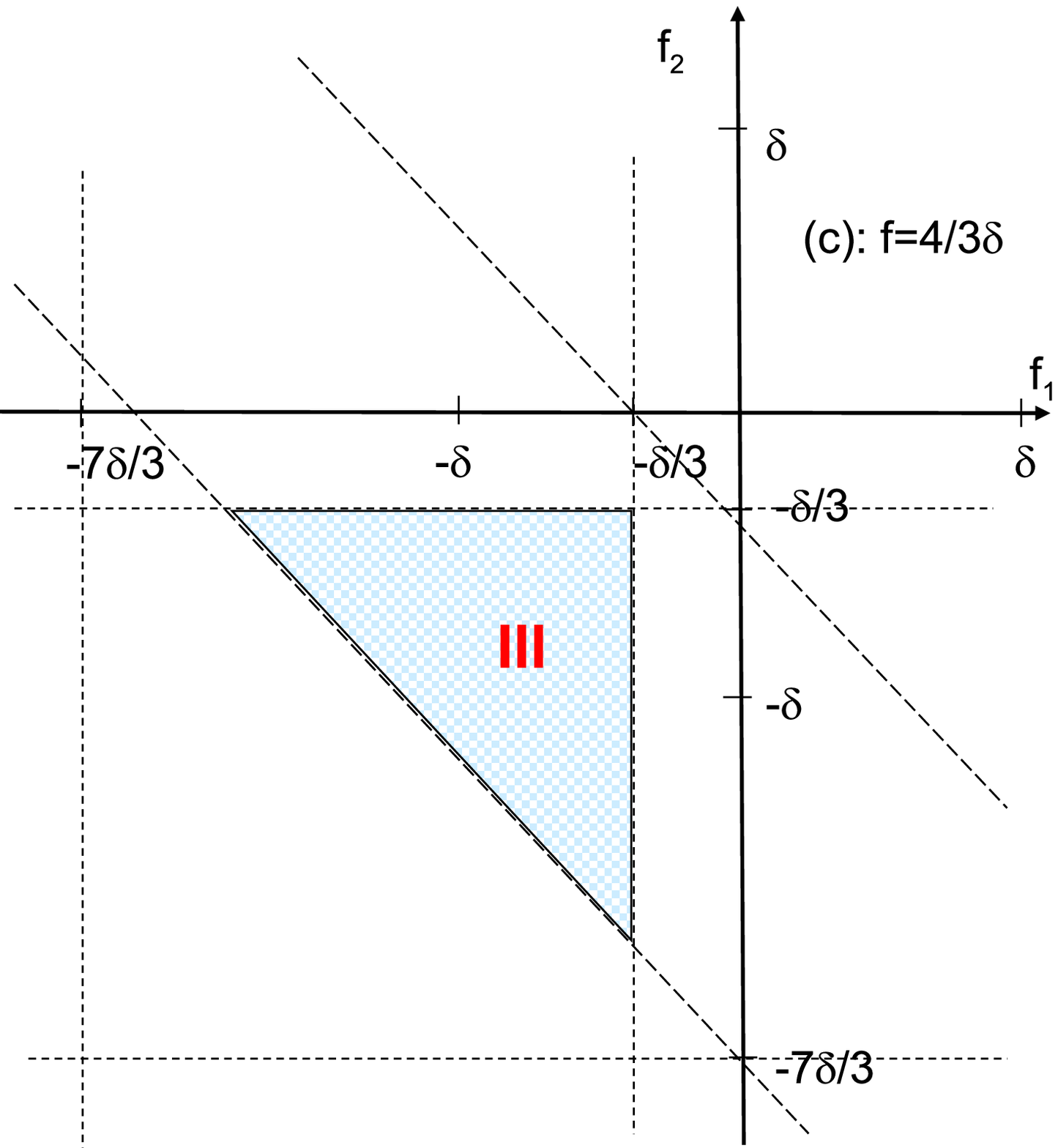}}
\caption{\label{fig:domains_abc} Domains over which integrand in (\ref{eq:Glausgangsbasis})
is non-zero when input PSD is a gate over $f\in[-\delta,\delta]$.
Integration over domains I trough IV yields the four lines (\ref{equno})-(\ref{eq:quattro})
in that order.}
 
\end{figure}
We tackle here the rectangular-spectrum single-channel case, or equivalently
the WDM case where no bandwidth gaps are present between neighboring
channels, known as the Nyquist-WDM case. The total power is $P$ and
the input PSD $G(f)=\frac{P}{2\delta}\mbox{rect}_{2\delta}(f+\delta)$
is a rectangular gate centered at $f=0$ with total two-sided bandwidth
$2\delta$. The integrand in (\ref{eq:Glausgangsbasis}) is non-zero
only over the shaded domains in quadrants I through IV shown in Fig.
\ref{fig:domains_abc} at several values of $f$. For instance, since
$|\mathcal{K}(v)|^{2}=|\mathcal{K}(-v)|^{2}$ for any kernel (\ref{eq:frequency-kernel}),
then the squared kernel is the same over the 4 quadrants, hence the
integral (\ref{eq:Glausgangsbasis}) over quadrants II and IV has
always the same value. Also, it is easy to see that the integrand
support disappears if $f>3\delta$. 

We can now state our main result on the PSD of the single-channel
interference (SCI):\\

\textbf{SCI Theorem} If the input channel has a rectangular PSD $G\left(f\right)=\frac{P}{2\delta}\mbox{rect}_{2\delta}(f+\delta)$
with bandwidth $2\delta$ and power $P$, then the PSD of the SCI
is given by (\ref{eq:Glausgangsbasis}). The normalized double integral
$\mathcal{I}(f):=I(f)/(P/(2\delta))^{3}$ can be exactly derived from
the $I(f)$ expression in (\ref{equno})-(\ref{eq:quattro}) as follows:

If $|f|<\delta$:
\begin{equation}
\begin{array}{rcl}
\mathcal{I}(f) & = & \int_{0}^{(\frac{\delta-f}{2})^{2}}|\mathcal{K}(v)|^{2}\ln\left(\frac{\frac{\delta-f}{2}+\sqrt{(\frac{\delta-f}{2})^{2}-v}}{\frac{\delta-f}{2}-\sqrt{(\frac{\delta-f}{2})^{2}-v}}\right)\mbox{d}v+2\int_{0}^{\delta^{2}-f^{2}}|\mathcal{K}(v)|^{2}\ln\left(\frac{\delta^{2}-f^{2}}{v}\right)\mbox{d}v\\
 & + & \int_{0}^{(\frac{\delta+f}{2})^{2}}|\mathcal{K}(v)|^{2}\ln\left(\frac{\frac{\delta+f}{2}+\sqrt{(\frac{\delta+f}{2})^{2}-v}}{\frac{\delta+f}{2}-\sqrt{(\frac{\delta+f}{2})^{2}-v}}\right)\mbox{d}v
\end{array}\label{eq:nyq_1}
\end{equation}

else if $\delta\leq|f|<3\delta$:
\begin{equation}
\begin{array}{rcl}
\mathcal{I}(f)=\int_{(|f|-\delta)^{2}}^{2\delta(|f|-\delta)}|\mathcal{K}(v)|^{2}\ln\left(\frac{v}{(|f|-\delta)^{2}}\right)\mbox{d}v+\int_{2\delta(|f|-\delta)}^{(\frac{\delta+|f|}{2})^{2}}|\mathcal{K}(v)|^{2}\ln\left(\frac{\frac{\delta+|f|}{2}+\sqrt{(\frac{\delta+|f|}{2})^{2}-v}}{\frac{\delta+|f|}{2}-\sqrt{(\frac{\delta+|f|}{2})^{2}-v}}\right)\mbox{d}v\end{array}\label{eq:nyq_2}
\end{equation}
otherwise \foreignlanguage{english}{$\mathcal{I}(f)=0$.}\\

For $f>0$, the first integral in (\ref{eq:nyq_1}) corresponds to
integration over domain I in Fig. \ref{fig:domains_abc}, the second
term to integration over domains II+IV, and the last term over domain
III. When \foreignlanguage{english}{$0<\delta\leq f<3\delta$ only
integration over domain III is nonzero. }

\selectlanguage{english}%
The proof is provided in Appendix A.

\selectlanguage{american}%

\subsection{Value at f=0}

\selectlanguage{english}%
From (\ref{eq:nyq_1}), the value at $f=0$ is found as:
\begin{equation}
\mathcal{I}(0)=2\int_{0}^{(\frac{\delta}{2})^{2}}|\mathcal{K}(v)|^{2}\ln\left(\frac{\frac{\delta}{2}+\sqrt{(\frac{\delta}{2})^{2}-v}}{\frac{\delta}{2}-\sqrt{(\frac{\delta}{2})^{2}-v}}\right)\mbox{d}v+2\int_{0}^{\delta^{2}}|\mathcal{K}(v)|^{2}\ln\left(\frac{\delta^{2}}{v}\right)\mbox{d}v.\label{eq:nyq_f0}
\end{equation}
Referring to \cite[Fig. 1]{poggio_solo} or \foreignlanguage{american}{Fig.
\ref{fig:domains_abc}}(a), the first term in the above sum corresponds
to integration of $|\mathcal{K}(f_{1}f_{2})|^{2}$ over the triangular
domains in the I+III quadrants of the $(f_{1},f_{2})$ plane, while
the second term to the square domains in quadrants II+IV.

\selectlanguage{american}%

\subsection{Examples and Cross-Checks}

\subsubsection{A theoretical cross check}

A simple theoretical example may be constructed by assuming the quadratic
kernel function to have a constant value $\mathcal{K}(0)\equiv1$
at all $f$. This physically corresponds to the zero dispersion case.
In this case 
\begin{equation}
\begin{split}I(f) & :=\iint_{-\infty}^{\infty}G(f+f_{1})G(f+f_{2})G(f+f_{1}+f_{2})\mbox{d}f_{1}\mbox{d}f_{2}\end{split}
\label{Glausgangsbasisexample-1}
\end{equation}
and the value of $\mathcal{I}(f):=I(f)/\left(\frac{P}{2\delta}\right)^{3}$
corresponds exactly to the areas of the integration domains sketched
in Fig. \ref{fig:domains_abc}. It can be readily seen from simple
geometrical considerations on Fig. \ref{fig:domains_abc} that 
\begin{equation}
\begin{split}\mathcal{I}(f)=\left\{ \begin{array}{ccc}
\frac{(\delta-|f|)^{2}}{2}+2\cdot(\delta^{2}-|f|^{2})+\frac{(\delta+|f|)^{2}}{2} & \text{if} & |f|\leq\delta\\
\frac{(3\delta-|f|)^{2}}{2} & \text{if} & \delta<|f|\leq3\delta\\
0 & \text{if} & |f|>3\delta.
\end{array}\right.\end{split}
\label{k1_example}
\end{equation}

Let's verify that the above expression indeed coincides with (\ref{eq:nyq_1})-(\ref{eq:nyq_2}).
Let's start with the following general result valid for $a>0$: 
\begin{equation}
\begin{split}\int\limits _{0}^{a^{2}}\mathrm{ln}\left(\frac{a+\sqrt{a^{2}-v}}{a-\sqrt{a^{2}-v}}\right)\,\mbox{d}v & =\left.2a^{2}-2a\sqrt{a^{2}-v}+\mathrm{ln}\left(\frac{a+\sqrt{a^{2}-v}}{a-\sqrt{a^{2}-v}}\right)\cdot v\right|_{0}^{a^{2}}\\
 & =2a^{2}+\mathrm{ln}\left(1\right)a^{2}-2a^{2}+2a\sqrt{a^{2}}=2a^{2}.
\end{split}
\label{Gl1int}
\end{equation}

By setting $a=\left(\frac{\delta-f}{2}\right)^{2}$, we get $\frac{(\delta-|f|)^{2}}{2}$
for the first integral in equation (\ref{eq:nyq_1}). Furthermore,
for $a>0$ we have 
\begin{equation}
\begin{split}\int\limits _{0}^{a}\mathrm{ln}\left(\frac{a}{v}\right)\,\mbox{d}v=\left.\mathrm{ln}\left(\frac{a}{v}\right)\cdot v+v\right|_{0}^{a}=a.\end{split}
\label{Gl2int}
\end{equation}

By setting $a=\delta^{2}-f^{2}$, we get $2(\delta^{2}-|f|^{2})$
for the second integral in equation (\ref{eq:nyq_1}). Finally by
setting $a=\frac{\delta+f}{2}$, the result in (\ref{Gl1int}) leads
to the value $\frac{(\delta+|f|)^{2}}{2}$ for the third integral.
All together, these lead to formula (\ref{k1_example}) for the case
$|f|<\delta$.

\rule{0pt}{1ex}

For the case $\delta<|f|\leq3\delta$ we may again use the integration
result (\ref{Gl1int}). For the first partial integral in (\ref{eq:nyq_2})
we have 
\begin{equation}
\begin{split}\int\limits _{b}^{a^{2}}\mathrm{ln}\left(\frac{a+\sqrt{a^{2}-v}}{a-\sqrt{a^{2}-v}}\right)\,\mbox{d}v & =\left.2a^{2}-2a\sqrt{a^{2}-v}+\mathrm{ln}\left(\frac{a+\sqrt{a^{2}-v}}{a-\sqrt{a^{2}-v}}\right)\cdot v\right|_{b}^{a^{2}}\\
 & =2a^{2}+\mathrm{ln}\left(1\right)a^{2}-2a^{2}+2a\sqrt{a^{2}-b}-\mathrm{ln}\left(\frac{a+\sqrt{a^{2}-b}}{a-\sqrt{a^{2}-b}}\right)\cdot b\\
 & =2a\sqrt{a^{2}-b}-\mathrm{ln}\left(\frac{a+\sqrt{a^{2}-b}}{a-\sqrt{a^{2}-b}}\right)\cdot b.
\end{split}
\label{Gl4int}
\end{equation}

Now 
\begin{equation}
\begin{split}a^{2}-b & =\left(\frac{\delta+|f|}{2}\right)^{2}-2\delta(|f|-\delta)\\
 & =\frac{1}{4}\left(\delta+2\delta|f|+|f|^{2}-8\delta|f|+8\delta^{2}\right)=\frac{1}{4}\left(|f|^{2}-6\delta|f|+9\delta^{2}\right)\\
 & =\left(\frac{1}{2}(3\delta-|f|)\right)^{2}
\end{split}
\label{Gl5int}
\end{equation}
and we thus derive from (\ref{Gl4int}): 
\begin{equation}
\begin{split} & \int\limits _{2\delta(|f|-\delta)}^{\left(\frac{\delta+|f|}{2}\right)^{2}}\mathrm{ln}\left(\frac{\frac{\delta+|f|}{2}+\sqrt{\left(\frac{\delta+|f|}{2}\right)^{2}-v}}{\frac{\delta+|f|}{2}-\sqrt{\left(\frac{\delta+|f|}{2}\right)^{2}-v}}\right)\,\mbox{d}v\\
 & \quad=2\frac{\delta+|f|}{2}\frac{1}{2}(3\delta-|f|)-\mathrm{ln}\left(\frac{\frac{\delta+|f|}{2}+\frac{1}{2}(3\delta-|f|)}{\frac{\delta+|f|}{2}-\frac{1}{2}(3\delta-|f|)}\right)\cdot2\delta(|f|-\delta)\\
 & =\frac{1}{2}(\delta+|f|)(3\delta-|f|)-2\mathrm{ln}\left(\frac{2\delta}{|f|-\delta}\right)\delta(|f|-\delta).
\end{split}
\label{Gl6int}
\end{equation}

For the second partial integral in (\ref{eq:nyq_2}) we get by setting
$a=(|f|-\delta)^{2}$ and $b=2\delta(|f|-\delta)$: 
\begin{equation}
\begin{split}\int\limits _{a}^{b}\mathrm{ln}\left(\frac{v}{a}\right)\,\mbox{d}v & =\left.\mathrm{ln}\left(\frac{v}{a}\right)\cdot v-v\right|_{a}^{b}=\mathrm{ln}\left(\frac{b}{a}\right)\cdot b-b-\mathrm{ln}\left(\frac{a}{a}\right)\cdot a+a\\
 & =\mathrm{ln}\left(\frac{b}{a}\right)\cdot b-b+a=2\mathrm{ln}\left(\frac{2\delta}{|f|-\delta}\right)\delta(|f|-\delta)-2\delta(|f|-\delta)+(|f|-\delta)^{2}.
\end{split}
\label{Gl7int}
\end{equation}

The sum of (\ref{Gl6int}) and (\ref{Gl7int}) gives: 
\begin{equation}
\begin{split} & \frac{1}{2}(\delta+|f|)(3\delta-|f|)-2\delta(|f|-\delta)+(|f|-\delta)^{2}\\
 & \quad=\frac{1}{2}(3\delta^{2}+2\delta|f|-|f|^{2})+3\delta^{2}-4\delta|f|+|f|^{2}\\
 & =\frac{1}{2}(9\delta^{2}-6\delta|f|+|f|^{2})=\frac{(3\delta-|f|)^{2}}{2}.
\end{split}
\label{Gl8int}
\end{equation}

This yields the formula (\ref{k1_example}) for the case $\delta<|f|<3\delta$.

\begin{figure}[h!]
\centering{}\includegraphics[width=0.35\textwidth]{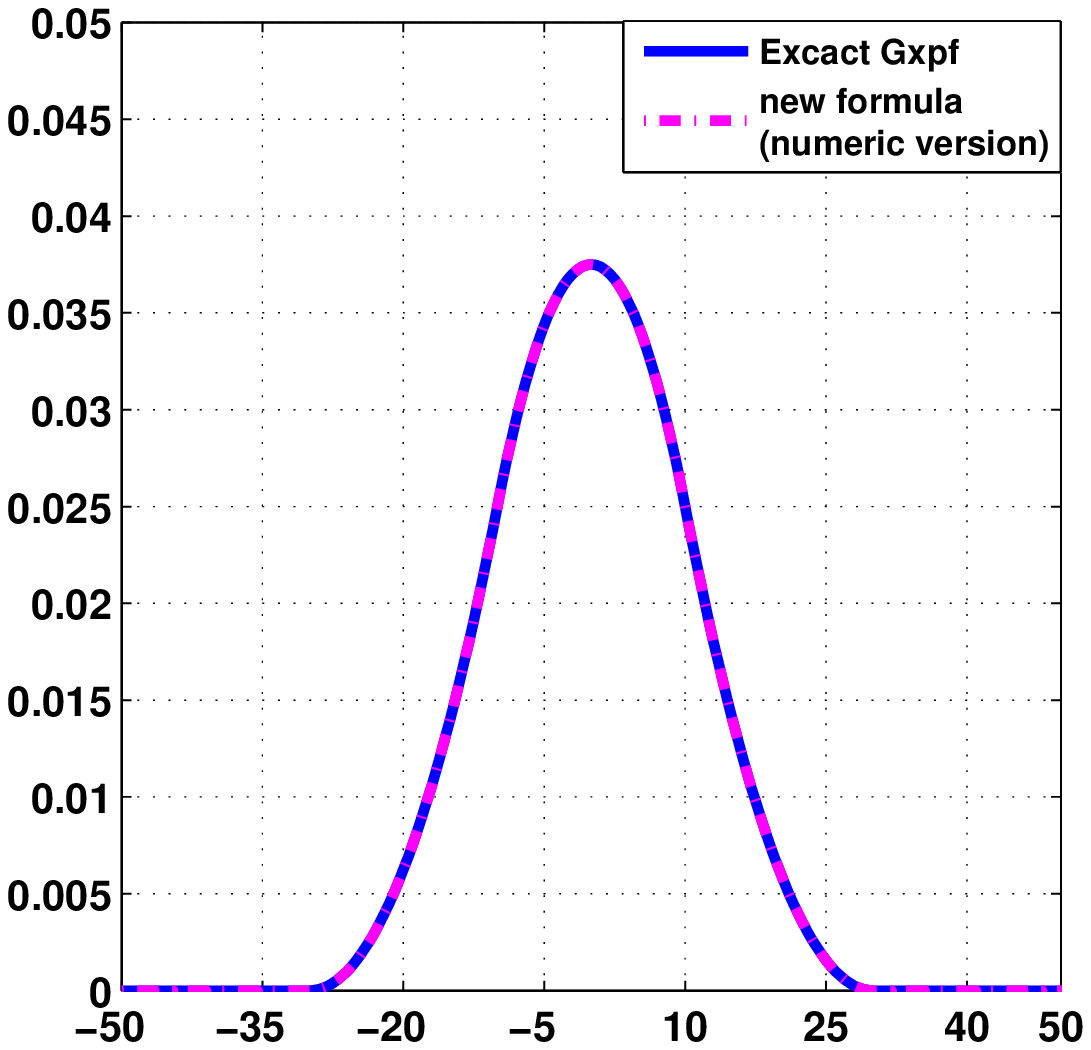}
\includegraphics[width=0.35\textwidth]{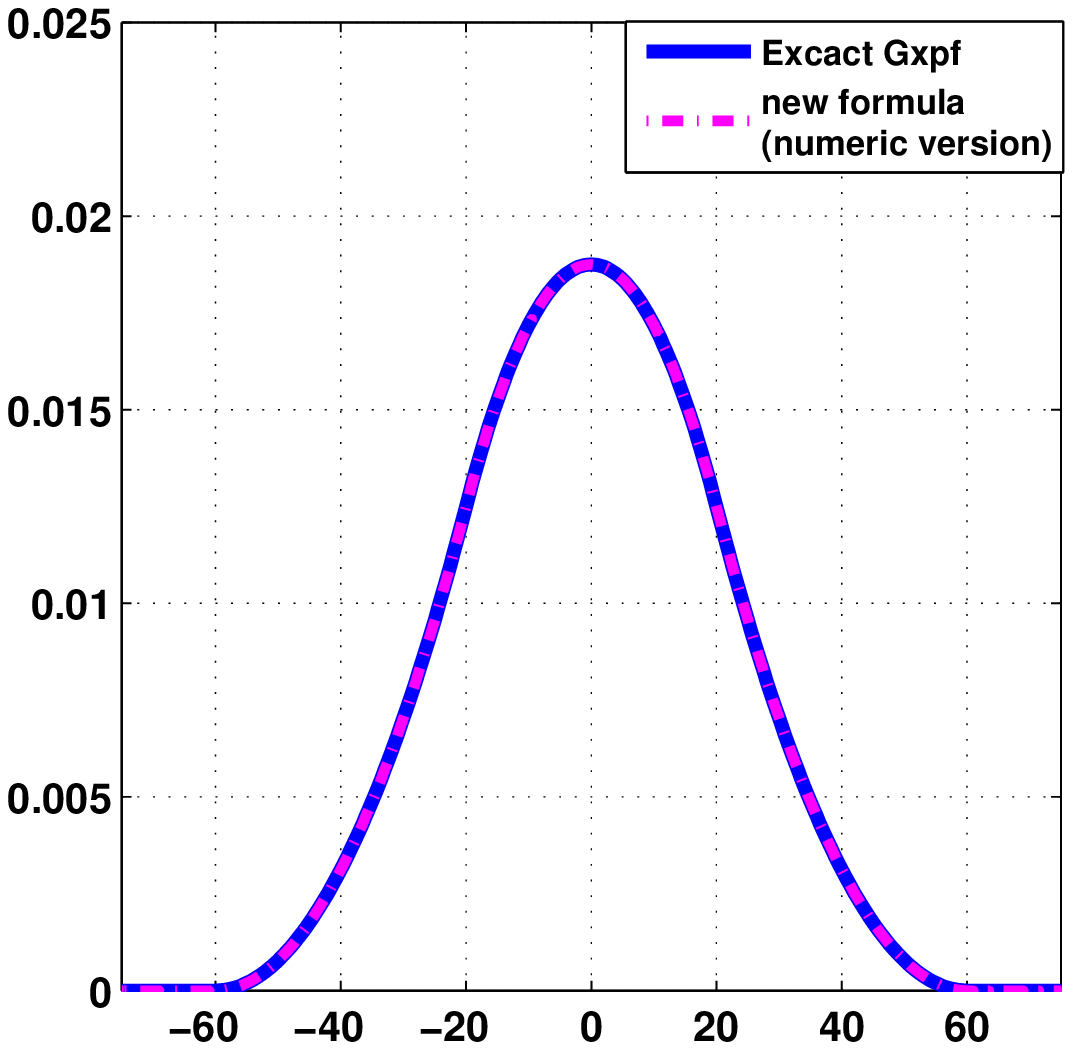} \caption{\label{fig:Examples-for-a}Plot of  $I(f)$ eq. (\ref{eq:Glausgangsbasis})
{[}mW/GHz{]} versus frequency {[}GHz{]} for a constant unit quadratic
kernel and rectangular input signal with $P=1$mW and support $[-10,10]$\foreignlanguage{english}{GHz}
(left) and support $[-20,20]$GHz (right). Label ``new formula'':
$\mathcal{I}(f)\cdot\left(\frac{P}{2\delta}\right)^{3}$, with $\mathcal{I}(f)$
as in (\ref{k1_example}). Label ``exact Gxpf'': direct numerical
evaluation of frequency double integral (\ref{eq:Glausgangsbasis}).}
\end{figure}

\rule{0pt}{1ex}

Fig. \ref{fig:Examples-for-a} illustrates this result for a rectangular
spectrum with support $[-10,10]\,\mathrm{GHz}$ (left figure) and
support $[-20,20]\,\mathrm{GHz}$ (right figure). The theoretical
result (\ref{eq:nyq_1})-(\ref{eq:nyq_2}) (labeled ``new formula'')
was cross-checked in these figures with an ad-hoc numerical double-integration
routine that we separately developed (labeled ``exact Gxpf'' in
the figures). The numerical routine greatly benefited from the explicit
formulas (\ref{eq:nyq_1})-(\ref{eq:nyq_2}) for debugging purposes.

\subsubsection{Numerical cross-checks}

The formulas (\ref{eq:nyq_1})-(\ref{eq:nyq_2}) have been cross-checked
also against numerical double-integration for realistic kernel functions.

We used a single-channel transmission over a 5-span dispersion-uncompensated
(DU) terrestrial link with 100 km fiber spans with dispersion \textbf{$17\,\mathrm{ps/nm/km}$
}(standard single mode (SMF) fiber) and attenuation \textbf{$0.2\,\mathrm{dB/km}$}.
The power was $P=1$ mW. 

\begin{figure}[h!]
\centering{}\includegraphics[width=0.35\textwidth]{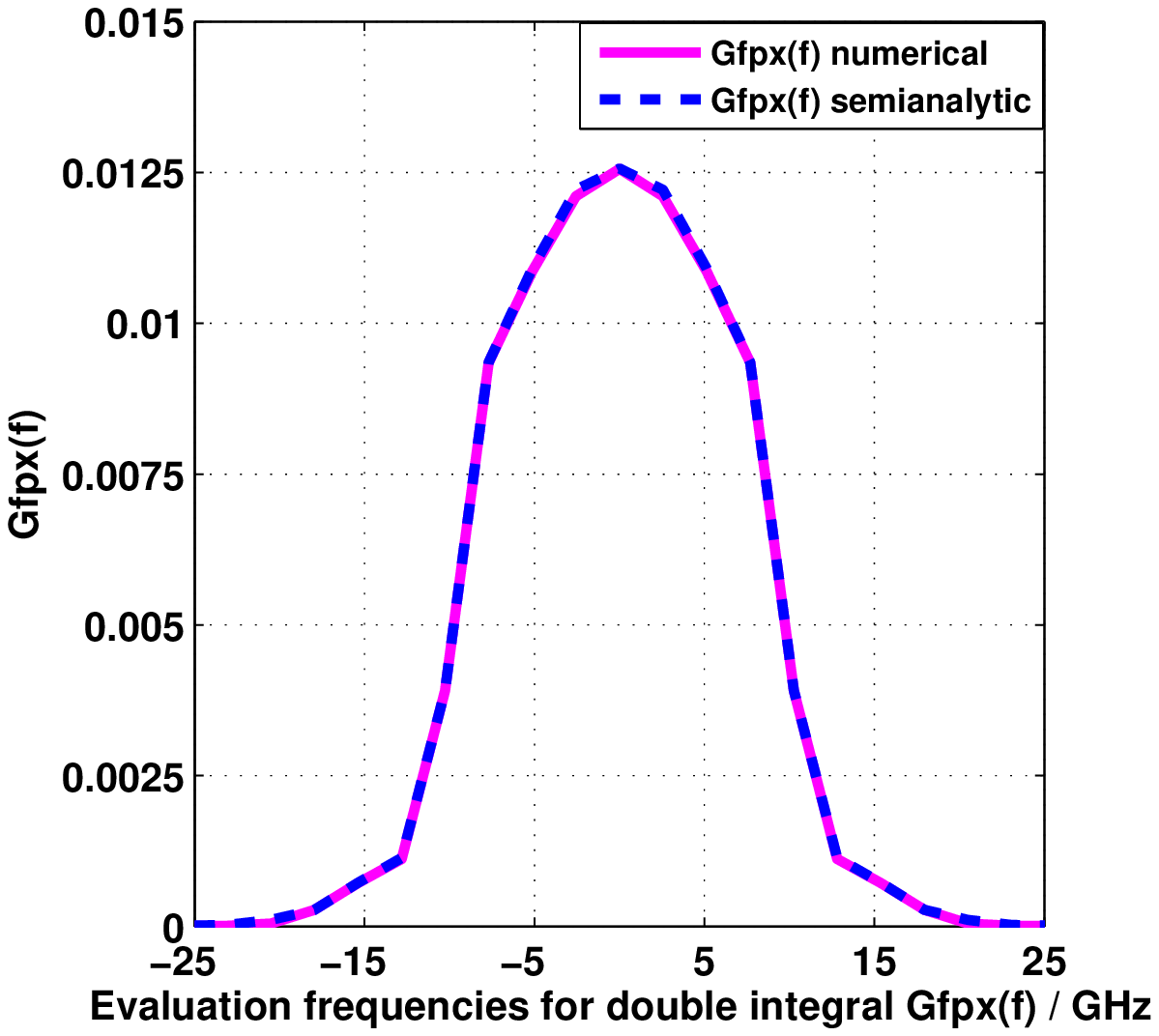}
\includegraphics[width=0.35\textwidth]{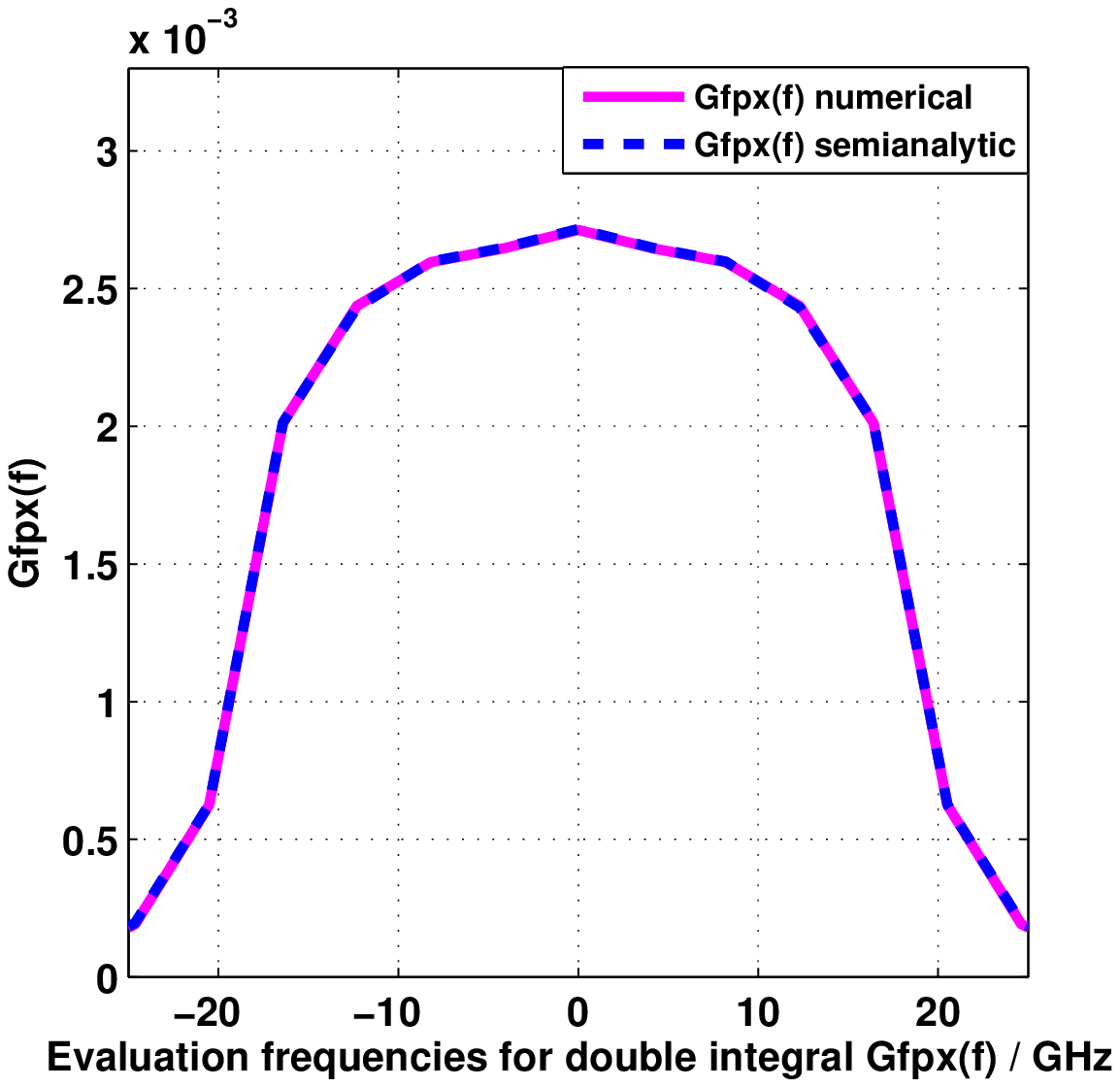} \caption{\label{fig:SCI-PSD-vs}$I(f)$ {[}mW/GHz{]} vs frequency {[}GHz{]}
for a rectangular input spectrum with $P=1$mW and support $[-10,10]\,\mathrm{GHz}$
(left) and support $[-20,20]\,\mathrm{GHz}$ (right) over a 5x100km
SMF DU link.}
\end{figure}

\begin{figure}[h!]
\centering{}\includegraphics[width=0.35\textwidth]{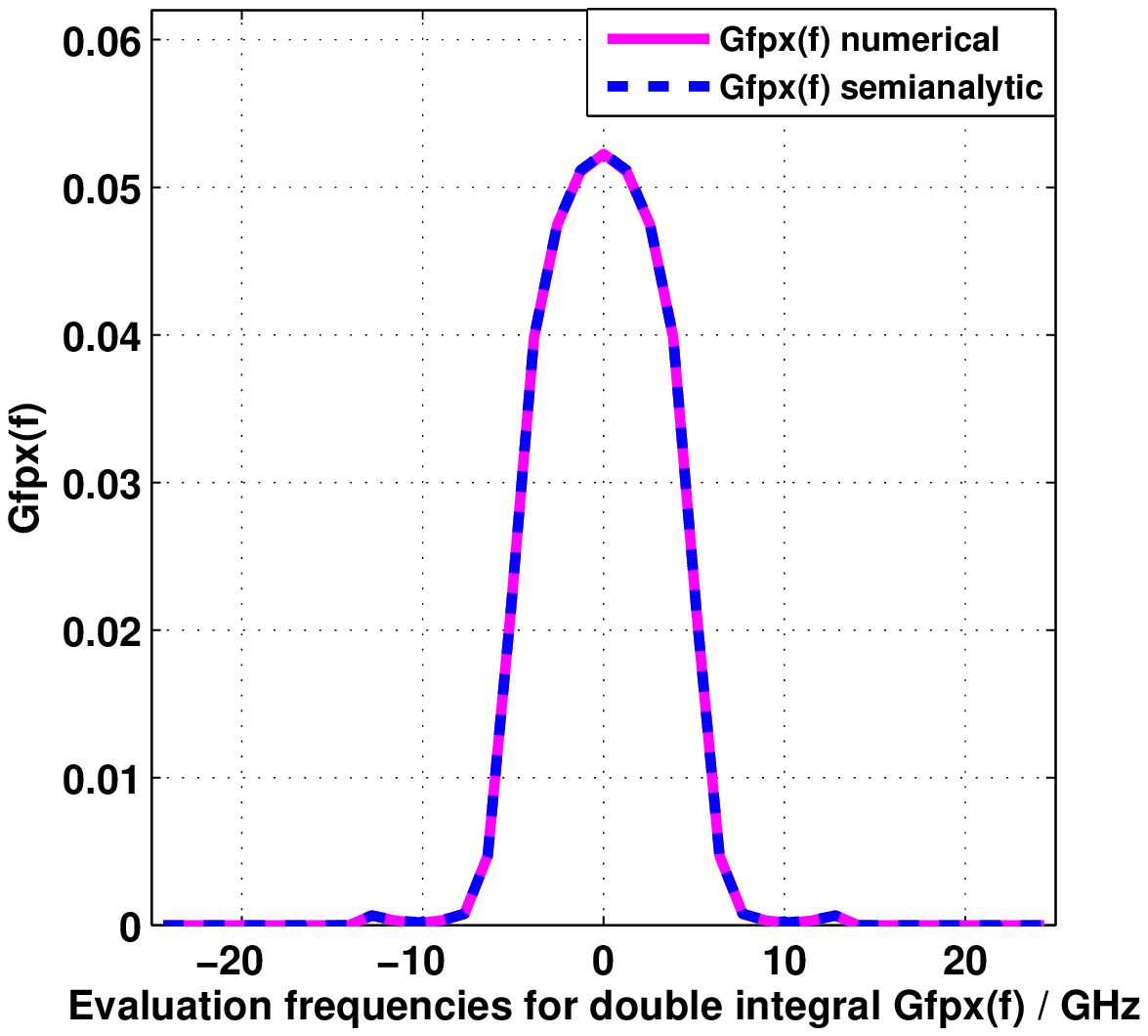}
\includegraphics[width=0.35\textwidth]{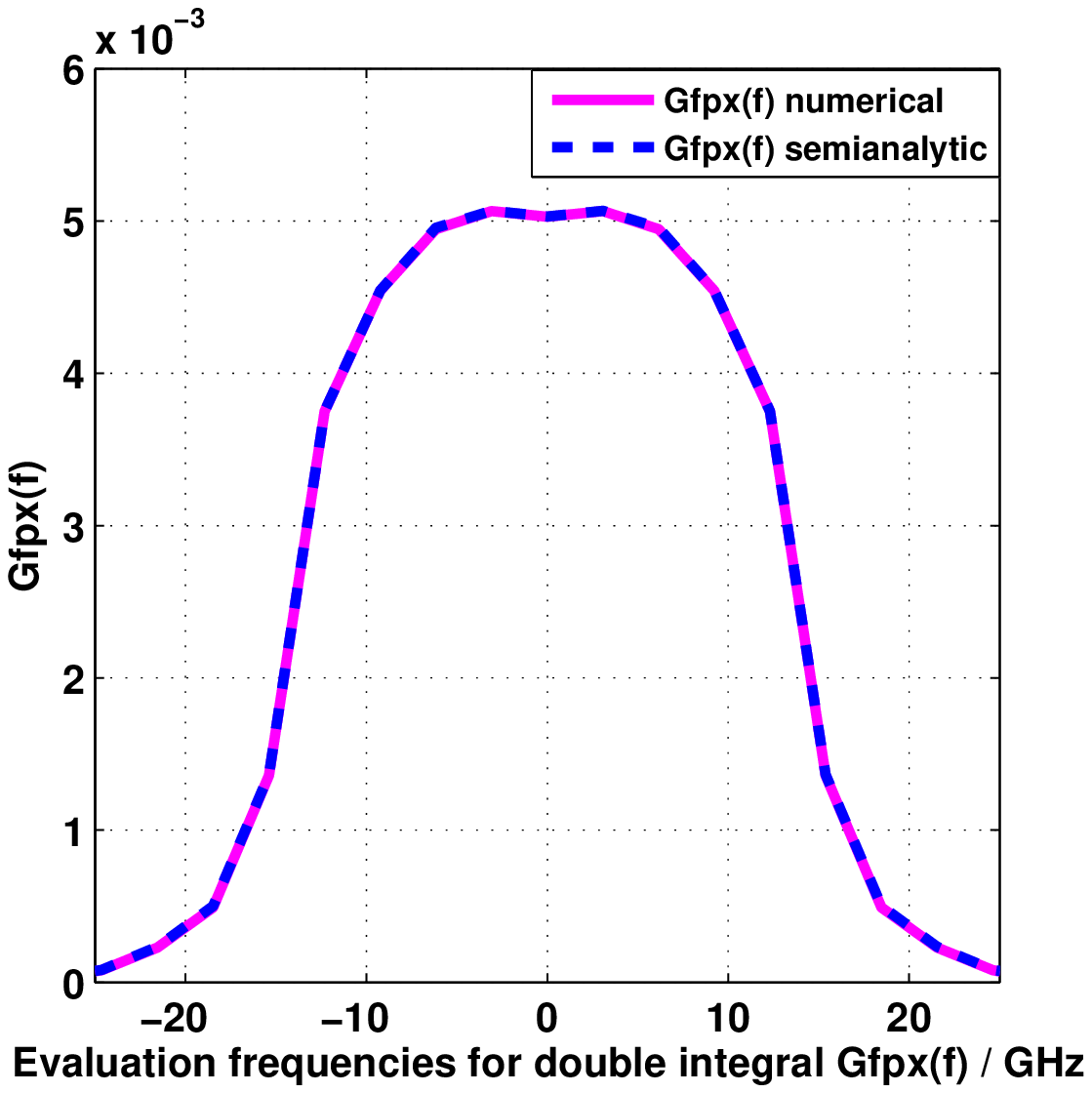} \caption{\label{fig:SCI-PSD-vs2}$I(f)$ {[}mW/GHz{]} vs frequency {[}GHz{]}
for a rectangular input spectrum with $P=1$mW and support $[-5,5]\,\mathrm{GHz}$
(left) and support $[-15,15]\,\mathrm{GHz}$ (right) over a 5x100km
SMF DU link.}
\end{figure}

Fig. \ref{fig:SCI-PSD-vs} and \ref{fig:SCI-PSD-vs2} show the SCI
PSD $G_{NLI}(f)/\frac{16}{27}=I(f)$ {[}mW/GHz{]} for a unit-power
rectangular input spectrum with various bandwidths. Again, theory
using (\ref{eq:nyq_1})-(\ref{eq:nyq_2}) (label ``semianalytical'')
was checked against direct numerical double-integration (label ``numerical'').

The examples show perfect coincidence between the numerical results
and the theory. Note that in all examples the numerical evaluation
of $G_{NLI}(f)$ was done at $39$ equidistant frequencies and took
between $230$ and $280$ seconds. The evaluation of $G_{NLI}(f)$
with the new semi-analytic formulas (\ref{eq:nyq_1})-(\ref{eq:nyq_2})
however took only between $0.3$ and $0.8$ seconds.

\subsubsection{Check of Nyquist-WDM}

\begin{figure}[h!]
\centering{}\includegraphics[width=0.5\textwidth]{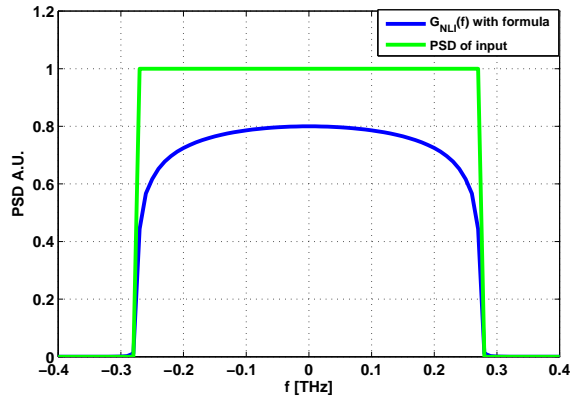} \caption{\label{fig:NY-SMF-system,-with}NY-SMF system, with 20 spans. Green
line: PSD of the transmitted signal $G_{WDM}$(f), equivalent to 17
Nyquist-WDM channels at 32 Gbaud. Blue line: PSD of NLI noise $G_{NLI}$
(f). Spectra arbitrarily rescaled as in \cite{poggio_solo}. }
\end{figure}

In \cite{poggio_solo}, Poggiolini presents an example of $G_{NLI}(f)$
in the Nyquist-WDM case over standard single-mode fiber (denoted as
NY-SMF in \cite{poggio_solo}) with 20 spans, an overall optical bandwidth
$B_{WDM}=544\,\mathrm{GHz}$, equivalent to $17$ Nyquist-WDM channels
at $32\,\mathrm{Gbaud}$. He used an WDM-input signal with an all
flat, i.e.\ rectangular shaped, PSD in the frequency band of $[-272,272]\,\mathrm{GHz}$.
Clearly, even if formulas (\ref{eq:nyq_1})-(\ref{eq:nyq_2}), as
stated above, are conceived for a single channel, they can be applied
to that particular case as well, because the $17$ Nyquist-WDM channels
may be identified with one single channel in the frequency band $[-272,272]\,\mathrm{GHz}$.

The NLI PSD $G_{NLI}(f)$ has been calculated with the new semi-analytic
formula and the result is depicted in Fig. \ref{fig:NY-SMF-system,-with}.
This result coincides exactly with that in \cite[Fig. 5]{poggio_solo}.
Once more, the result confirms the correctness of formulas (\ref{eq:nyq_1})-(\ref{eq:nyq_2}).

\selectlanguage{english}%

\section{Non-Nyquist WDM systems}

\selectlanguage{american}%
We assume here a WDM system with a reference central channel, $N_{c}$
channels to its left and $N_{c}$ channels to its right on the frequency
axis, with uniform frequency spacing $\Delta$. The WDM comb has input
PSD 
\begin{equation}
\begin{split}G\left(f\right)=\sum\limits _{k=-N_{c}}^{N_{c}}G_{k}(f):=\sum\limits _{k=-N_{c}}^{N_{c}}S(f-k\Delta).\end{split}
\label{GlrectshapedPSDWDM-1}
\end{equation}
where \emph{each} lowpass equivalent channel envelope has power $P$
and a rectangular PSD with bandwidth $2\delta$, namely $S(f)=\frac{P}{2\delta}\mbox{rect}_{2\delta}(f+\delta)$.
The Nyquist-WDM case has $2\delta=\Delta$. When channels do not spectrally
overlap and have guard-bands, we have the traditional Non-Nyquist
WDM system, for which $2\delta<\Delta$. 

Substitution of (\ref{GlrectshapedPSDWDM-1}) in (\ref{eq:Glausgangsbasis})
yields: 
\begin{equation}
\begin{split}I(f) & =\iint_{-\infty}^{\infty}\left|\mathcal{K}(f_{1}f_{2})\right|^{2}G(f+f_{1})G(f+f_{2})G(f+f_{1}+f_{2})\mbox{d}f_{1}\mbox{d}f_{2}\\
 & =\iint_{-\infty}^{\infty}\left|\mathcal{K}(f_{1}f_{2})\right|^{2}\sum\limits _{k=-N_{c}}^{N_{c}}\sum\limits _{l=-N_{c}}^{N_{c}}\sum\limits _{m=-N_{c}}^{N_{c}}G_{k}(f+f_{1})G_{l}(f+f_{2})G_{m}(f+f_{1}+f_{2})\mbox{d}f_{1}\mbox{d}f_{2}\\
 & =\sum\limits _{k,l,m=-N_{c}}^{N_{c}}\iint_{-\infty}^{\infty}\left|\mathcal{K}(f_{1}f_{2})\right|^{2}G_{k}(f+f_{1})G_{l}(f+f_{2})G_{m}(f+f_{1}+f_{2})\mbox{d}f_{1}\mbox{d}f_{2}.
\end{split}
\label{Glausgangsbasisfolow-1}
\end{equation}
\begin{figure}[h!]
\centering{}\includegraphics[width=0.7\textwidth]{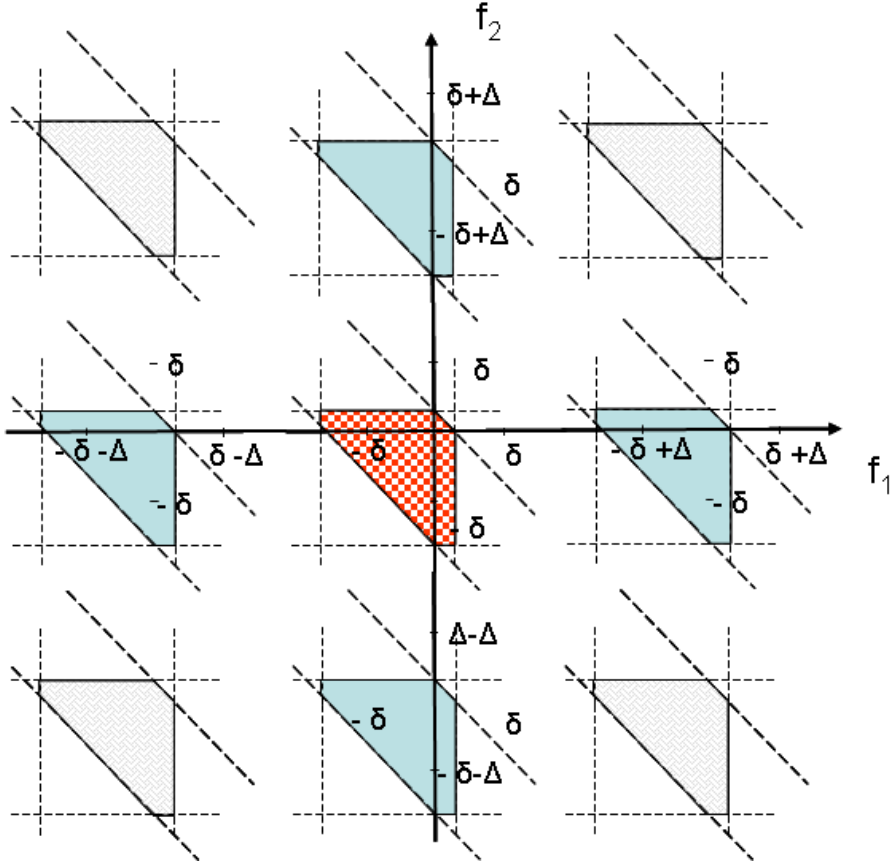} \caption{\label{AbbXPMBeispiel-1} Example of XCI (blue) and SCI (red) integration
domains for a 3-channel WDM system with spacing $\Delta$ and rectangular
channel spectra with bandwidth $2\delta$. The off-axes domains correspond
to MCI (i.e., FWM).}
\end{figure}
 In practice we have broken up the global integral into the sum of
partial integrals over special integration domains or ``islands''.
Fig. \ref{AbbXPMBeispiel-1} shows such domains, where the integrand
$G(\cdot)G(\cdot)G(\cdot)>0$ for rectangular channel spectra, a channel
spacing $\Delta=50\,\mathrm{GHz}$, a per-channel bandwidth $2\delta$
of $40\,\mathrm{GHz}$, a frequency $f=\frac{2\delta}{3}\,\mathrm{GHz}$
and $N_{c}=1$ adjacent channel, i.e., a 3-channel WDM system. The
set of integration 'islands' for rectangular spectra is also presented
in the special case $f=0$ in \foreignlanguage{english}{\cite[Fig. 3]{poggio_solo}}.
Since integration is additive over the islands, the NLI PSD may be
decomposed as the sum of single-channel interference (SCI), cross-channel
interference (XCI) and multi-channel interference (MCI, also known
as four-wave mixing (FWM)) \foreignlanguage{english}{\cite{poggio_solo}}:
\begin{equation}
G_{NLI}(f)=G_{SCI}(f)+G_{XCI}(f)+G_{MCI}(f).\label{eq:_NL_decomposition-1}
\end{equation}
Integration over the central red island in \foreignlanguage{english}{\cite[Fig. 3]{poggio_solo}}
corresponds to the SCI and can directly be obtained from (\ref{eq:nyq_1}),(\ref{eq:nyq_2}).

\subsection{Cross-Channel-Interference (XCI)}

Consider now only the case $k=0$ and its symmetric case $l=0$. For
this portion of the NLI we get: 
\begin{equation}
\begin{split}I(f)= & 2\sum\limits _{l,m=-N_{c}}^{N_{c}}\iint_{-\infty}^{\infty}\left|\mathcal{K}(f_{1}f_{2})\right|^{2}G_{0}(f+f_{1})G_{l}(f+f_{2})G_{m}(f+f_{1}+f_{2})\mbox{d}f_{1}\mbox{d}f_{2}.\end{split}
\label{GlausgangsbasisfolowXPM-1}
\end{equation}
Note that if $m\neq l$ the support of $G_{m}(f+f_{1}+f_{2})$ never
intersects the support of the other two terms, and thus the contribution
is zero. So we may simplify (\ref{GlausgangsbasisfolowXPM-1}) to:
\begin{equation}
\begin{split}I(f) & =2\sum\limits _{m=-N_{c}}^{N_{c}}\iintop_{-\infty}^{\infty}\left|\mathcal{K}(f_{1}f_{2})\right|^{2}G_{0}(f+f_{1})G_{m}(f+f_{2})G_{m}(f+f_{1}+f_{2})\mbox{d}f_{1}\mbox{d}f_{2}.\end{split}
\label{GlausgangsbasisfolowXPMsimlp-1}
\end{equation}

If we also exclude the term for $m=0$ (which represents the SCI),
then we get the cross-channel interference (XCI \cite{poggio_solo})
contribution to $I(f)$. XCI  encompasses both scalar cross-phase
modulation and cross-polarization modulation \cite{bon_ofc_11}. In
summary, the XCI PSD is given by 
\begin{equation}
\begin{array}{rcl}
G_{XCI}(f) & = & \frac{16}{27}I_{XCI}(f)\\
I_{XCI}(f) & := & 2\sum_{m=1}^{N_{c}}(I_{m}(f)+I_{-m}(f))
\end{array}\label{eq:G_XCI-1}
\end{equation}
where $I_{m}$ is defined as 
\begin{equation}
I_{m}(f):=\iintop_{-\infty}^{\infty}|\mathcal{K}(f_{1}f_{2})|^{2}G_{0}(f+f_{1})G_{m}(f+f_{2})G_{m}(f+f_{1}+f_{2})\mbox{d}f_{1}\mbox{d}f_{2}.\label{eq:Im_f-1}
\end{equation}

After the usual change of variable, such an integral can be written
as

\begin{eqnarray}
I_{m}(f)= & \int_{0}^{\infty}|\mathcal{K}(v)|^{2} & \left[\int_{0}^{\infty}\frac{1}{u}G_{0}(f+u)G_{m}(f+\frac{v}{u})G_{m}(f+u+\frac{v}{u})\mbox{d}u\right.\label{equno-1}\\
 &  & +\int_{0}^{\infty}\frac{1}{u}G_{0}(f-u)G_{m}(f+\frac{v}{u})G_{m}(f-u+\frac{v}{u})\mbox{d}u\label{eqdue-1}\\
 &  & +\int_{0}^{\infty}\frac{1}{u}G_{0}(f-u)G_{m}(f-\frac{v}{u})G_{m}(f-u-\frac{v}{u})\mbox{d}u\label{eq:tre-1}\\
 &  & \left.+\int_{0}^{\infty}\frac{1}{u}G_{0}(f+u)G_{m}(f-\frac{v}{u})G_{m}(f+u-\frac{v}{u})\mbox{d}u\right]\mbox{d}v.\label{eq:quattro-1}
\end{eqnarray}
We can now state our main result on the XCI spectrum.\\

\textbf{XCI Theorem} If the input WDM system has a symmetric PSD $G\left(f\right)=\sum\limits _{k=-N_{c}}^{N_{c}}S(f-k\Delta)$
with channel spacing $\Delta$, a rectangular per-channel spectrum
$S(f)=\frac{P}{2\delta}\mbox{rect}_{2\delta}(f+\delta)$ with bandwidth
$2\delta$ and per-channel power $P$, then for any integer $m>0$
the normalized double integral $\mathcal{I}_{m}(f)\triangleq(I_{m}(f)+I_{-m}(f))/(P/2\delta)^{3}$
can be written as follows. Define
\begin{equation}
\begin{array}{ccc}
\eta:=\delta-|f| & \mbox{and} & \epsilon:=\delta+|f|\\
\eta_{m}^{+}:=m\Delta+\eta & \mbox{and} & \epsilon_{m}^{+}:=m\Delta+\epsilon\\
\eta_{m}^{-}:=m\Delta-\eta & \mbox{and} & \epsilon_{m}^{-}:=m\Delta-\epsilon.
\end{array}\label{eq:Defs}
\end{equation}

Then, if $|f|<\delta$:
\begin{equation}
\begin{array}{rcl}
\mathcal{I}_{m}(f) & = & \int_{0}^{\eta\epsilon_{m}^{-}}|\mathcal{K}(v)|^{2}\ln\left(\frac{\frac{v}{\epsilon_{m}^{-}}}{\frac{\eta_{m}^{+}}{2}-\sqrt{(\frac{\eta_{m}^{+}}{2})^{2}-v}}\right)\mbox{d}v+\int_{\eta\epsilon_{m}^{-}}^{\eta m\Delta}|\mathcal{K}(v)|^{2}\ln\left(\frac{\eta}{\frac{\eta_{m}^{+}}{2}-\sqrt{(\frac{\eta_{m}^{+}}{2})^{2}-v}}\right)\mbox{d}v\\
 & + & \int_{0}^{\eta m\Delta}|\mathcal{K}(v)|^{2}\ln\left(\frac{-\frac{\eta_{m}^{-}}{2}+\sqrt{(\frac{\eta_{m}^{-}}{2})^{2}+v}}{\frac{v}{\epsilon_{m}^{+}}}\right)\mbox{d}v+\int_{\eta m\Delta}^{\eta\epsilon_{m}^{+}}|\mathcal{K}(v)|^{2}\ln\left(\frac{\eta\epsilon_{m}^{+}}{v}\right)\mbox{d}v\\
 & + & \int_{0}^{\epsilon m\Delta}|\mathcal{K}(v)|^{2}\ln\left(\frac{-\frac{\epsilon_{m}^{-}}{2}+\sqrt{(\frac{\epsilon_{m}^{-}}{2})^{2}+v}}{\frac{v}{\eta_{m}^{+}}}\right)\mbox{d}v+\int_{\epsilon m\Delta}^{\epsilon\eta_{m}^{+}}|\mathcal{K}(v)|^{2}\ln\left(\frac{\epsilon\eta_{m}^{+}}{v}\right)\mbox{d}v\\
 & + & \int_{0}^{\epsilon\eta_{m}^{-}}|\mathcal{K}(v)|^{2}\ln\left(\frac{\frac{v}{\eta_{m}^{-}}}{\frac{\epsilon_{m}^{+}}{2}-\sqrt{(\frac{\epsilon_{m}^{+}}{2})^{2}-v}}\right)\mbox{d}v+\int_{\epsilon\eta_{m}^{-}}^{\epsilon m\Delta}|\mathcal{K}(v)|^{2}\ln\left(\frac{\epsilon}{\frac{\epsilon_{m}^{+}}{2}-\sqrt{(\frac{\epsilon_{m}^{+}}{2})^{2}-v}}\right)\mbox{d}v
\end{array}\label{eq:n_nyq_1}
\end{equation}
else if $\delta\leq|f|<3\delta$:
\begin{equation}
\begin{array}{rcl}
\mathcal{I}_{m}(f)=\int_{-\eta(\epsilon_{m}^{-}-\eta)}^{-\eta\eta_{m}^{+}}|\mathcal{K}(v)|^{2}\ln\left(\frac{\frac{\epsilon_{m}^{-}}{2}-\sqrt{(\frac{\epsilon_{m}^{-}}{2})^{2}+v}}{\eta}\right)\mbox{d}v+\int_{-\eta\eta_{m}^{+}}^{2\delta\eta_{m}^{+}}|\mathcal{K}(v)|^{2}\ln\left(\frac{-\frac{\epsilon_{m}^{-}}{2}+\sqrt{(\frac{\epsilon_{m}^{-}}{2})^{2}+v}}{\frac{v}{\eta_{m}^{+}}}\right)\mbox{d}v\\
+\int_{-\eta\eta_{m}^{-}}^{-\eta(\epsilon_{m}^{+}+\eta)}|\mathcal{K}(v)|^{2}\ln\left(\frac{v}{-\eta\eta_{m}^{-}}\right)\mbox{d}v+\int_{-\eta(\epsilon_{m}^{+}+\eta)}^{2\delta\eta_{m}^{-}}|\mathcal{K}(v)|^{2}\ln\left(\frac{\frac{v}{\eta_{m}^{-}}}{\frac{\epsilon_{m}^{+}}{2}-\sqrt{(\frac{\epsilon_{m}^{+}}{2})^{2}-v}}\right)\mbox{d}v
\end{array}\label{eq:n_nyq_2}
\end{equation}

otherwise \foreignlanguage{english}{$\mathcal{I}_{m}(f)=0$. }\\

\selectlanguage{english}%
The details of the proof can be found in\foreignlanguage{american}{
Appendix B.}

\selectlanguage{american}%

\subsection{Value at f=0}

\selectlanguage{english}%
As a corollary, the value at $f=0$ is found from (\ref{eq:n_nyq_1})
as follows.\foreignlanguage{american}{ Define
\[
\begin{array}{ccc}
\Delta_{m}^{+}:=m\Delta+\delta & \mbox{and} & \Delta_{m}^{-}:=m\Delta-\delta.\end{array}
\]
Then, 
\begin{equation}
\begin{array}{r}
\mathcal{I}_{m}(0)=2\{\int_{0}^{\delta\Delta_{m}^{-}}|\mathcal{K}(v)|^{2}\ln\left(\frac{\frac{v}{\Delta_{m}^{-}}}{\frac{\Delta_{m}^{+}}{2}-\sqrt{(\frac{\Delta_{m}^{+}}{2})^{2}-v}}\right)\mbox{d}v+\int_{\delta\Delta_{m}^{-}}^{\delta m\Delta}|\mathcal{K}(v)|^{2}\ln\left(\frac{\delta}{\frac{\Delta_{m}^{+}}{2}-\sqrt{(\frac{\Delta_{m}^{+}}{2})^{2}-v}}\right)\mbox{d}v\\
+\int_{0}^{\delta m\Delta}|\mathcal{K}(v)|^{2}\ln\left(\frac{-\frac{\Delta_{m}^{-}}{2}+\sqrt{(\frac{\Delta_{m}^{-}}{2})^{2}+v}}{\frac{v}{\Delta_{m}^{+}}}\right)\mbox{d}v+\int_{\delta m\Delta}^{\delta\Delta_{m}^{+}}|\mathcal{K}(v)|^{2}\ln\left(\frac{\delta\Delta_{m}^{+}}{v}\right)\mbox{d}v\}.
\end{array}\label{eq:Im0}
\end{equation}
}

\selectlanguage{american}%
\begin{figure}
\centering \foreignlanguage{english}{\includegraphics[width=0.4\linewidth]{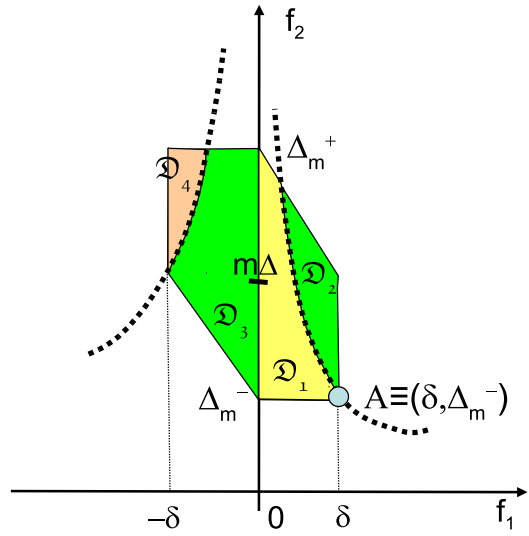}}
\caption{\label{fig:4_regions} The four domains corresponding to the four
terms in (\foreignlanguage{english}{\ref{eq:Im0}}) in the order they
appear. For instance, $\iint_{\mathcal{D}_{4}}|\mathcal{K}(f_{1}f_{2})|^{2}\mbox{d}f_{1}\mbox{d}f_{2}=\int_{\delta m\Delta}^{\delta\Delta_{m}^{+}}|\mathcal{K}(v)|^{2}\ln\left(\frac{\delta\Delta_{m}^{+}}{v}\right)\mbox{d}v.$
Dashed curves are hyperbolas $f_{1}f_{2}=const$. }
 
\end{figure}
\foreignlanguage{english}{Fig. \ref{fig:4_regions} shows a geometric
interpretation of the 4 terms in the curly bracket in (\ref{eq:Im0})
as the integral of $|\mathcal{K}(f_{1}f_{2})|^{2}$ over the shown
domains $\mathcal{D}_{1}$ through $\mathcal{D}_{4}$ in the $(f_{1},f_{2})$
plane (in the order they appear in eq. (\ref{eq:Im0})).}

\subsection{Examples and Cross Checks}

\subsubsection{A theoretical cross check}

\begin{figure}[h!]
\centering{}\includegraphics[width=0.35\textwidth]{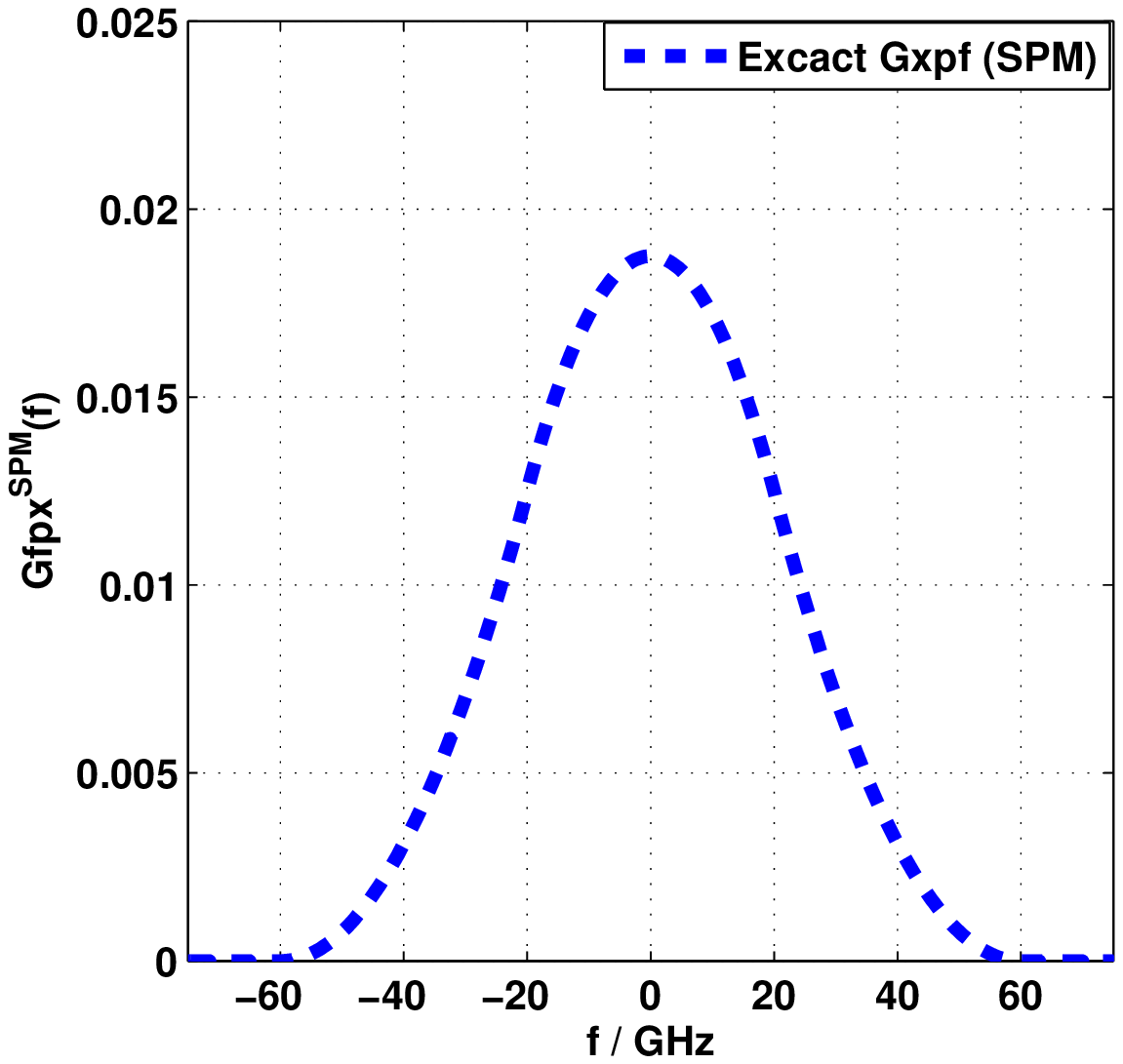}
\includegraphics[width=0.365\textwidth]{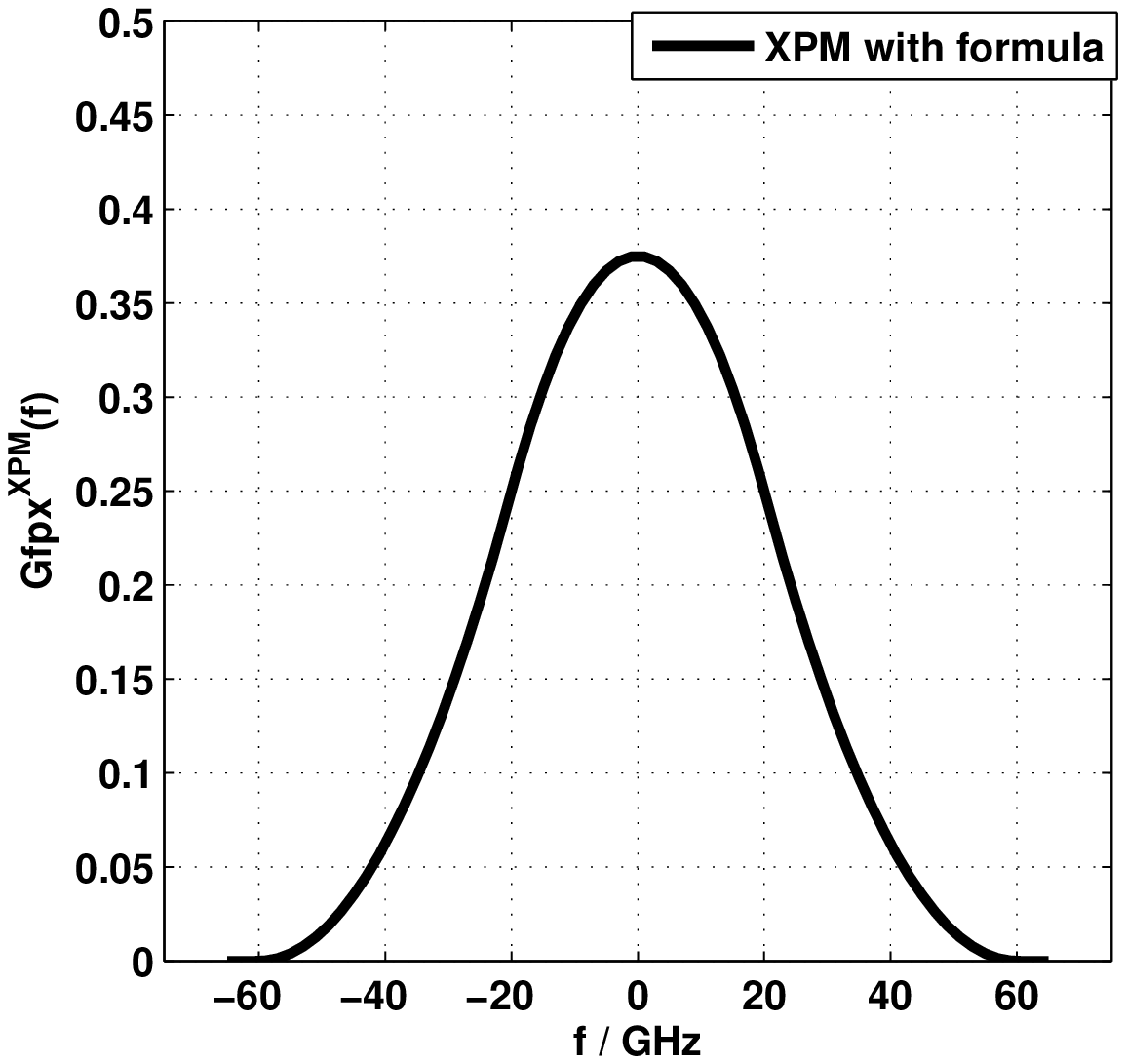} \caption{\label{AbbXPMBeispiel2} (left) Function $G_{SCI}(f)$ and (right)
function $G_{XCI}(f)$ for $N_{c}=5$ adjacent channels.}
\end{figure}

When the quadratic kernel has a constant value $1$, then the double
integral is proportional to the area of the integration islands. As
seen in Fig. \ref{AbbXPMBeispiel-1}, such islands all have the same
area. Hence $G_{XCI}(f)$ in this case is simply $4N_{c}$-times the
value $G_{SCI}(f)=\left(\frac{P}{2\delta}\right)^{3}\mathcal{I}(f)$,
with $\mathcal{I}(f)$ as given in (\ref{k1_example}), since there
are $2N_{c}$ XCI islands on every axis. Fig. \ref{AbbXPMBeispiel2}
shows the calculation of the theoretical $G_{SPM}(f)$ and $G_{XPM}(f)$
with a unity squared kernel and $N_{c}=5$ adjacent channels. Note
that the scale on the y-axis of the XPM-figure is $20=4N_{c}$-times
larger than that of the SPM-figure.

\subsubsection{Numerical cross-checks}

The formulas (\ref{eq:n_nyq_1})-(\ref{eq:n_nyq_2}) have been cross-checked
also against numerical double-integration for realistic kernel functions.

\begin{figure}[h!]
\begin{centering}
\includegraphics[width=0.33\textwidth]{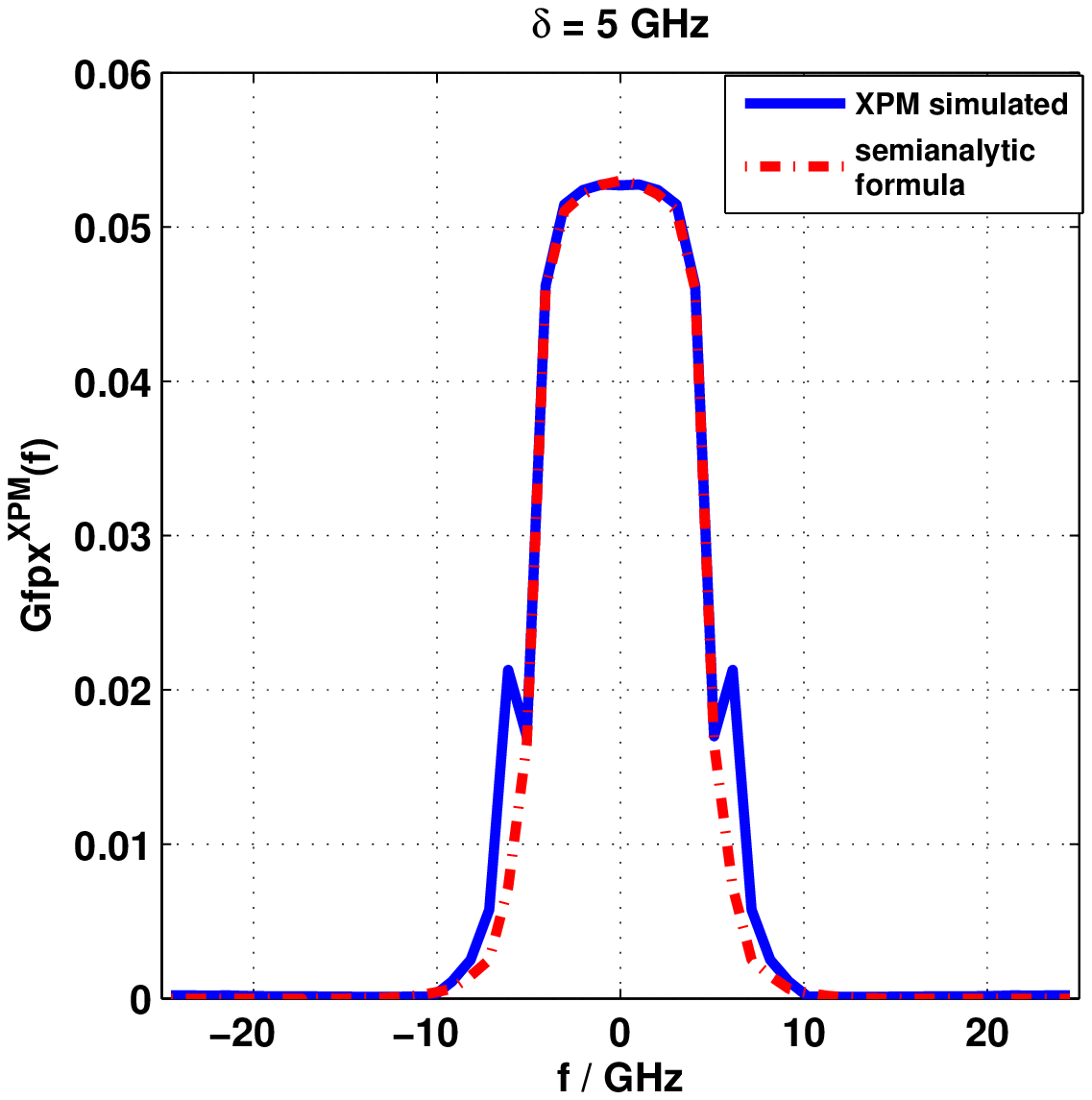}
\includegraphics[width=0.33\textwidth]{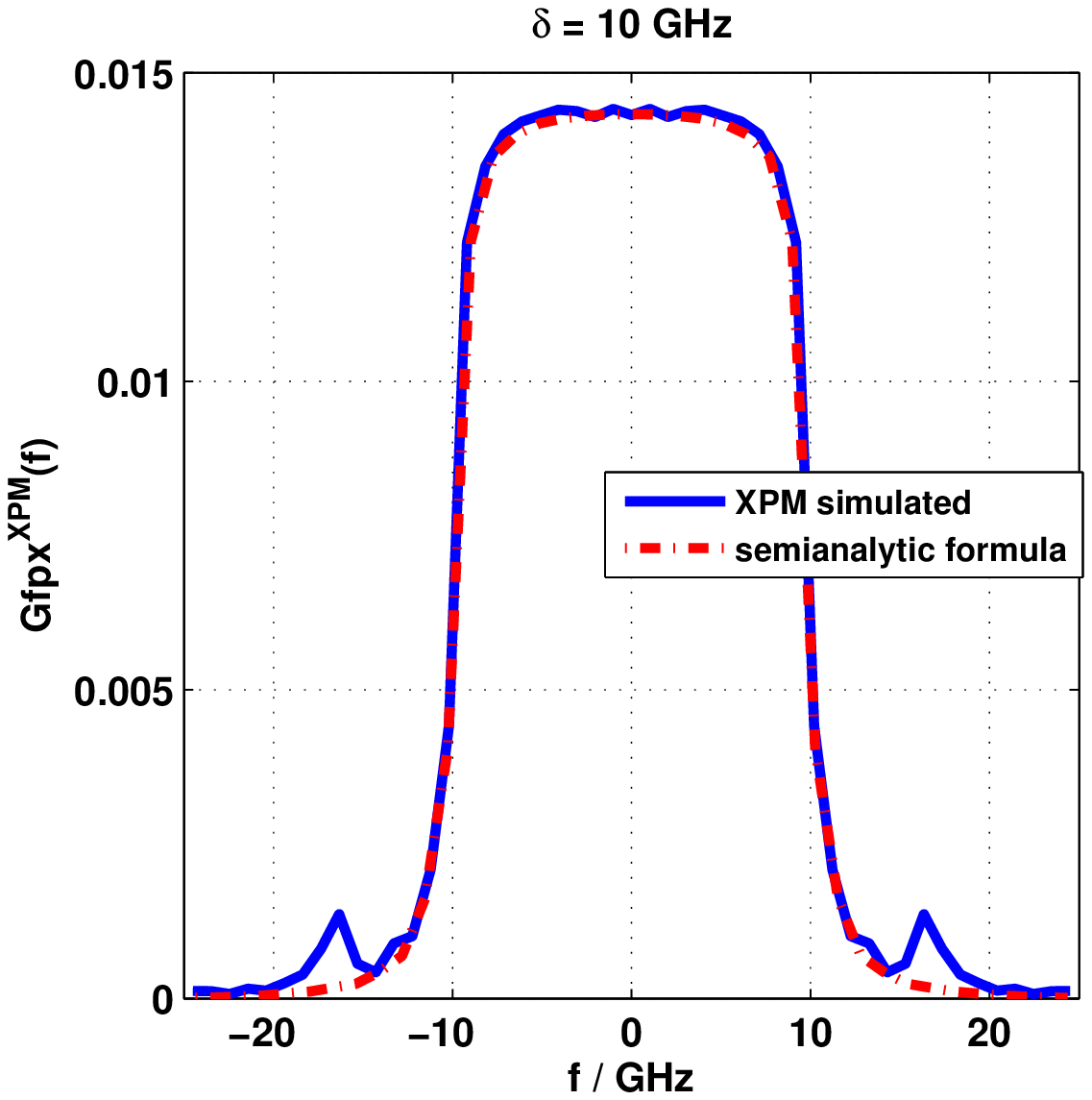}
\par\end{centering}

\centering{}\includegraphics[width=0.33\textwidth]{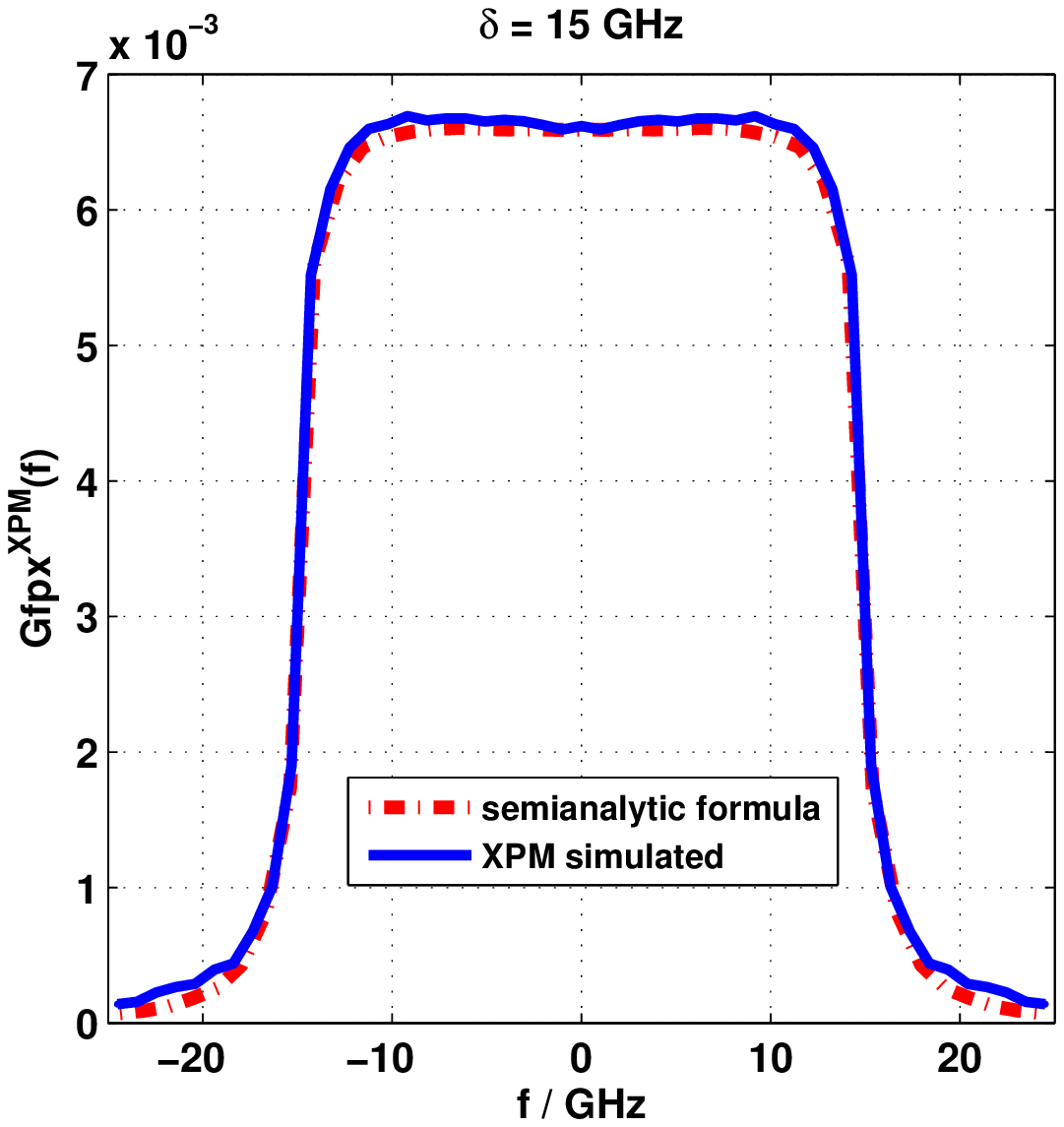}
\includegraphics[width=0.33\textwidth]{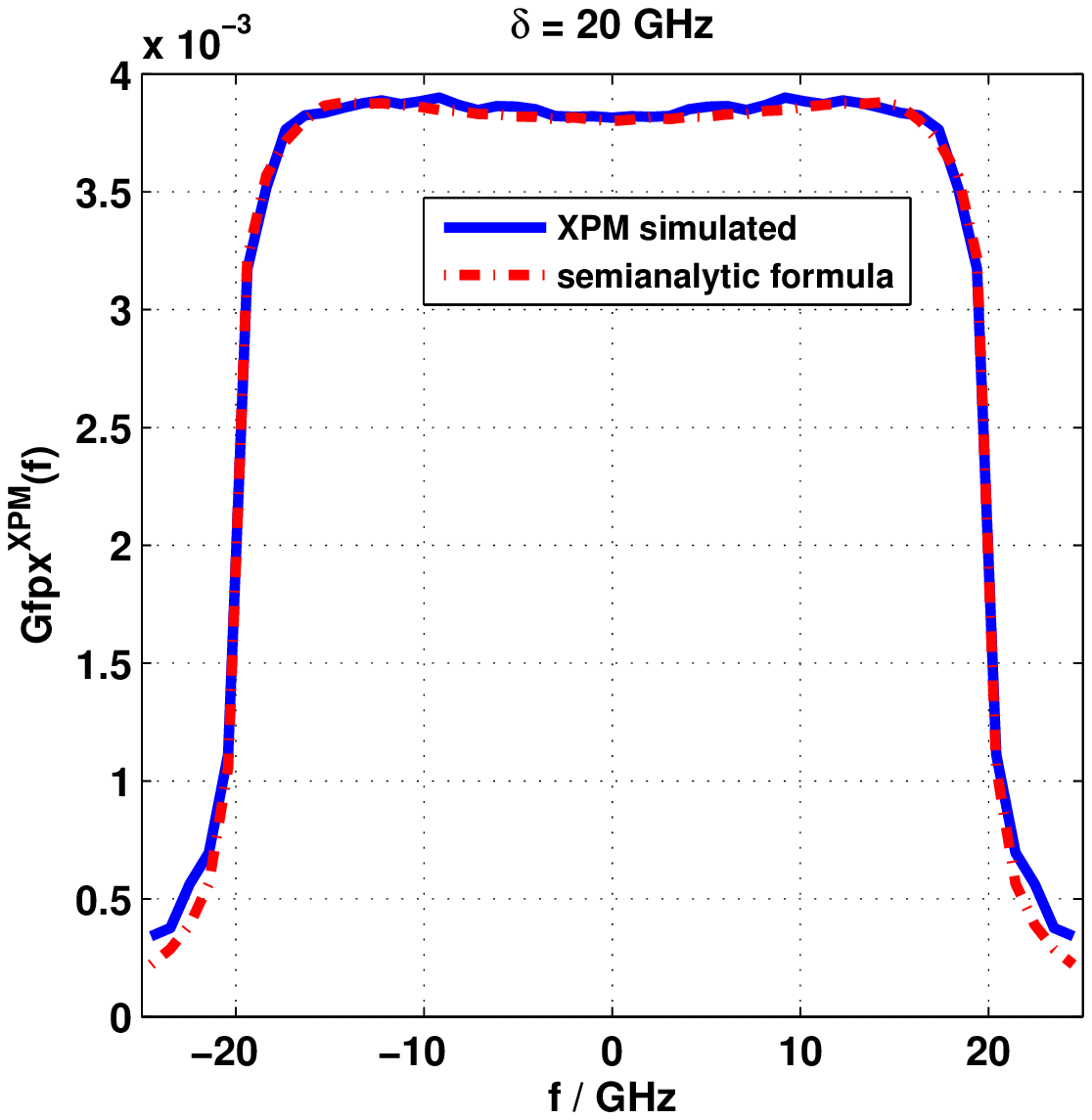}
\caption{\label{fig:-XCI-PSD} XCI PSD on central channel vs frequency for
rectangular signal spectra, with support $[-5,5]\,\mathrm{GHz}$ (left)
and support $[-20,20]\,\mathrm{GHz}$ (right) over a 5x100km SMF DU
link. Spacing $\Delta=50$GHz, 11 channels ($N_{c}=5$).}
\end{figure}

We used an 11-channel ($N_{c}=5$) WDM non-Nyquist transmission with
spacing $\Delta=50$GHz over a 5-span dispersion-uncompensated terrestrial
link with 100 km fiber spans with dispersion \textbf{$17\,\mathrm{ps/nm/km}$
}and attenuation \textbf{$0.2\,\mathrm{dB/km}$}. The power per channel
was $P=1$ mW. Figure \ref{fig:-XCI-PSD} shows the XCI PSD $G_{XCI}(f)/\frac{16}{27}=I_{XCI}(f)$
{[}mW/GHz{]} for rectangular per-channel input spectra with various
bandwidths. Theory using (\ref{eq:n_nyq_1})-(\ref{eq:n_nyq_2}) (label
``semianalytic formula'') was checked against direct numerical double-integration
(label ``XPM simulated''). Some discrepancies between theory and
numerical double integration are visible in the figures. We later
found that the double integration routine had mis-convergence problems,
that were finally fixed to perfectly match with theory. 

\section{Conclusions}

We have presented new semi-analytical power spectral density formulas
of the received nonlinear interference, both for single-channel and
cross-channel interference. The great value of these formulas is twofold: 

1) they represent a benchmark against which more general GNRF solvers
can be tested; 

2) it is now possible to easily analyze the separate behavior of SCI
and XCI in order to quickly find out the dominant nonlinear effect
\cite{bon_ofc_11} in highly-dispersed nonlinear coherent transmissions.
This second aspect will be developed in a future publication.

\section*{Acknowledgments}

The present paper is a synthesis due to the first author of two reports
\cite{ottmar_spm,ottmar_xpm} due to the second author, both written
at the end of his 6-month sabbatical leave at the Department of Information
Engineering of Parma University, Italy. The authors gladly acknowledge
discussions on the developments of this work with Dr. Paolo Serena
and Dr. Nicolaos Mantzoukis.

\section*{Appendix A: Proof of SCI Integral $\mathcal{I}(f)$}

In this Appendix we prove the expressions of the SCI integral $\mathcal{I}(f)$
given in (\ref{eq:nyq_1})-(\ref{eq:nyq_2}). By the symmetry (\ref{Glausgangsymmetrie-1-1})
we only need calculations at $f\geq0$.

\subsection*{Calculation of partial integral (\ref{equno}), quadrant I:}

Regarding the integrand of the inner integral (\ref{equno}) we deduce:
\begin{equation}
\begin{split}G\left(f+u\right)\neq0\quad\Longleftrightarrow\quad f+u\leq\delta\quad\Longleftrightarrow\quad u\leq\delta-f:=\tilde{\delta}.\end{split}
\label{GlrectshapedPSDded1-1}
\end{equation}

Note that this implies for the following analysis of (\ref{equno})
that $\tilde{\delta}:=\delta-f>0$ since by definition $u\geq0$.
Otherwise the factor $G\left(f+u\right)$ is zero and the integral
(\ref{equno}) disappears. This implies also that integral (\ref{equno})
disappears for $f>\delta$ (Cfr Fig. \ref{fig:domains_abc}c). For
the second factor we have: 
\begin{equation}
\begin{split}G\left(f+\frac{v}{u}\right)\neq0\quad\Longleftrightarrow\quad f+\frac{v}{u}\leq\delta\quad\Longleftrightarrow\quad u\geq\frac{v}{\delta-f}=\frac{v}{\tilde{\delta}}.\end{split}
\label{GlrectshapedPSDded2-1}
\end{equation}

Note that we used $\tilde{\delta}:=\delta-f>0$ in the last transformation
of the inequality. For the third factor we have (note $u>0$): 
\begin{equation}
\begin{split}G\left(f+u+\frac{v}{u}\right)\neq0\quad\Longleftrightarrow\quad u+\frac{v}{u}\leq\tilde{\delta}\quad\Longleftrightarrow\quad u^{2}-\tilde{\delta}u+v\leq0.\end{split}
\label{GlrectshapedPSDded3}
\end{equation}

Since 
\begin{equation}
\begin{split}u^{2}-\tilde{\delta}u+v\leq0\quad & \Longleftrightarrow\quad\left(u-\frac{\tilde{\delta}}{2}\right)^{2}-\left(\frac{\tilde{\delta}}{2}\right)^{2}+v\leq0\\
\quad & \Longleftrightarrow\quad\left(u-\frac{\tilde{\delta}}{2}\right)^{2}\leq\left(\frac{\tilde{\delta}}{2}\right)^{2}-v
\end{split}
\label{GlrectshapedPSDded3b}
\end{equation}
the factor $G\left(f+u+\frac{v}{u}\right)$ is always $0$ if $\left(\frac{\tilde{\delta}}{2}\right)^{2}<v$.
If $\left(\frac{\tilde{\delta}}{2}\right)^{2}\geq v$ then (\ref{GlrectshapedPSDded3b})
has solutions and 
\begin{equation}
\begin{split}\left(u-\frac{\tilde{\delta}}{2}\right)^{2} & \leq\left(\frac{\tilde{\delta}}{2}\right)^{2}-v\\
\quad\Longleftrightarrow\quad u & \leq\sqrt{\left(\frac{\tilde{\delta}}{2}\right)^{2}-v}+\frac{\tilde{\delta}}{2}:=u^{(1)}\quad\text{and}\quad u\geq-\sqrt{\left(\frac{\tilde{\delta}}{2}\right)^{2}-v}+\frac{\tilde{\delta}}{2}:=u^{(0)}.
\end{split}
\label{GlrectshapedPSDded3c}
\end{equation}

Thus using (\ref{GlrectshapedPSDded1-1}), (\ref{GlrectshapedPSDded2-1})
and (\ref{GlrectshapedPSDded3c}) the partial integral (\ref{equno})
reads for $f<\delta$: 
\begin{equation}
\begin{split}\int\limits _{0}^{\infty} & \left|\mathcal{K}(v)\right|^{2}\left[\int\limits _{0}^{\infty}\frac{1}{u}\cdot G\left(f+u\right)G\left(f+\frac{v}{u}\right)G\left(f+u+\frac{v}{u}\right)\,\mbox{d}u\right]\,\mbox{d}v\\
 & =\int\limits _{0}^{\left(\frac{\tilde{\delta}}{2}\right)^{2}}\left|\mathcal{K}(v)\right|^{2}\left[\int\limits _{\max\left\{ {u^{(0)},\frac{v}{\tilde{\delta}}}\right\} }^{\min\left\{ {u^{(1)},\tilde{\delta}}\right\} }\frac{1}{u}\cdot G\left(f+u\right)G\left(f+\frac{v}{u}\right)G\left(f+u+\frac{v}{u}\right)\,\mbox{d}u\right]\,\mbox{d}v.
\end{split}
\label{GlpartintegIvereinfacht}
\end{equation}

Note once more that for $f\geq\delta$ the partial integral (\ref{equno})
is zero. Since 
\begin{equation}
\begin{split}u^{(1)}=\sqrt{\left(\frac{\tilde{\delta}}{2}\right)^{2}-v}+\frac{\tilde{\delta}}{2}\leq\frac{\tilde{\delta}}{2}+\frac{\tilde{\delta}}{2}=\tilde{\delta}\end{split}
\label{GlGrenzenteilint1}
\end{equation}
and 
\begin{equation}
\begin{split}u^{(0)} & =-\sqrt{\left(\frac{\tilde{\delta}}{2}\right)^{2}-v}+\frac{\tilde{\delta}}{2}\geq\frac{v}{\tilde{\delta}}\quad\Longleftrightarrow\quad\left(\frac{\tilde{\delta}}{2}-\frac{v}{\tilde{\delta}}\right)^{2}\geq\left(\frac{\tilde{\delta}}{2}\right)^{2}-v\\
\quad & \Longleftrightarrow\quad-v+\left(\frac{v}{\tilde{\delta}}\right)^{2}\geq-v
\end{split}
\label{GlGrenzenteilint1p2}
\end{equation}
is always true, then the integral limits for the first inner integral
are $u^{(0)}$ and $u^{(1)}$. We thus get for $f<\delta$: 
\begin{equation}
\begin{split}\int\limits _{0}^{\infty} & \left|\mathcal{K}(v)\right|^{2}\left[\int\limits _{0}^{\infty}\frac{1}{u}\cdot G\left(f+u\right)G\left(f+\frac{v}{u}\right)G\left(f+u+\frac{v}{u}\right)\,\mbox{d}u\right]\,\mbox{d}v\\
 & =\left(\frac{P}{2\delta}\right)^{3}\cdot\int\limits _{0}^{\left(\frac{\tilde{\delta}}{2}\right)^{2}}\left|\mathcal{K}(v)\right|^{2}\left[\int\limits _{u^{(0)}}^{u^{(1)}}\frac{1}{u}\,\mbox{d}u\right]\,\mbox{d}v=\left(\frac{P}{2\delta}\right)^{3}\cdot\int\limits _{0}^{\left(\frac{\tilde{\delta}}{2}\right)^{2}}\left|\mathcal{K}(v)\right|^{2}\mathrm{ln}\left(u^{(1)}/u^{(0)}\right)\,\mbox{d}v\\
 & =\left(\frac{P}{2\delta}\right)^{3}\cdot\int\limits _{0}^{\left(\frac{\tilde{\delta}}{2}\right)^{2}}\left|\mathcal{K}(v)\right|^{2}\mathrm{ln}\left(\frac{\frac{\tilde{\delta}}{2}+\sqrt{\left(\frac{\tilde{\delta}}{2}\right)^{2}-v}}{\frac{\tilde{\delta}}{2}-\sqrt{\left(\frac{\tilde{\delta}}{2}\right)^{2}-v}}\right)\,\mbox{d}v.
\end{split}
\label{GlpartintegIvereinfachtteil2}
\end{equation}

\subsection*{Calculation of partial integral (\ref{eq:quattro}), quadrant IV:}

Regarding the integrand of the inner integral (\ref{eq:quattro})
we deduce according to (\ref{GlrectshapedPSDded1-1}): 
\begin{equation}
\begin{split}G\left(f+u\right)\neq0\quad\Longleftrightarrow\quad u\leq\tilde{\delta}.\end{split}
\label{GlrectshapedPSDded1Int2}
\end{equation}

Again this implies for the following analysis of (\ref{eq:quattro})
that $\tilde{\delta}>0$ and that integral (\ref{eq:quattro}) disappears
for $f>\delta$. For the second factor we have: 
\begin{equation}
\begin{split}G\left(f-\frac{v}{u}\right)\neq0\quad & \Longleftrightarrow\quad f-\frac{v}{u}\leq\delta\quad\text{and}\quad f-\frac{v}{u}\geq-\delta\\
\quad & \Longleftrightarrow\quad-\frac{v}{u}\leq\tilde{\delta}\quad\text{and}\quad-\frac{v}{u}\geq-\delta-f=-(\delta+f):=-\overline{\delta}\\
\quad & \Longleftrightarrow\quad\frac{v}{u}\geq-\tilde{\delta}\quad\text{and}\quad\frac{v}{u}\leq\overline{\delta}.
\end{split}
\label{GlrectshapedPSDded2Int2-1}
\end{equation}

Since $\tilde{\delta}>0$ and $u,v\geq0$ the first inequality doesn't
represent a constraint. So we have: 
\begin{equation}
\begin{split}G\left(f-\frac{v}{u}\right)\neq0\quad & \Longleftrightarrow\quad\frac{v}{u}\leq\overline{\delta}\quad\Longleftrightarrow\quad u\geq\frac{v}{\overline{\delta}}.\end{split}
\label{GlrectshapedPSDded2Int2-1}
\end{equation}

Note that we may exclude the special case $\delta=0$ since the whole
double integral will be zero in this case. The quotient $\frac{v}{\overline{\delta}}$
is therefore well defined. For the third factor we have: 
\begin{equation}
\begin{split}G\left(f+u-\frac{v}{u}\right)\neq0\quad & \Longleftrightarrow\quad u-\frac{v}{u}\leq\tilde{\delta}\quad\text{and}\quad u-\frac{v}{u}\geq-\overline{\delta}.\end{split}
\label{GlrectshapedPSDded3Int2-1}
\end{equation}

For the first inequality we deduce: 
\begin{equation}
\begin{split}u-\frac{v}{u}\leq\tilde{\delta}\quad & \Longleftrightarrow\quad u^{2}-\tilde{\delta}u-v\leq0\\
\quad & \Longleftrightarrow\quad\left(u-\frac{\tilde{\delta}}{2}\right)^{2}-\left(\frac{\tilde{\delta}}{2}\right)^{2}-v\leq0\quad\Longleftrightarrow\quad\left(u-\frac{\tilde{\delta}}{2}\right)^{2}\leq\left(\frac{\tilde{\delta}}{2}\right)^{2}+v.
\end{split}
\label{GlrectshapedPSDded3Int2first-1}
\end{equation}

This implies 
\begin{equation}
\begin{split}u\leq\frac{\tilde{\delta}}{2}+\sqrt{\left(\frac{\tilde{\delta}}{2}\right)^{2}+v}\quad\text{and}\quad-u\leq-\frac{\tilde{\delta}}{2}+\sqrt{\left(\frac{\tilde{\delta}}{2}\right)^{2}+v}.\end{split}
\label{GlrectshapedPSDded3Int2firstb-1}
\end{equation}

Since $u,v\geq0$ the last inequality is always fulfilled and doesn't
represent a constraint. So finally the first inequality implies 
\begin{equation}
\begin{split}u\leq u^{(3)}:=\frac{\tilde{\delta}}{2}+\sqrt{\left(\frac{\tilde{\delta}}{2}\right)^{2}+v}.\end{split}
\label{GlrectshapedPSDded3Int2firstbrev}
\end{equation}

For the second inequality in (\ref{GlrectshapedPSDded3Int2-1}) we
deduce: 
\begin{equation}
\begin{split}u-\frac{v}{u}\geq-\overline{\delta}\quad & \Longleftrightarrow\quad u^{2}+\overline{\delta}u-v\geq0\\
\quad & \Longleftrightarrow\quad\left(u+\frac{\overline{\delta}}{2}\right)^{2}-\left(\frac{\overline{\delta}}{2}\right)^{2}-v\geq0\quad\Longleftrightarrow\quad\left(u+\frac{\overline{\delta}}{2}\right)^{2}\geq\left(\frac{\overline{\delta}}{2}\right)^{2}+v.
\end{split}
\label{GlrectshapedPSDded3Int2second-1}
\end{equation}

This implies

\begin{equation}
\begin{split}u\geq u^{(2)}:=-\frac{\overline{\delta}}{2}+\sqrt{\left(\frac{\overline{\delta}}{2}\right)^{2}+v}.\end{split}
\label{GlrectshapedPSDded3Int2secondbrev}
\end{equation}

Thus using (\ref{GlrectshapedPSDded1Int2}), (\ref{GlrectshapedPSDded2Int2-1}),
(\ref{GlrectshapedPSDded3Int2firstbrev}) and (\ref{GlrectshapedPSDded3Int2secondbrev})
the partial integral (\ref{eq:quattro}) reads for $f<\delta$: 
\begin{equation}
\begin{split}\int\limits _{0}^{\infty} & \left|\mathcal{K}(v)\right|^{2}\left[\int\limits _{0}^{\infty}\frac{1}{u}\cdot G\left(f+u\right)G\left(f-\frac{v}{u}\right)G\left(f+u-\frac{v}{u}\right)\,\mbox{d}u\right]\,\mbox{d}v\\
 & =\int\limits _{0}^{\infty}\left|\mathcal{K}(v)\right|^{2}\left[\int\limits _{\max\left\{ {u^{(2)},\frac{v}{\overline{\delta}}}\right\} }^{\min\left\{ {u^{(3)},\tilde{\delta}}\right\} }\frac{1}{u}\cdot G\left(f+u\right)G\left(f-\frac{v}{u}\right)G\left(f+u-\frac{v}{u}\right)\,\mbox{d}u\right]\,\mbox{d}v.
\end{split}
\label{GlpartintegIIvereinfacht}
\end{equation}

Since 
\begin{equation}
\begin{split}u^{(3)}=\frac{\tilde{\delta}}{2}+\sqrt{\left(\frac{\tilde{\delta}}{2}\right)^{2}+v}\geq\frac{\tilde{\delta}}{2}+\sqrt{\left(\frac{\tilde{\delta}}{2}\right)^{2}}=2\frac{\tilde{\delta}}{2}=\tilde{\delta}\end{split}
\label{GlrectshapedPSDded3Int2firstabsch}
\end{equation}
we have 
\begin{equation}
\begin{split}\min\left\{ {u^{(3)},\tilde{\delta}}\right\} =\tilde{\delta}.\end{split}
\label{GlrectshapedPSDded3Int2firstabsch1}
\end{equation}

Additionally since 
\begin{equation}
\begin{split}\frac{v}{\overline{\delta}}\geq u^{(2)} & =-\frac{\overline{\delta}}{2}+\sqrt{\left(\frac{\overline{\delta}}{2}\right)^{2}+v}\quad\Longleftrightarrow\quad\frac{v}{\overline{\delta}}+\frac{\overline{\delta}}{2}\geq+\sqrt{\left(\frac{\overline{\delta}}{2}\right)^{2}+v}\\
\quad & \Longleftrightarrow\quad\left(\frac{v}{\overline{\delta}}\right)^{2}+v+\left(\frac{\overline{\delta}}{2}\right)^{2}\geq\left(\frac{\overline{\delta}}{2}\right)^{2}+v
\end{split}
\label{GlrectshapedPSDded3Int2secondabsch}
\end{equation}
we have 
\begin{equation}
\begin{split}\max\left\{ u^{(2)},\frac{v}{\overline{\delta}}\right\} =\frac{v}{\overline{\delta}}.\end{split}
\label{GlrectshapedPSDded3Int2secondabsch1}
\end{equation}

Thus 
\begin{equation}
\begin{split}\frac{v}{\overline{\delta}}\leq u\leq\tilde{\delta}\end{split}
\label{GlrectshapedPSDded3Int2secondabsch1bx}
\end{equation}
otherwise the partial integral (\ref{eq:quattro}) disappears. This
however imposes a restriction on $v$, because it implies $v\leq\tilde{\delta}\cdot\overline{\delta}$!
Thus finally the partial integral (\ref{eq:quattro}) reads for $f<\delta$:
\begin{equation}
\begin{split}\int\limits _{0}^{\infty} & \left|\mathcal{K}(v)\right|^{2}\left[\int\limits _{0}^{\infty}\frac{1}{u}\cdot G\left(f+u\right)G\left(f-\frac{v}{u}\right)G\left(f+u-\frac{v}{u}\right)\,\mbox{d}u\right]\,\mbox{d}v\\
 & =\left(\frac{P}{2\delta}\right)^{3}\cdot\int\limits _{0}^{\tilde{\delta}\cdot\overline{\delta}}\left|\tilde{\eta}(v)\right|^{2}\left[\int\limits _{\frac{v}{\overline{\delta}}}^{\tilde{\delta}}\frac{1}{u}\,\mbox{d}u\right]\,\mbox{d}v=\left(\frac{P}{2\delta}\right)^{3}\cdot\int\limits _{0}^{\tilde{\delta}\cdot\overline{\delta}}\left|\mathcal{K}(v)\right|^{2}\mathrm{ln}\left(\frac{\tilde{\delta}\cdot\overline{\delta}}{v}\right)\,\mbox{d}v.
\end{split}
\label{GlpartintegIIvereinfachtschluss}
\end{equation}


\subsection*{Calculation of partial integral (\ref{eqdue}), quadrant II:}

Regarding the integrand of the inner integral (\ref{eqdue}) we deduce:
\begin{equation}
\begin{split}G\left(f-u\right)\neq0\quad & \Longleftrightarrow\quad-u\leq\tilde{\delta}\quad\text{and}\quad-u\geq-\overline{\delta}\\
\quad & \Longleftrightarrow\quad u\leq\overline{\delta}\quad\text{and}\quad u\geq-\tilde{\delta}.
\end{split}
\label{GlrectshapedPSDded1Int3-1}
\end{equation}

For the second factor we have like in (\ref{GlrectshapedPSDded2-1}):
\begin{equation}
\begin{split}G\left(f+\frac{v}{u}\right)\neq0\quad\Longleftrightarrow\quad\frac{v}{u}\leq\tilde{\delta}\quad\Longleftrightarrow\quad u\geq\frac{v}{\tilde{\delta}}.\end{split}
\label{GlrectshapedPSDded2Int3-1}
\end{equation}

Note that we may suppose $\tilde{\delta}>0$ since otherwise because
of $\frac{v}{u}\leq\tilde{\delta}$ and the fact that $u,v\geq0$
the factor $G\left(f+\frac{v}{u}\right)$ and consequently the whole
integral would be zero. Hence once more the partial integral (\ref{eqdue})
disappears if $f>\delta$! Since $\tilde{\delta}>0$ the second inequality
in (\ref{GlrectshapedPSDded1Int3-1}) doesn't represent a constraint.
So we have 
\begin{equation}
\begin{split}G\left(f-u\right)\neq0\quad & \Longleftrightarrow\quad u\leq\overline{\delta}.\end{split}
\label{GlrectshapedPSDded1Int3rev}
\end{equation}
and 
\begin{equation}
\begin{split}G\left(f+\frac{v}{u}\right)\neq0\quad\Longleftrightarrow\quad u\geq\frac{v}{\tilde{\delta}}.\end{split}
\label{GlrectshapedPSDded2Int3rev}
\end{equation}

For the third factor we have: 
\begin{equation}
\begin{split}G\left(f-u+\frac{v}{u}\right)\neq0\quad & \Longleftrightarrow\quad-u+\frac{v}{u}\leq\tilde{\delta}\quad\text{and}\quad-u+\frac{v}{u}\geq-\overline{\delta}.\end{split}
\label{GlrectshapedPSDded3Int3-1}
\end{equation}

For the first inequality we deduce: 
\begin{equation}
\begin{split}-u+\frac{v}{u}\leq\tilde{\delta}\quad & \Longleftrightarrow\quad-u^{2}-\tilde{\delta}u+v\leq0\quad\Longleftrightarrow\quad u^{2}+\tilde{\delta}u-v\geq0\\
\quad & \Longleftrightarrow\quad\left(u+\frac{\tilde{\delta}}{2}\right)^{2}-\left(\frac{\tilde{\delta}}{2}\right)^{2}-v\geq0\quad\Longleftrightarrow\quad\left(u+\frac{\tilde{\delta}}{2}\right)^{2}\geq\left(\frac{\tilde{\delta}}{2}\right)^{2}+v.
\end{split}
\label{GlrectshapedPSDded3Int3first}
\end{equation}

This implies 
\begin{equation}
\begin{split}u\geq u^{(4)}:=-\frac{\tilde{\delta}}{2}+\sqrt{\left(\frac{\tilde{\delta}}{2}\right)^{2}+v}.\end{split}
\label{GlrectshapedPSDded3Int3firstbrev}
\end{equation}

For the second inequality we have: 
\begin{equation}
\begin{split}-u+\frac{v}{u}\geq-\overline{\delta}\quad & \Longleftrightarrow\quad-u^{2}+\overline{\delta}u+v\geq0\\
\quad & \Longleftrightarrow\quad\left(u-\frac{\overline{\delta}}{2}\right)^{2}-\left(\frac{\overline{\delta}}{2}\right)^{2}-v\leq0\quad\Longleftrightarrow\quad\left(u-\frac{\overline{\delta}}{2}\right)^{2}\leq\left(\frac{\overline{\delta}}{2}\right)^{2}+v.
\end{split}
\label{GlrectshapedPSDded3Int3second}
\end{equation}

This implies

\begin{equation}
\begin{split}u\leq u^{(5)}:=\frac{\overline{\delta}}{2}+\sqrt{\left(\frac{\overline{\delta}}{2}\right)^{2}+v}.\end{split}
\label{GlrectshapedPSDded3Int3secondbrev}
\end{equation}

Using (\ref{GlrectshapedPSDded1Int3rev}), (\ref{GlrectshapedPSDded2Int3rev}),
(\ref{GlrectshapedPSDded3Int3firstbrev}) and (\ref{GlrectshapedPSDded3Int3secondbrev})
the partial integral (\ref{eqdue}) reads for $f<\delta$: 
\begin{equation}
\begin{split}\int\limits _{0}^{\infty} & \left|\mathcal{K}(v)\right|^{2}\left[\int\limits _{0}^{\infty}\frac{1}{u}\cdot G\left(f+u\right)G\left(f-\frac{v}{u}\right)G\left(f+u-\frac{v}{u}\right)\,\mbox{d}u\right]\,\mbox{d}v\\
 & =\int\limits _{0}^{\infty}\left|\mathcal{K}(v)\right|^{2}\left[\int\limits _{\max\left\{ {u^{(4)},\frac{v}{\tilde{\delta}}}\right\} }^{\min\left\{ {u^{(5)},\overline{\delta}}\right\} }\frac{1}{u}\cdot G\left(f+u\right)G\left(f-\frac{v}{u}\right)G\left(f+u-\frac{v}{u}\right)\,\mbox{d}u\right]\,\mbox{d}v.
\end{split}
\label{GlpartintegIIIvereinfacht}
\end{equation}

Since 
\begin{equation}
\begin{split}u^{(5)}=\frac{\overline{\delta}}{2}+\sqrt{\left(\frac{\overline{\delta}}{2}\right)^{2}+v}\geq\frac{\overline{\delta}}{2}+\sqrt{\left(\frac{\overline{\delta}}{2}\right)^{2}}=2\frac{\overline{\delta}}{2}=\overline{\delta}\end{split}
\label{GlrectshapedPSDded3Int3firstabsch}
\end{equation}
we have 
\begin{equation}
\begin{split}\min\left\{ {u^{(5)},\overline{\delta}}\right\} =\overline{\delta}.\end{split}
\label{GlrectshapedPSDded3Int3firstabsch1}
\end{equation}

Additionally since 
\begin{equation}
\begin{split}\frac{v}{\tilde{\delta}}\geq u^{(4)} & =-\frac{\tilde{\delta}}{2}+\sqrt{\left(\frac{\tilde{\delta}}{2}\right)^{2}+v}\quad\Longleftrightarrow\quad\frac{v}{\tilde{\delta}}+\frac{\tilde{\delta}}{2}\geq+\sqrt{\left(\frac{\tilde{\delta}}{2}\right)^{2}+v}\\
\quad & \Longleftrightarrow\quad\left(\frac{v}{\tilde{\delta}}\right)^{2}+v+\left(\frac{\tilde{\delta}}{2}\right)^{2}\geq\left(\frac{\tilde{\delta}}{2}\right)^{2}+v
\end{split}
\label{GlrectshapedPSDded3Int3secondabsch}
\end{equation}
we have 
\begin{equation}
\begin{split}\max\left\{ u^{(4)},\frac{v}{\tilde{\delta}}\right\} =\frac{v}{\tilde{\delta}}.\end{split}
\label{GlrectshapedPSDded3Int3secondabsch1}
\end{equation}

Thus 
\begin{equation}
\begin{split}\frac{v}{\tilde{\delta}}\leq u\leq\overline{\delta}\end{split}
\label{GlrectshapedPSDded3Int2secondabsch1bx}
\end{equation}
otherwise the partial integral (\ref{eqdue}) disappears. This however
imposes a restriction on $v$, because it implies again $v\leq\tilde{\delta}\cdot\overline{\delta}$!
Thus finally the partial integral (\ref{eqdue}) reads for $f<\delta$:
\begin{equation}
\begin{split}\int\limits _{0}^{\infty} & \left|\mathcal{K}(v)\right|^{2}\left[\int\limits _{0}^{\infty}\frac{1}{u}\cdot G\left(f+u\right)G\left(f-\frac{v}{u}\right)G\left(f+u-\frac{v}{u}\right)\,\mbox{d}u\right]\,\mbox{d}v\\
 & =\left(\frac{P}{2\delta}\right)^{3}\cdot\int\limits _{0}^{\tilde{\delta}\cdot\overline{\delta}}\left|\mathcal{K}(v)\right|^{2}\left[\int\limits _{\frac{v}{\tilde{\delta}}}^{\overline{\delta}}\frac{1}{u}\,\mbox{d}u\right]\,\mbox{d}v=\left(\frac{P}{2\delta}\right)^{3}\cdot\int\limits _{0}^{\tilde{\delta}\cdot\overline{\delta}}\left|\mathcal{K}(v)\right|^{2}\mathrm{ln}\left(\frac{\tilde{\delta}\cdot\overline{\delta}}{v}\right)\,\mbox{d}v.
\end{split}
\label{GlpartintegIIIvereinfachtschluss}
\end{equation}

\subsection*{Calculation of partial integral (\ref{eq:tre}), quadrant III}

The forth integral is the only one for which $f<\delta$ doesn't follow
necessarily as a condition for not being zero.

So we have to make a distinction between the two cases $f<\delta$
and $f\geq\delta$.

\subsubsection*{The partial integral (\ref{eq:tre}) for $f<\delta$}

Regarding the integrand of the inner integral (\ref{eq:tre}) we have
according to (\ref{GlrectshapedPSDded1Int3-1}): 
\begin{equation}
\begin{split}G\left(f-u\right)\neq0\quad & \Longleftrightarrow\quad-u\leq\tilde{\delta}\quad\text{and}\quad-u\geq-\overline{\delta}\\
\quad & \Longleftrightarrow\quad u\leq\overline{\delta}\quad\text{and}\quad u\geq-\tilde{\delta}.
\end{split}
\label{GlrectshapedPSDded1Int4-1}
\end{equation}

Since by assumption $\tilde{\delta}=\delta-f>0$, then the condition
$u\geq-\tilde{\delta}$ is always fulfilled and the only remaining
restriction is: 
\begin{equation}
\begin{split}u\leq\overline{\delta}.\end{split}
\label{GlrectshapedPSDded1Int4b}
\end{equation}

For the second factor we have according to (\ref{GlrectshapedPSDded2Int2-1}):
\begin{equation}
\begin{split}G\left(f-\frac{v}{u}\right)\neq0\quad & \Longleftrightarrow\quad f-\frac{v}{u}\leq\delta\quad\text{and}\quad f-\frac{v}{u}\geq-\delta\\
\quad & \Longleftrightarrow\quad\frac{v}{u}\geq-\tilde{\delta}\quad\text{and}\quad\frac{v}{u}\leq\overline{\delta}.
\end{split}
\label{GlrectshapedPSDded2Int4-1}
\end{equation}

Again since $\tilde{\delta}>0$ the first inequality is always fulfilled
and we have (note that $\overline{\delta}=f+\delta>0$ by definition):
\begin{equation}
\begin{split}u\geq\frac{v}{\overline{\delta}}.\end{split}
\label{GlrectshapedPSDded2Int4b}
\end{equation}

For the third factor we have: 
\begin{equation}
\begin{split}G\left(f-u-\frac{v}{u}\right)\neq0\quad & \Longleftrightarrow\quad-u-\frac{v}{u}\leq\tilde{\delta}\quad\text{and}\quad-u-\frac{v}{u}\geq-\overline{\delta}\\
\quad & \Longleftrightarrow\quad u+\frac{v}{u}\geq-\tilde{\delta}\quad\text{and}\quad u+\frac{v}{u}\leq\overline{\delta}.
\end{split}
\label{GlrectshapedPSDded3Int4-1}
\end{equation}

Again since $u,v,\tilde{\delta}>0$ the first inequality doesn't deliver
a restriction and we get for the second one: 
\begin{equation}
\begin{split}u^{2}-\overline{\delta}u+v\leq0\quad & \Longleftrightarrow\quad\left(u-\frac{\overline{\delta}}{2}\right)^{2}-\left(\frac{\overline{\delta}}{2}\right)^{2}+v\leq0\\
\quad & \Longleftrightarrow\quad\left(u-\frac{\overline{\delta}}{2}\right)^{2}\leq\left(\frac{\overline{\delta}}{2}\right)^{2}-v.
\end{split}
\label{GlrectshapedPSDded3bInt4secineq}
\end{equation}

The partial integral (\ref{eq:tre}) therefore disappears if $v>\left(\frac{\overline{\delta}}{2}\right)^{2}$.
For $v<\left(\frac{\overline{\delta}}{2}\right)^{2}$ we have 
\begin{equation}
\begin{split}u\leq u^{(7)}:=\frac{\overline{\delta}}{2}+\sqrt{\left(\frac{\overline{\delta}}{2}\right)^{2}-v}\end{split}
\label{GlrectshapedPSDded3bInt4secineq2}
\end{equation}
and 
\begin{equation}
\begin{split}\frac{\overline{\delta}}{2}-u & \leq\sqrt{\left(\frac{\overline{\delta}}{2}\right)^{2}-v}\quad\Longleftrightarrow\quad u\geq u^{(6)}:=\frac{\overline{\delta}}{2}-\sqrt{\left(\frac{\overline{\delta}}{2}\right)^{2}-v}.\end{split}
\label{GlrectshapedPSDded3bInt4secineq3}
\end{equation}

Using (\ref{GlrectshapedPSDded1Int4b}), (\ref{GlrectshapedPSDded2Int4b}),
(\ref{GlrectshapedPSDded3bInt4secineq2}) and (\ref{GlrectshapedPSDded3bInt4secineq3})
the partial integral (\ref{eq:tre}) reads for $f<\delta$: 
\begin{equation}
\begin{split}\int\limits _{0}^{\infty} & \left|\mathcal{K}(v)\right|^{2}\left[\int\limits _{0}^{\infty}\frac{1}{u}\cdot G\left(f-u\right)G\left(f-\frac{v}{u}\right)G\left(f-u-\frac{v}{u}\right)\,\mbox{d}u\right]\,\mbox{d}v.\\
 & =\int\limits _{0}^{\left(\frac{\overline{\delta}}{2}\right)^{2}}\left|\mathcal{K}(v)\right|^{2}\left[\int\limits _{\max\left\{ {u^{(6)},\frac{v}{\tilde{\delta}}}\right\} }^{\min\left\{ {u^{(7)},\overline{\delta}}\right\} }\frac{1}{u}\cdot G\left(f-u\right)G\left(f-\frac{v}{u}\right)G\left(f-u-\frac{v}{u}\right)\,\mbox{d}u\right]\,\mbox{d}v.
\end{split}
\label{GlpartintegIVvereinfacht}
\end{equation}

It is easy to see that 
\begin{equation}
\begin{split}\min\left\{ {u^{(7)},\overline{\delta}}\right\} =u^{(7)}=\frac{\overline{\delta}}{2}+\sqrt{\left(\frac{\overline{\delta}}{2}\right)^{2}-v}.\end{split}
\label{GlrectshapedPSDded3Int4firstabsch1}
\end{equation}

Since $v<\left(\frac{\overline{\delta}}{2}\right)^{2}=\frac{1}{4}\overline{\delta}\,^{2}$
we deduce: 
\begin{equation}
\begin{split}2v\leq\overline{\delta}^{2}\quad\Longleftrightarrow\frac{v}{\overline{\delta}}\leq\frac{\overline{\delta}}{2}\quad\Longleftrightarrow\quad\frac{v}{\overline{\delta}}-\frac{\overline{\delta}}{2}\leq0\quad\Longleftrightarrow\quad\frac{\overline{\delta}}{2}-\frac{v}{\overline{\delta}}\geq0.\end{split}
\label{GlrectshapedPSDded3Int4firstabsch3}
\end{equation}

Then 
\begin{equation}
\begin{split}u^{(6)}\geq\frac{v}{\overline{\delta}}\quad & \Longleftrightarrow\quad\frac{\overline{\delta}}{2}-\frac{v}{\overline{\delta}}\geq\sqrt{\left(\frac{\overline{\delta}}{2}\right)^{2}-v}\quad\Longleftrightarrow\quad\left(\frac{\overline{\delta}}{2}-\frac{v}{\overline{\delta}}\right)^{2}\geq\left(\frac{\overline{\delta}}{2}\right)^{2}-v\\
\quad & \Longleftrightarrow\quad\left(\frac{\overline{\delta}}{2}\right)^{2}-v+\left(\frac{v}{\overline{\delta}}\right)^{2}\geq\left(\frac{\overline{\delta}}{2}\right)^{2}-v\quad\Longleftrightarrow\quad\left(\frac{v}{\overline{\delta}}\right)^{2}\geq0.
\end{split}
\label{GlrectshapedPSDded3bInt4secineq5}
\end{equation}

Thus 
\begin{equation}
\begin{split}\max\left\{ u^{(6)},\frac{v}{\overline{\delta}}\right\} =u^{(6)}=\frac{\overline{\delta}}{2}-\sqrt{\left(\frac{\overline{\delta}}{2}\right)^{2}-v}.\end{split}
\label{GlrectshapedPSDded3Int4firstabsch6}
\end{equation}

Finally in the case $f<\delta$ for the partial integral (\ref{eq:tre})
follows: 
\begin{equation}
\begin{split}\int\limits _{0}^{\infty} & \left|\mathcal{K}(v)\right|^{2}\left[\int\limits _{0}^{\infty}\frac{1}{u}\cdot G\left(f-u\right)G\left(f-\frac{v}{u}\right)G\left(f-u-\frac{v}{u}\right)\,\mbox{d}u\right]\,\mbox{d}v\\
 & =\left(\frac{P}{2\delta}\right)^{3}\cdot\int\limits _{0}^{\left(\frac{\overline{\delta}}{2}\right)^{2}}\left|\mathcal{K}(v)\right|^{2}\left[\int\limits _{u^{(6)}}^{u^{(7)}}\frac{1}{u}\,\mbox{d}u\right]\,\mbox{d}v=\left(\frac{P}{2\delta}\right)^{3}\cdot\int\limits _{0}^{\left(\frac{\overline{\delta}}{2}\right)^{2}}\left|\mathcal{K}(v)\right|^{2}\mathrm{ln}\left(u^{(7)}/u^{(6)}\right)\,\mbox{d}v\\
 & =\left(\frac{P}{2\delta}\right)^{3}\cdot\int\limits _{0}^{\left(\frac{\overline{\delta}}{2}\right)^{2}}\left|\mathcal{K}(v)\right|^{2}\mathrm{ln}\left(\frac{\frac{\overline{\delta}}{2}+\sqrt{\left(\frac{\overline{\delta}}{2}\right)^{2}-v}}{\frac{\overline{\delta}}{2}-\sqrt{\left(\frac{\overline{\delta}}{2}\right)^{2}-v}}\right)\,\mbox{d}v.
\end{split}
\label{GlpartintegIVvereinfachtteil2}
\end{equation}

Together with (\ref{GlpartintegIvereinfachtteil2}), (\ref{GlpartintegIIvereinfachtschluss})
and (\ref{GlpartintegIIIvereinfachtschluss}) this proves equation
(\ref{eq:nyq_1}).\\

\subsubsection*{The partial integral (\ref{eq:tre}) for $f>\delta$}

Again we have according to (\ref{GlrectshapedPSDded1Int3-1}): 
\begin{equation}
\begin{split}G\left(f-u\right)\neq0\quad & \Longleftrightarrow\quad-u\leq\tilde{\delta}\quad\text{and}\quad-u\geq-\overline{\delta}\\
\quad & \Longleftrightarrow\quad u\leq\overline{\delta}\quad\text{and}\quad u\geq-\tilde{\delta}.
\end{split}
\label{GlrectshapedPSDded1Int4-1}
\end{equation}

This time $u\geq-\tilde{\delta}$ is a genuine restriction because
$-\tilde{\delta}=f-\delta>0$ by assumption. For the second factor
we have like in (\ref{GlrectshapedPSDded2Int4-1}): 
\begin{equation}
\begin{split}G\left(f-\frac{v}{u}\right)\neq0\quad & \Longleftrightarrow\quad f-\frac{v}{u}\leq\delta\quad\text{and}\quad f-\frac{v}{u}\geq-\delta\\
\quad & \Longleftrightarrow\quad\frac{v}{u}\geq-\tilde{\delta}\quad\text{and}\quad\frac{v}{u}\leq\overline{\delta}.
\end{split}
\label{GlrectshapedPSDded2Int4x}
\end{equation}

Since $\overline{\delta}>0,-\tilde{\delta}>0$ this leads to: 
\begin{equation}
\begin{split}u\geq\frac{v}{\overline{\delta}}\quad\text{and}\quad u\leq-\frac{v}{\tilde{\delta}}=\frac{v}{-\tilde{\delta}}.\end{split}
\label{GlrectshapedPSDded2Int4bspec}
\end{equation}

Especially (\ref{GlrectshapedPSDded1Int4-1}) and (\ref{GlrectshapedPSDded2Int4bspec})
imply the following restrictions on $v$: 
\begin{equation}
\begin{split}\frac{v}{\overline{\delta}}\leq\overline{\delta}\quad\text{and}\quad-\tilde{\delta}\leq\frac{v}{-\tilde{\delta}}.\end{split}
\label{GlrectshapedPSDded2Int4bspecrestonv1}
\end{equation}

Thus 
\begin{equation}
\begin{split}\left(-\tilde{\delta}\right)^{2}=\tilde{\delta}^{2}\leq v\leq\overline{\delta}\,^{2}.\end{split}
\label{GlrectshapedPSDded2Int4bspecrestonv2}
\end{equation}

So the the partial integral (\ref{eq:tre}) reads for $f>\delta$:
\begin{equation}
\begin{split}\int\limits _{0}^{\infty} & \left|\mathcal{K}(v)\right|^{2}\left[\int\limits _{0}^{\infty}\frac{1}{u}\cdot G\left(f-u\right)G\left(f-\frac{v}{u}\right)G\left(f-u-\frac{v}{u}\right)\,\mbox{d}u\right]\,\mbox{d}v\\
 & =\left(\frac{P}{2\delta}\right)^{2}\cdot\int\limits _{\tilde{\delta}^{2}}^{\overline{\delta}^{2}}\left|\mathcal{K}(v)\right|^{2}\left[\int\limits _{\max\left\{ \frac{v}{\overline{\delta}},-\tilde{\delta}\right\} }^{\min\left\{ \frac{v}{-\tilde{\delta}},\overline{\delta}\right\} }\frac{1}{u}\cdot G\left(f-u-\frac{v}{u}\right)\,\mbox{d}u\right]\,\mbox{d}v.
\end{split}
\label{GlpartintegIVvereinfachtspecialint4}
\end{equation}

For the third factor we have like in (\ref{GlrectshapedPSDded3Int4-1}):
\begin{equation}
\begin{split}G\left(f-u-\frac{v}{u}\right)\neq0\quad & \Longleftrightarrow\quad-u-\frac{v}{u}\leq\tilde{\delta}\quad\text{and}\quad-u-\frac{v}{u}\geq-\overline{\delta}\\
\quad & \Longleftrightarrow\quad u+\frac{v}{u}\geq-\tilde{\delta}\quad\text{and}\quad u+\frac{v}{u}\leq\overline{\delta}.
\end{split}
\label{GlrectshapedPSDded3Int4spec}
\end{equation}

For the first inequality we get equivalently: 
\begin{equation}
\begin{split}\left(u+\frac{\tilde{\delta}}{2}\right)^{2}\geq\left(\frac{\tilde{\delta}}{2}\right)^{2}-v.\end{split}
\label{GlrectshapedPSDded3Int4specfirst}
\end{equation}

For the second inequality we have: 
\begin{equation}
\begin{split}\left(u-\frac{\overline{\delta}}{2}\right)^{2}\leq\left(\frac{\overline{\delta}}{2}\right)^{2}-v.\end{split}
\label{GlrectshapedPSDded3Int4specsecond}
\end{equation}

The last inequality implies that the whole integral is zero if $v$
exceeds $\left(\frac{\overline{\delta}}{2}\right)^{2}$. So the upper
limit of the first integral of (\ref{GlpartintegIVvereinfachtspecialint4})
is $\left(\frac{\overline{\delta}}{2}\right)^{2}$ instead of $\overline{\delta}\,^{2}$.
Since by (\ref{GlrectshapedPSDded2Int4bspecrestonv2}) $v\geq\tilde{\delta}\,^{2}$
inequality (\ref{GlrectshapedPSDded3Int4specfirst}) is no constraint.
Inequality (\ref{GlrectshapedPSDded3Int4specsecond}) is obviously
fulfilled iff 
\begin{equation}
\begin{split}\frac{\overline{\delta}}{2}-\sqrt{\left(\frac{\overline{\delta}}{2}\right)^{2}-v}\leq u\leq\frac{\overline{\delta}}{2}+\sqrt{\left(\frac{\overline{\delta}}{2}\right)^{2}-v}.\end{split}
\label{GlrectshapedPSDded3Int4specsecondlsg}
\end{equation}

The partial integral (\ref{eq:tre}) for $f>\delta$ is consequently:
\begin{equation}
\begin{split}\int\limits _{0}^{\infty} & \left|\mathcal{K}(v)\right|^{2}\left[\int\limits _{0}^{\infty}\frac{1}{u}\cdot G\left(f-u\right)G\left(f-\frac{v}{u}\right)G\left(f-u-\frac{v}{u}\right)\,\mbox{d}u\right]\,\mbox{d}v\\
 & =\left(\frac{P}{2\delta}\right)^{2}\cdot\int\limits _{\tilde{\delta}^{2}}^{\left(\frac{\overline{\delta}}{2}\right)^{2}}\left|\mathcal{K}(v)\right|^{2}\left[\int\limits _{\max\left\{ \frac{v}{\overline{\delta}},-\tilde{\delta},\frac{\overline{\delta}}{2}-\sqrt{\left(\frac{\overline{\delta}}{2}\right)^{2}-v}\right\} }^{\min\left\{ \frac{v}{-\tilde{\delta}},\overline{\delta},\frac{\overline{\delta}}{2}+\sqrt{\left(\frac{\overline{\delta}}{2}\right)^{2}-v}\right\} }\frac{1}{u}\cdot G\left(f-u-\frac{v}{u}\right)\,\mbox{d}u\right]\,\mbox{d}v\\
 & =\left(\frac{P}{2\delta}\right)^{2}\cdot\int\limits _{\tilde{\delta}^{2}}^{\left(\frac{\overline{\delta}}{2}\right)^{2}}\left|\mathcal{K}(v)\right|^{2}\left[\int\limits _{\max\left\{ \frac{v}{\overline{\delta}},-\tilde{\delta},\frac{\overline{\delta}}{2}-\sqrt{\left(\frac{\overline{\delta}}{2}\right)^{2}-v}\right\} }^{\min\left\{ \frac{v}{-\tilde{\delta}},\frac{\overline{\delta}}{2}+\sqrt{\left(\frac{\overline{\delta}}{2}\right)^{2}-v}\right\} }\frac{1}{u}\cdot G\left(f-u-\frac{v}{u}\right)\,\mbox{d}u\right]\,\mbox{d}v.
\end{split}
\label{GlpartintegIVvereinfachtspecialint6}
\end{equation}

Further we deduce 
\begin{equation}
\begin{split}\frac{v}{\overline{\delta}}\leq\frac{\overline{\delta}}{2}-\sqrt{\left(\frac{\overline{\delta}}{2}\right)^{2}-v}\quad & \Longleftrightarrow\quad\frac{\overline{\delta}}{2}-\frac{v}{\overline{\delta}}\geq\sqrt{\left(\frac{\overline{\delta}}{2}\right)^{2}-v}\\
\quad & \Longleftrightarrow\quad\frac{v^{2}}{\overline{\delta}\,^{2}}-v+\left(\frac{\overline{\delta}}{2}\right)^{2}\geq\left(\frac{\overline{\delta}}{2}\right)^{2}-v\\
\quad & \Longleftrightarrow\quad\frac{v^{2}}{\overline{\delta}\,^{2}}\geq0.
\end{split}
\label{GlrectshapedPSDded3Int4absh1analog}
\end{equation}

Since this condition is always fulfilled we get: 
\begin{equation}
\begin{split}\int\limits _{0}^{\infty} & \left|\mathcal{K}(v)\right|^{2}\left[\int\limits _{0}^{\infty}\frac{1}{u}\cdot G\left(f-u\right)G\left(f-\frac{v}{u}\right)G\left(f-u-\frac{v}{u}\right)\,\mbox{d}u\right]\,\mbox{d}v\\
 & =\left(\frac{P}{2\delta}\right)^{2}\cdot\int\limits _{\tilde{\delta}^{2}}^{\left(\frac{\overline{\delta}}{2}\right)^{2}}\left|\mathcal{K}(v)\right|^{2}\left[\int\limits _{\max\left\{ -\tilde{\delta},\frac{\overline{\delta}}{2}-\sqrt{\left(\frac{\overline{\delta}}{2}\right)^{2}-v}\right\} }^{\min\left\{ \frac{v}{-\tilde{\delta}},\frac{\overline{\delta}}{2}+\sqrt{\left(\frac{\overline{\delta}}{2}\right)^{2}-v}\right\} }\frac{1}{u}\cdot G\left(f-u-\frac{v}{u}\right)\,\mbox{d}u\right]\,\mbox{d}v.
\end{split}
\label{GlpartintegIVvereinfachtspecialint7}
\end{equation}

We further have: 
\begin{equation}
\begin{split}\frac{v}{-\tilde{\delta}}\leq\frac{\overline{\delta}}{2}+\sqrt{\left(\frac{\overline{\delta}}{2}\right)^{2}-v}\quad & \Longleftrightarrow\quad\left(\pm\frac{v}{-\tilde{\delta}}\mp\frac{\overline{\delta}}{2}\right)^{2}\leq\left(\frac{\overline{\delta}}{2}\right)^{2}-v\\
\quad & \Longleftrightarrow\quad\frac{v^{2}}{\tilde{\delta}^{2}}+\frac{\overline{\delta}}{\tilde{\delta}}v+\left(\frac{\overline{\delta}}{2}\right)^{2}\leq\left(\frac{\overline{\delta}}{2}\right)^{2}-v\\
\quad & \Longleftrightarrow\quad v\leq-\tilde{\delta}^{2}-\tilde{\delta}\overline{\delta}=-(\delta-f)^{2}-\left(\delta^{2}-f^{2}\right)\\
\quad & \Longleftrightarrow\quad v\leq-\delta^{2}+2\delta^{2}f-f^{2}-\delta^{2}+f^{2}=-2\delta^{2}+2\delta^{2}f=2\delta(f-\delta).
\end{split}
\label{GlrectshapedPSDded3Int4absh1}
\end{equation}

So if $v\leq2\delta(f-\delta)$ then $\frac{v}{-\tilde{\delta}}$
is the upper limit of the inner integral of (\ref{GlpartintegIVvereinfachtspecialint6})
else $\frac{\overline{\delta}}{2}+\sqrt{\left(\frac{\overline{\delta}}{2}\right)^{2}-v}$
is the upper limit. Note that 
\begin{equation}
\begin{split}2\delta(f-\delta)\leq\left(\frac{\overline{\delta}}{2}\right)^{2}\quad & \Longleftrightarrow\quad8\delta f-8\delta^{2}\leq f^{2}+2\delta f+\delta^{2}\\
\quad & \Longleftrightarrow\quad0\leq9\delta^{2}-6\delta f+f^{2}=\left(3\delta-f\right)^{2}\\
\quad & \Longleftrightarrow\quad f\leq3\delta.
\end{split}
\label{GlrectshapedPSDded3Int4absh2}
\end{equation}

This condition is always fulfilled since we may restrict the analysis
to that case, knowing that the Nonlinearity Double Integral is always
$0$ for $f>3\delta$. We also have 
\begin{equation}
\begin{split}-\tilde{\delta}\leq\frac{\overline{\delta}}{2}-\sqrt{\left(\frac{\overline{\delta}}{2}\right)^{2}-v}\quad & \Longleftrightarrow\quad\left(\pm(-\tilde{\delta})\mp\frac{\overline{\delta}}{2}\right)^{2}\leq\left(\frac{\overline{\delta}}{2}\right)^{2}-v\\
\quad & \Longleftrightarrow\quad(-\tilde{\delta})^{2}+\overline{\delta}\tilde{\delta}+\left(\frac{\overline{\delta}}{2}\right)^{2}\leq\left(\frac{\overline{\delta}}{2}\right)^{2}-v\\
\quad & \Longleftrightarrow\quad v\leq-\tilde{\delta}^{2}-\tilde{\delta}\overline{\delta}=2\delta(f-\delta).
\end{split}
\label{GlrectshapedPSDded3Int4abshxy10}
\end{equation}

So if $v\leq2\delta(f-\delta)$ then $-\tilde{\delta}$ is the lower
limit of the inner integral of (\ref{GlpartintegIVvereinfachtspecialint6})
else $\frac{\overline{\delta}}{2}-\sqrt{\left(\frac{\overline{\delta}}{2}\right)^{2}-v}$
is the lower limit. This leads to 
\begin{equation}
\begin{split}\int\limits _{0}^{\infty} & \left|\mathcal{K}(v)\right|^{2}\left[\int\limits _{0}^{\infty}\frac{1}{u}\cdot G\left(f-u\right)G\left(f-\frac{v}{u}\right)G\left(f-u-\frac{v}{u}\right)\,\mbox{d}u\right]\,\mbox{d}v\\
 & =\left(\frac{P}{2\delta}\right)^{2}\cdot\int\limits _{\tilde{\delta}^{2}}^{2\delta(f-\delta)}\left|\mathcal{K}(v)\right|^{2}\left[\int\limits _{-\tilde{\delta}}^{\frac{v}{-\tilde{\delta}}}\frac{1}{u}\,\mbox{d}u\right]\,\mbox{d}v\\
 & +\left(\frac{P}{2\delta}\right)^{2}\cdot\int\limits _{2\delta(f-\delta)}^{\left(\frac{\overline{\delta}}{2}\right)^{2}}\left|\mathcal{K}(v)\right|^{2}\left[\int\limits _{\frac{\overline{\delta}}{2}-\sqrt{\left(\frac{\overline{\delta}}{2}\right)^{2}-v}}^{\frac{\overline{\delta}}{2}+\sqrt{\left(\frac{\overline{\delta}}{2}\right)^{2}-v}}\frac{1}{u}\,\mbox{d}u\right]\,\mbox{d}v.
\end{split}
\label{GlpartintegIVvereinfachtspecialint8}
\end{equation}
and proves together with the remark following equation (\ref{GlrectshapedPSDded3Int4absh2})
the equation (\ref{eq:nyq_2}). The result for negative $f<-\delta$
follows from the symmetry property (\ref{Glausgangsymmetrie-1-1}).\\

\subsubsection*{The partial integral (\ref{eq:tre}) for $f=\delta$}

The value of the partial integral (\ref{eq:tre}) for $f=\delta$
is simply deduced by letting $|f|$ tend to $\delta$ in (\ref{eq:nyq_1})
or (\ref{eq:nyq_2}). It can be easily seen that in both cases the
limit value is: 
\begin{align}
\notag\mathcal{I}(f) & =\int\limits _{0}^{\delta^{2}}\left|\mathcal{K}(v)\right|^{2}\mathrm{ln}\left(\frac{\delta+\sqrt{\delta^{2}-v}}{\delta-\sqrt{\delta^{2}-v}}\right)\,\mbox{d}v.
\end{align}

\section*{Appendix B: Proof of the XCI integrals $\mathcal{I}_{m}(f)$}

In this Appendix we prove the expressions of the XCI integrals $\mathcal{I}_{m}(f)$
given in (\ref{eq:n_nyq_1})-(\ref{eq:n_nyq_2}). By the symmetry
(\ref{Glausgangsymmetrie-1-1}) we only need calculations at $f\geq0$.

\subsection{\label{ProofA} Proof for $f<\delta$ (resp.\ $|f|<\delta$)}

\subsection*{Calculation of partial integral (\ref{equno-1}), quadrant I:}

Regarding the integrand of the inner integral (\ref{equno-1}) we
deduce: 
\begin{equation}
\begin{split}G_{0}\left(f+u\right)\neq0\quad\Longleftrightarrow\quad f+u\leq\delta\quad\Longleftrightarrow\quad u\leq\delta-f:=\eta.\end{split}
\label{GlrectshapedPSDded1}
\end{equation}

Note that this implies for the following analysis of (\ref{equno-1})
that $\eta>0$ since by definition $u\geq0$. Otherwise the factor
$G_{0}\left(f+u\right)$ is zero and the integral (\ref{equno-1})
vanishes. This implies also that integral (\ref{equno-1}) vanishes
for $f>\delta$. For the second factor we have: 
\begin{equation}
\begin{split}G_{m}\left(f+\frac{v}{u}\right)\neq0\quad & \Longleftrightarrow\quad f+\frac{v}{u}-m\Delta\leq\delta\quad\text{and}\quad f+\frac{v}{u}-m\Delta\geq-\delta\\
\quad & \Longleftrightarrow\quad\frac{v}{u}\leq\eta+m\Delta\quad\text{and}\quad\frac{v}{u}\geq m\Delta-(\delta+f)\\
\quad & \Longleftrightarrow\quad\frac{v}{u}\leq\eta_{m}^{+}\quad\text{and}\quad\frac{v}{u}\geq\varepsilon_{m}^{-}
\end{split}
\label{GlrectshapedPSDded2-2}
\end{equation}
where we defined 
\begin{equation}
\begin{split}\eta_{m}^{+}:=m\Delta+\eta\quad\text{and}\quad\varepsilon_{m}^{-}:=m\Delta-(\delta+f).\end{split}
\label{GlrectshapedPSDded2Defdeltas}
\end{equation}

Since $0<\eta\leq\delta<\Delta$ we see that the first inequality
of (\ref{GlrectshapedPSDded2-2}) is never fulfilled for $m<0$. So
the integral (\ref{equno-1}) is zero for $m<0$. For $m>0$ we get
(since all terms are positive) 
\begin{equation}
\begin{split}G_{m}\left(f+\frac{v}{u}\right)\neq0\quad & \Longleftrightarrow\quad\frac{u}{v}\geq\frac{1}{\eta_{m}^{+}}\quad\text{and}\quad\frac{u}{v}\leq\frac{1}{\varepsilon_{m}^{-}}\\
\quad & \Longleftrightarrow\quad u\geq\frac{v}{\eta_{m}^{+}}\quad\text{and}\quad u\leq\frac{v}{\varepsilon_{m}^{-}}.
\end{split}
\label{GlrectshapedPSDded2b}
\end{equation}

Putting (\ref{GlrectshapedPSDded1}) and (\ref{GlrectshapedPSDded2b})
together this leads to the restrictions: 
\begin{equation}
\begin{split}\frac{v}{\eta_{m}^{+}}\leq u\leq\min\left\{ \eta,\frac{v}{\varepsilon_{m}^{-}}\right\} .\end{split}
\label{GlrectshapedPSDded2bR2}
\end{equation}

Note that this implies: 
\begin{equation}
\begin{split}\quad & \min\left\{ \eta,\frac{v}{\varepsilon_{m}^{-}}\right\} =\eta\quad\text{iff}\quad v\geq\varepsilon_{m}^{-}\eta\\
\quad\text{and}\quad\quad & \min\left\{ \eta,\frac{v}{\varepsilon_{m}^{-}}\right\} =\frac{v}{\varepsilon_{m}^{-}}\quad\text{iff}\quad v<\varepsilon_{m}^{-}\eta.
\end{split}
\label{GlrectshapedPSDded2bR3-1}
\end{equation}

For the third factor we have (note $u>0$): 
\begin{equation}
\begin{split}G_{m}\left(f+u+\frac{v}{u}\right)\neq0\quad & \Longleftrightarrow\quad u+\frac{v}{u}\leq\eta_{m}^{+}\quad\text{and}\quad u+\frac{v}{u}\geq\varepsilon_{m}^{-}\\
\quad & \Longleftrightarrow\quad u^{2}-\eta_{m}^{+}u+v\leq0\quad\text{and}\quad u^{2}-\varepsilon_{m}^{-}u+v\geq0.
\end{split}
\label{GlrectshapedPSDded3-1}
\end{equation}

Note that for $m>0$ the second inequality is a genuine restriction
because $\varepsilon_{m}^{-}>0$. Since 
\begin{equation}
\begin{split}u^{2}-\eta_{m}^{+}u+v\leq0\quad & \Longleftrightarrow\quad\left(u-\frac{\eta_{m}^{+}}{2}\right)^{2}\leq\left(\frac{\eta_{m}^{+}}{2}\right)^{2}-v\end{split}
\label{GlrectshapedPSDded3b-1}
\end{equation}
the factor $G_{m}\left(f+u+\frac{v}{u}\right)$ is always $0$ if
$\left(\frac{\eta_{m}^{+}}{2}\right)^{2}<v$. If $\left(\frac{\eta_{m}^{+}}{2}\right)^{2}\geq v$
then (\ref{GlrectshapedPSDded3b-1}) has solutions and 
\begin{equation}
\begin{split}\left(u-\frac{\eta_{m}^{+}}{2}\right)^{2} & \leq\left(\frac{\eta_{m}^{+}}{2}\right)^{2}-v\\
\quad\Longleftrightarrow\quad u & \leq\frac{\eta_{m}^{+}}{2}+\sqrt{\left(\frac{\eta_{m}^{+}}{2}\right)^{2}-v}:=u^{(1)}\quad\text{or}\quad u\geq\frac{\eta_{m}^{+}}{2}-\sqrt{\left(\frac{\eta_{m}^{+}}{2}\right)^{2}-v}:=u^{(0)}.
\end{split}
\label{GlrectshapedPSDded3c-1}
\end{equation}

Since $\frac{\eta_{m}^{+}}{2}>\frac{m\Delta}{2}\geq\eta$ taking into
account (\ref{GlrectshapedPSDded2bR2}) the only remaining restriction
is $u^{(0)}$. For the second inequality we get: 
\begin{equation}
\begin{split}u^{2}-\varepsilon_{m}^{-}u+v\geq0\quad & \Longleftrightarrow\quad\left(u-\frac{\varepsilon_{m}^{-}}{2}\right)^{2}\geq\left(\frac{\varepsilon_{m}^{-}}{2}\right)^{2}-v.\end{split}
\label{GlrectshapedPSDded3bX}
\end{equation}

This is no restriction if $v>\left(\frac{\varepsilon_{m}^{-}}{2}\right)^{2}$.
If $v\leq\left(\frac{\varepsilon_{m}^{-}}{2}\right)^{2}$ then the
condition is equivalent to: 
\begin{equation}
\begin{split}u\leq u^{(k)}\triangleq\frac{\varepsilon_{m}^{-}}{2}-\sqrt{\left(\frac{\varepsilon_{m}^{-}}{2}\right)^{2}-v}\quad\text{or}\quad u\geq u^{(n)}\triangleq\frac{\varepsilon_{m}^{-}}{2}+\sqrt{\left(\frac{\varepsilon_{m}^{-}}{2}\right)^{2}-v}.\end{split}
\label{GlrectshapedPSDded3bXb}
\end{equation}

Since $\frac{\varepsilon_{m}^{-}}{2}>\frac{m\Delta}{2}\geq\eta$ taking
again into account (\ref{GlrectshapedPSDded2bR2}) the only remaining
restriction is $u^{(k)}$. Note however that in general 
\begin{equation}
\begin{split}a-\sqrt{a^{2}-x}\leq b-\sqrt{b^{2}-x}.\end{split}
\label{GlrectshapedPSDded3bXgeneral}
\end{equation}

if $a\geq b$. Since $\frac{\varepsilon_{m}^{-}}{2}\geq\frac{\eta_{m}^{+}}{2}$
this leads to 
\begin{equation}
\begin{split}u^{(k)}\leq u^{(0)}\end{split}
\label{GlrectshapedPSDded3bXgeneralcos}
\end{equation}
and consequently $u^{(0)}$ is the lower limit for $u$. Thus using
all this the terms of the partial integral (\ref{equno-1}) read for
$f<\delta$ and $m>0$: 
\begin{equation}
\begin{split}\int\limits _{0}^{\infty} & \left|\mathcal{K}(v)\right|^{2}\left[\int\limits _{0}^{\infty}\frac{1}{u}\cdot G_{0}\left(f+u\right)G_{m}\left(f+\frac{v}{u}\right)G_{m}\left(f+u+\frac{v}{u}\right)\,\mbox{d}u\right]\,\mbox{d}v\\
 & =\int\limits _{0}^{\varepsilon_{m}^{-}\eta}\left|\mathcal{K}(v)\right|^{2}\left[\int\limits _{u^{(0)}}^{\frac{v}{\varepsilon_{m}^{-}}}\frac{1}{u}\,\mbox{d}u\right]\,\mbox{d}v+\int\limits _{\varepsilon_{m}^{-}\eta}^{\left(\frac{\eta_{m}^{+}}{2}\right)^{2}}\left|\mathcal{K}(v)\right|^{2}\left[\int\limits _{u^{(0)}}^{\eta}\frac{1}{u}\,\mbox{d}u\right]\,\mbox{d}v.
\end{split}
\label{GlpartXPMintegIvereinfacht}
\end{equation}

Finally we should take into account that 
\begin{equation}
\begin{split}\frac{\eta_{m}^{+}}{2}-\sqrt{\left(\frac{\eta_{m}^{+}}{2}\right)^{2}-v}\leq\eta\quad & \Longleftrightarrow\quad\sqrt{\left(\frac{\eta_{m}^{+}}{2}\right)^{2}-v}\geq\frac{\eta_{m}^{+}}{2}-\eta\\
\quad & \Longleftrightarrow\quad\left(\frac{\eta_{m}^{+}}{2}\right)^{2}-v\geq\left(\frac{\eta_{m}^{+}}{2}\right)^{2}-\eta\eta_{m}^{+}+\eta^{2}\\
\quad & \Longleftrightarrow\quad v\leq\eta\left(\eta_{m}^{+}-\eta\right)=m\Delta\eta
\end{split}
\label{GlGrenzenteilint1x}
\end{equation}
which imposes an upper restriction on the admissible values of $v$.
In the end we get for $f<\delta$ and $m>0$: 
\begin{equation}
\begin{split}\int\limits _{0}^{\infty} & \left|\mathcal{K}(v)\right|^{2}\left[\int\limits _{0}^{\infty}\frac{1}{u}\cdot G_{0}\left(f+u\right)G_{m}\left(f+\frac{v}{u}\right)G_{m}\left(f+u+\frac{v}{u}\right)\,\mbox{d}u\right]\,\mbox{d}v\\
 & =\int\limits _{0}^{\varepsilon_{m}^{-}\eta}\left|\mathcal{K}(v)\right|^{2}\left[\int\limits _{u^{(0)}}^{\frac{v}{\varepsilon_{m}^{-}}}\frac{1}{u}\,\mbox{d}u\right]\,\mbox{d}v+\int\limits _{\varepsilon_{m}^{-}\eta}^{m\Delta\eta}\left|\mathcal{K}(v)\right|^{2}\left[\int\limits _{u^{(0)}}^{\eta}\frac{1}{u}\,\mbox{d}u\right]\,\mbox{d}v\\
 & =\int\limits _{0}^{\varepsilon_{m}^{-}\eta}\left|\mathcal{K}(v)\right|^{2}\mathrm{ln}\left(\frac{\frac{v}{\eta_{m}^{+}}}{\frac{\eta_{m}^{+}}{2}-\sqrt{\left(\frac{\eta_{m}^{+}}{2}\right)^{2}-v}}\right)\,\mbox{d}v+\int\limits _{\varepsilon_{m}^{-}\eta}^{m\Delta\eta}\left|\mathcal{K}(v)\right|^{2}\mathrm{ln}\left(\frac{\eta}{\frac{\eta_{m}^{+}}{2}-\sqrt{\left(\frac{\eta_{m}^{+}}{2}\right)^{2}-v}}\right)\,\mbox{d}v.
\end{split}
\label{GlpartXPMintegIvereinfacht2}
\end{equation}

\rule{0pt}{1ex}

The first part (\ref{equno-1}) of $I_{XCI}(f)$ now reads:

\begin{equation}
\begin{split}2\left(\frac{P}{2\delta}\right)^{3}\sum\limits _{m=1}^{N_{c}}\left[\int\limits _{0}^{\varepsilon_{m}^{-}\eta}\left|\mathcal{K}(v)\right|^{2}\mathrm{ln}\left(\frac{\frac{v}{\varepsilon_{m}^{-}}}{\frac{\eta_{m}^{+}}{2}-\sqrt{\left(\frac{\eta_{m}^{+}}{2}\right)^{2}-v}}\right)\,\mbox{d}v+\int\limits _{\varepsilon_{m}^{-}\eta}^{m\Delta\eta}\left|\mathcal{K}(v)\right|^{2}\mathrm{ln}\left(\frac{\eta}{\frac{\eta_{m}^{+}}{2}-\sqrt{\left(\frac{\eta_{m}^{+}}{2}\right)^{2}-v}}\right)\,\mbox{d}v\right].\end{split}
\label{GlpartXPMintegIvereinfacht3}
\end{equation}
{\small{\rule{0pt}{2ex}}}{\small \par}

{\small{}}
\begin{figure}[h!]
\centering{}{\small{\includegraphics[width=0.5\textwidth]{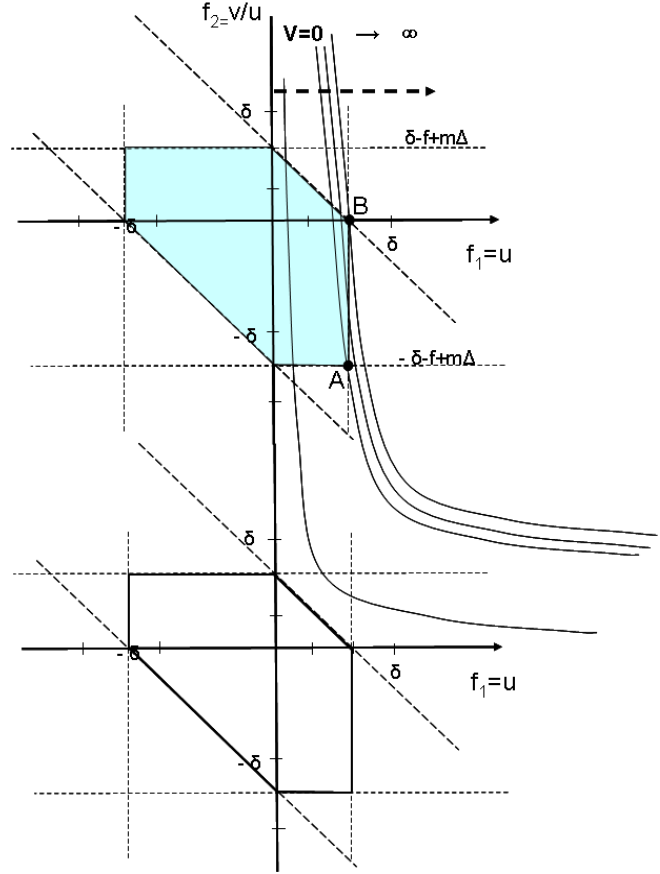}
\caption{{\small{\label{Abbfiggeointerp} Geometric interpretation for result
(\ref{GlpartXPMintegIvereinfacht3})}}}
}}
\end{figure}
{\small \par}

{\small{There is a useful geometrical interpretation for this result.
Fig. \ref{Abbfiggeointerp} depicts the corresponding situation in
the $(f_{1},f_{2})$-plane. Note that the transformation to the $(u,v)$-plane
is such that $u=f_{1}$ and $f_{2}=\frac{v}{u}$. Integration with
respect to $u$ geometrically means integration along the depicted
equipotential lines $\frac{v}{u}$. For a fixed small $v\approx0$
those lines intersect the lozenge-shaped domain of $G(\cdot)G(\cdot)G(\cdot)$
between the upper limit 
\begin{equation}
\begin{split}m\Delta+(\delta-f)-f_{1}=\eta_{m}^{+}-u\end{split}
\label{Glgeoint1}
\end{equation}
and the lower limit 
\begin{equation}
\begin{split}m\Delta-(\delta+f)=\varepsilon_{m}^{-}\end{split}
\label{Glgeoint2}
\end{equation}
until the point $A$ is reached. At this point 
\begin{equation}
\begin{split}\frac{v}{\delta-f}=\varepsilon_{m}^{-}\quad & \Longleftrightarrow\quad v=\eta\varepsilon_{m}^{-}.\end{split}
\label{Glgeoint2}
\end{equation}
}}{\small \par}

{\small{As long as $0<v\leq\eta\varepsilon_{m}^{-}$ for a given $v$
the equipotential line intersects first at the solution of 
\begin{equation}
\begin{split}\eta_{m}^{+}-u=\frac{v}{u}\end{split}
\label{Glgeoint3}
\end{equation}
which is 
\begin{equation}
\begin{split}\frac{\eta_{m}^{+}}{2}-\sqrt{\left(\frac{\eta_{m}^{+}}{2}\right)^{2}-v}\end{split}
\label{Glgeoint4}
\end{equation}
and the solution of 
\begin{equation}
\begin{split}\varepsilon_{m}^{-}=\frac{v}{u}\end{split}
\label{Glgeoint5}
\end{equation}
which is 
\begin{equation}
\begin{split}\frac{v}{\varepsilon_{m}^{-}}.\end{split}
\label{Glgeoint6}
\end{equation}
}}{\small \par}

{\small{This explains the first integral in (\ref{GlpartXPMintegIvereinfacht3}).
If $v$ increases and the equipotential line passes point $A$, it
intersects still at the solution (\ref{Glgeoint4}) of $\eta_{m}^{+}-u=\frac{v}{u}$
and then at the right limit line. In this case at $u=\eta$. This
is true until point $B$ is reached. At this point the equation 
\begin{equation}
\begin{split}\frac{v}{\eta}=m\Delta\quad & \Longleftrightarrow\quad v=m\Delta\eta\end{split}
\label{Glgeoint7}
\end{equation}
holds. All this explains second integral in (\ref{GlpartXPMintegIvereinfacht3}).}}\\
{\small \par}

\subsection*{Calculation of partial integral (\ref{eq:quattro-1}), quadrant IV:}

For the integrand of the inner integral (\ref{eq:quattro-1}) we deduce
according to (\ref{GlrectshapedPSDded1}): 
\begin{equation}
\begin{split}G_{0}\left(f+u\right)\neq0\quad\Longleftrightarrow\quad f+u\leq\delta\quad\Longleftrightarrow\quad u\leq\delta-f:=\eta.\end{split}
\label{GlrectshapedPSDded2}
\end{equation}

Note that this again implies for the following analysis of (\ref{eq:quattro-1})
that $\eta>0$ since by definition $u\geq0$. Otherwise the factor
$G_{0}\left(f+u\right)$ is zero and the integral (\ref{eq:quattro-1})
disappears. This implies also that integral (\ref{eq:quattro-1})
disappears for $f>\delta$.

For the second factor we have: 
\begin{equation}
\begin{split}G_{m}\left(f-\frac{v}{u}\right)\neq0\quad & \Longleftrightarrow\quad f-\frac{v}{u}-m\Delta\leq\delta\quad\text{and}\quad f-\frac{v}{u}-m\Delta\geq-\delta\\
\quad & \Longleftrightarrow\quad-\frac{v}{u}\leq\eta_{m}^{+}\quad\text{and}\quad-\frac{v}{u}\geq\varepsilon_{m}^{-}\\
\quad & \Longleftrightarrow\quad\frac{v}{u}\geq-\eta_{m}^{+}\quad\text{and}\quad\frac{v}{u}\leq-\varepsilon_{m}^{-}.
\end{split}
\label{GlrectshapedPSDded2Int2}
\end{equation}

Since 
\begin{equation}
\begin{split}-\varepsilon_{m}^{-}=(\delta+f)-m\Delta\leq2\delta-m\Delta\leq\Delta-m\Delta\end{split}
\label{GlrectshapedPSDded2Int2b}
\end{equation}
we see that the second inequality of (\ref{GlrectshapedPSDded2Int2})
is never fulfilled for $m>0$. So the integral (\ref{eq:quattro-1})
is zero for $m>0$. We then consider only $m<0$ in the following.
For $m<0$ we get (since all terms are positive) 
\begin{equation}
\begin{split}G_{m}\left(f-\frac{v}{u}\right)\neq0\quad & \Longleftrightarrow\quad\frac{u}{v}\leq\frac{1}{-\eta_{m}^{+}}\quad\text{and}\quad\frac{u}{v}\geq\frac{1}{-\varepsilon_{m}^{-}}\\
\quad & \Longleftrightarrow\quad u\leq\frac{v}{-\eta_{m}^{+}}\quad\text{and}\quad u\geq\frac{v}{-\varepsilon_{m}^{-}}.
\end{split}
\label{GlrectshapedPSDded2bInt2}
\end{equation}

Now (\ref{GlrectshapedPSDded2}) and (\ref{GlrectshapedPSDded2bInt2})
together give: 
\begin{equation}
\begin{split}\frac{v}{-\varepsilon_{m}^{-}}\leq u\leq\min\left\{ \eta,\frac{v}{-\eta_{m}^{+}}\right\} .\end{split}
\label{GlrectshapedPSDded2bR2Int2}
\end{equation}

Note that this implies: 
\begin{equation}
\begin{split}\quad & \min\left\{ \eta,\frac{v}{-\eta_{m}^{+}}\right\} =\eta\quad\text{iff}\quad v\leq-\varepsilon_{m}^{-}\eta\\
\quad\text{and}\quad\quad & \min\left\{ \eta,\frac{v}{-\eta_{m}^{+}}\right\} =\frac{v}{-\eta_{m}^{+}}\quad\text{iff}\quad v>-\varepsilon_{m}^{-}\eta.
\end{split}
\label{GlrectshapedPSDded2bR3}
\end{equation}

For the third factor we have: 
\begin{equation}
\begin{split}G_{m}\left(f+u-\frac{v}{u}\right)\neq0\quad & \Longleftrightarrow\quad u-\frac{v}{u}\leq\eta_{m}^{+}\quad\text{and}\quad u-\frac{v}{u}\geq-\varepsilon_{m}^{-}\\
\quad & \Longleftrightarrow\quad u^{2}-\eta_{m}^{+}u-v\leq0\quad\text{and}\quad u^{2}+\varepsilon_{m}^{-}u-v\geq0.
\end{split}
\label{GlrectshapedPSDded3Int2}
\end{equation}

For the first inequality we deduce: 
\begin{equation}
\begin{split}u^{2}-\eta_{m}^{+}u-v\leq0\quad & \Longleftrightarrow\quad\left(u-\frac{\eta_{m}^{+}}{2}\right)^{2}-\left(\frac{\eta_{m}^{+}}{2}\right)^{2}-v\leq0\\
\quad & \Longleftrightarrow\quad\left(u-\frac{\eta_{m}^{+}}{2}\right)^{2}\leq\left(\frac{\eta_{m}^{+}}{2}\right)^{2}+v.
\end{split}
\label{GlrectshapedPSDded3Int2first}
\end{equation}

This implies 
\begin{equation}
\begin{split}u\leq\frac{\eta_{m}^{+}}{2}+\sqrt{\left(\frac{\eta_{m}^{+}}{2}\right)^{2}+v}:=u^{(1)}\quad\text{or}\quad u\geq\frac{\eta_{m}^{+}}{2}-\sqrt{\left(\frac{\eta_{m}^{+}}{2}\right)^{2}+v}:=u^{(0)}.\end{split}
\label{GlrectshapedPSDded3Int2firstb}
\end{equation}

However $\eta_{m}^{+}$ is negative since $m<0$ and so the last condition
doesn't represent a restriction and 
\begin{equation}
\begin{split}u\leq\frac{\eta_{m}^{+}}{2}+\sqrt{\left(\frac{\eta_{m}^{+}}{2}\right)^{2}+v}=u^{(1)}\end{split}
\label{GlrectshapedPSDded3Int2firstb2}
\end{equation}
remains. Now (note that $\eta-\frac{\eta_{m}^{+}}{2}$ is positive
since $m<0$): 
\begin{equation}
\begin{split}u^{(1)}=\frac{\eta_{m}^{+}}{2}+\sqrt{\left(\frac{\eta_{m}^{+}}{2}\right)^{2}+v}\leq\eta\quad & \Longleftrightarrow\quad\sqrt{\left(\frac{\eta_{m}^{+}}{2}\right)^{2}+v}\leq\eta-\frac{\eta_{m}^{+}}{2}\\
\quad & \Longleftrightarrow\quad\left(\frac{\eta_{m}^{+}}{2}\right)^{2}+v\leq\eta^{2}-\eta_{m}^{+}\eta\left(\frac{\eta_{m}^{+}}{2}\right)^{2}\\
\quad & \Longleftrightarrow\quad v\leq\eta\left(\eta-\eta_{m}^{+}\right)=-m\Delta\eta.
\end{split}
\label{GlrectshapedPSDded3Int2firstc}
\end{equation}

So for $v\leq-m\Delta\eta<-\varepsilon_{m}^{-}\eta$ the upper limit
of the inner integral is $u^{(1)}$ if $v\leq-m\Delta\eta$ else the
upper limit is $\eta$. For the second inequality in (\ref{GlrectshapedPSDded3Int2})
we deduce: 
\begin{equation}
\begin{split}u^{2}+\varepsilon_{m}^{-}u-v\geq0\quad & \Longleftrightarrow\quad\left(u+\frac{\varepsilon_{m}^{-}}{2}\right)^{2}-\left(\frac{\varepsilon_{m}^{-}}{2}\right)^{2}-v\geq0\\
\quad & \Longleftrightarrow\quad\left(u+\frac{\varepsilon_{m}^{-}}{2}\right)^{2}\geq\left(\frac{\varepsilon_{m}^{-}}{2}\right)^{2}+v.
\end{split}
\label{GlrectshapedPSDded3Int2second}
\end{equation}

This implies 
\begin{equation}
\begin{split}u\geq u^{(k)}\triangleq-\frac{\varepsilon_{m}^{-}}{2}+\sqrt{\left(\frac{\varepsilon_{m}^{-}}{2}\right)^{2}+v}\quad\text{or}\quad u\leq u^{(n)}\triangleq-\frac{\varepsilon_{m}^{-}}{2}-\sqrt{\left(\frac{\varepsilon_{m}^{-}}{2}\right)^{2}+v}.\end{split}
\label{GlrectshapedPSDded3bXbInt2}
\end{equation}

Since $-\frac{\varepsilon_{m}^{-}}{2}>-\frac{m\Delta}{2}\geq\eta$
and since $u^{(n)}$ is always negative, this doesn't impose new restrictions.
Consequently $\frac{v}{-\eta_{m}^{+}}$ is always the lower limit
for $u$. Using all this, the terms of the partial integral (\ref{eq:quattro-1})
read for $f<\delta$ and $m<0$: 
\begin{equation}
\begin{split}\int\limits _{0}^{\infty} & \left|\mathcal{K}(v)\right|^{2}\left[\int\limits _{0}^{\infty}\frac{1}{u}\cdot G_{0}\left(f+u\right)G_{m}\left(f-\frac{v}{u}\right)G_{m}\left(f+u-\frac{v}{u}\right)\,\mbox{d}u\right]\,\mbox{d}v\\
 & =\int\limits _{0}^{-m\Delta\eta}\left|\mathcal{K}(v)\right|^{2}\left[\int\limits _{\frac{v}{-\varepsilon_{m}^{-}}}^{\frac{\eta_{m}^{+}}{2}+\sqrt{\left(\frac{\eta_{m}^{+}}{2}\right)^{2}+v}}\frac{1}{u}\,\mbox{d}u\right]\,\mbox{d}v+\int\limits _{-m\Delta\eta}^{-\varepsilon_{m}^{-}\eta}\left|\mathcal{K}(v)\right|^{2}\left[\int\limits _{\frac{v}{-\varepsilon_{m}^{-}}}^{\eta}\frac{1}{u}\,\mbox{d}u\right]\,\mbox{d}v\\
 & =\int\limits _{0}^{-m\Delta\eta}\left|\mathcal{K}(v)\right|^{2}\mathrm{ln}\left(\frac{\frac{\eta_{m}^{+}}{2}+\sqrt{\left(\frac{\eta_{m}^{+}}{2}\right)^{2}+v}}{\frac{v}{-\varepsilon_{m}^{-}}}\right)\,\mbox{d}v+\int\limits _{-m\Delta\eta}^{-\varepsilon_{m}^{-}\eta}\left|\mathcal{K}(v)\right|^{2}\mathrm{ln}\left(\frac{\eta}{\frac{v}{-\varepsilon_{m}^{-}}}\right)\,\mbox{d}v.
\end{split}
\label{GlpartXPMintegIIvereinfacht}
\end{equation}

\rule{0pt}{1ex}

The second part (\ref{eq:quattro-1}) of $I_{XCI}(f)$ now reads:

\begin{equation}
\begin{split}2\left(\frac{P}{2\delta}\right)^{3}\sum\limits _{m=-N_{c}}^{-1}\left[\int\limits _{0}^{-m\Delta\eta}\left|\mathcal{K}(v)\right|^{2}\mathrm{ln}\left(\frac{\frac{\eta_{m}^{+}}{2}+\sqrt{\left(\frac{\eta_{m}^{+}}{2}\right)^{2}+v}}{\frac{v}{-\varepsilon_{m}^{-}}}\right)\,\mbox{d}v+\int\limits _{-m\Delta\eta}^{-\varepsilon_{m}^{-}\eta}\left|\mathcal{K}(v)\right|^{2}\mathrm{ln}\left(\frac{\eta}{\frac{v}{-\varepsilon_{m}^{-}}}\right)\,\mbox{d}v\right].\end{split}
\label{GlpartXPMintegIIvereinfacht3}
\end{equation}

\subsection*{Calculation of partial integral (\ref{eqdue-1}), quadrant II:}

{\small{From the inner integral (}}\ref{eqdue-1}{\small{) we deduce:
\begin{equation}
\begin{split}G_{0}\left(f-u\right)\neq0\quad & \Longleftrightarrow\quad-\delta\leq f-u\leq\delta\quad\Longleftrightarrow\quad-\varepsilon:=-(\delta+f)\leq-u\leq\eta\\
\quad & \Longleftrightarrow\quad-\eta\leq u\leq\varepsilon.
\end{split}
\label{GlrectshapedPSDded1Int3}
\end{equation}
}}{\small \par}

{\small{Note that since we }}\textbf{\small{are supposing}}{\small{
in this analysis that $\eta:=\delta-f>0$ the first inequality is
no restriction. So we get in this case: 
\begin{equation}
\begin{split}G_{0}\left(f-u\right)\neq0\quad & \Longleftrightarrow\quad u\leq\varepsilon.\end{split}
\label{GlrectshapedPSDded1Int3b}
\end{equation}
}}{\small \par}

{\small{
}}{\small \par}

{\small{For the second factor we have according to (\ref{GlrectshapedPSDded2}):
\begin{equation}
\begin{split}G_{m}\left(f+\frac{v}{u}\right)\neq0\quad & \Longleftrightarrow\quad\frac{v}{u}\leq\eta_{m}^{+}\quad\text{and}\quad\frac{v}{u}\geq\varepsilon_{m}^{-}.\end{split}
\label{GlrectshapedPSDded2Int3}
\end{equation}
}}{\small \par}

{\small{Reasoning analogously to (\ref{GlrectshapedPSDded2}) the
first inequality of (\ref{GlrectshapedPSDded2Int3}) is never fulfilled
for $m<0$. So the integral (}}\ref{eqdue-1}{\small{) is zero for
$m<0$. For $m>0$ we get again (since all terms are positive) 
\begin{equation}
\begin{split}G_{m}\left(f+\frac{v}{u}\right)\neq0\quad & \Longleftrightarrow\quad u\geq\frac{v}{\eta_{m}^{+}}\quad\text{and}\quad u\leq\frac{v}{\varepsilon_{m}^{-}}.\end{split}
\label{GlrectshapedPSDded2bInt3}
\end{equation}
}}{\small \par}

{\small{Putting (\ref{GlrectshapedPSDded1Int3b}) and (\ref{GlrectshapedPSDded2bInt3})
together this leads to the restrictions: 
\begin{equation}
\begin{split}\frac{v}{\eta_{m}^{+}}\leq u\leq\min\left\{ \varepsilon,\frac{v}{\varepsilon_{m}^{-}}\right\} .\end{split}
\label{GlrectshapedPSDded2bR2Int3}
\end{equation}
}}{\small \par}

{\small{Note that this immediately implies: 
\begin{equation}
\begin{split}\quad & \min\left\{ \varepsilon,\frac{v}{\varepsilon_{m}^{-}}\right\} =\varepsilon\quad\text{iff}\quad v\geq\varepsilon_{m}^{-}\varepsilon\\
\quad\text{and}\quad\quad & \min\left\{ \varepsilon,\frac{v}{\varepsilon_{m}^{-}}\right\} =\frac{v}{\varepsilon_{m}^{-}}\quad\text{iff}\quad v<\varepsilon_{m}^{-}\varepsilon.
\end{split}
\label{GlrectshapedPSDded2bR3Int}
\end{equation}
}}{\small \par}

{\small{
}}{\small \par}

{\small{For the third factor we have (note $u>0$): 
\begin{equation}
\begin{split}G_{m}\left(f-u+\frac{v}{u}\right)\neq0\quad & \Longleftrightarrow\quad-u+\frac{v}{u}\leq\eta_{m}^{+}\quad\text{and}\quad-u+\frac{v}{u}\geq\varepsilon_{m}^{-}\\
\quad & \Longleftrightarrow\quad-u^{2}-\eta_{m}^{+}u+v\leq0\quad\text{and}\quad-u^{2}-\varepsilon_{m}^{-}u+v\geq0\\
\quad & \Longleftrightarrow\quad u^{2}+\eta_{m}^{+}u-v\geq0\quad\text{and}\quad u^{2}+\varepsilon_{m}^{-}u-v\leq0.
\end{split}
\label{GlrectshapedPSDded3Int3}
\end{equation}
}}{\small \par}

{\small{We get 
\begin{equation}
\begin{split}u^{2}+\eta_{m}^{+}u-v\geq0\quad & \Longleftrightarrow\quad\left(u+\frac{\eta_{m}^{+}}{2}\right)^{2}\geq\left(\frac{\eta_{m}^{+}}{2}\right)^{2}+v\\
\quad\text{and}\quad u^{2}+\varepsilon_{m}^{-}u-v\leq0\quad & \Longleftrightarrow\quad\left(u+\frac{\varepsilon_{m}^{-}}{2}\right)^{2}\leq\left(\frac{\varepsilon_{m}^{-}}{2}\right)^{2}+v.
\end{split}
\label{GlrectshapedPSDded3bInt3}
\end{equation}
}}{\small \par}

{\small{This leads to the conditions: 
\begin{equation}
\begin{split}u & \geq u^{(0)}:=-\frac{\eta_{m}^{+}}{2}+\sqrt{\left(\frac{\eta_{m}^{+}}{2}\right)^{2}+v}\quad\text{or}\quad u\leq u^{(1)}:=-\frac{\eta_{m}^{+}}{2}-\sqrt{\left(\frac{\eta_{m}^{+}}{2}\right)^{2}+v}\end{split}
\label{GlrectshapedPSDded3cInt3f1}
\end{equation}
}}{\small \par}

{\small{and 
\begin{equation}
\begin{split}u & \geq u^{(k)}:=-\frac{\varepsilon_{m}^{-}}{2}-\sqrt{\left(\frac{\varepsilon_{m}^{-}}{2}\right)^{2}+v}\quad\text{or}\quad u\leq u^{(n)}:=-\frac{\varepsilon_{m}^{-}}{2}+\sqrt{\left(\frac{\varepsilon_{m}^{-}}{2}\right)^{2}+v}.\end{split}
\label{GlrectshapedPSDded3cInt3f2}
\end{equation}
}}{\small \par}

{\small{Since the expression $-\frac{\eta_{m}^{+}}{2}<0$ and $-\frac{\varepsilon_{m}^{-}}{2}<0$
for $m>0$ the first inequality of (\ref{GlrectshapedPSDded3bInt3})
is equivalent to 
\begin{equation}
\begin{split}u & \geq u^{(0)}:=-\frac{\eta_{m}^{+}}{2}+\sqrt{\left(\frac{\eta_{m}^{+}}{2}\right)^{2}+v}\end{split}
\label{GlrectshapedPSDded3cInt3f3}
\end{equation}
}}{\small \par}

{\small{and the remaining restriction in the second is: 
\begin{equation}
\begin{split}u\leq u^{(n)}:=-\frac{\varepsilon_{m}^{-}}{2}+\sqrt{\left(\frac{\varepsilon_{m}^{-}}{2}\right)^{2}+v}.\end{split}
\label{GlrectshapedPSDded3cInt3f4}
\end{equation}
}}{\small \par}

{\small{Consequently the terms of the partial integral (}}\ref{eqdue-1}{\small{)
read for $f<\delta$ and $m>0$: 
\begin{equation}
\begin{split}\int\limits _{0}^{\infty} & \left|\mathcal{K}(v)\right|^{2}\left[\int\limits _{0}^{\infty}\frac{1}{u}\cdot G_{0}\left(f-u\right)G_{m}\left(f+\frac{v}{u}\right)G_{m}\left(f-u+\frac{v}{u}\right)\,\mbox{d}u\right]\,\mbox{d}v\\
 & =\int\limits _{0}^{\varepsilon_{m}^{-}\varepsilon}\left|\mathcal{K}(v)\right|^{2}\left[\int\limits _{\max\left\{ \frac{v}{\eta_{m}^{+}},u^{(0)}\right\} }^{\min\left\{ \frac{v}{\varepsilon_{m}^{-}},u^{(n)}\right\} }\frac{1}{u}\,\mbox{d}u\right]\,\mbox{d}v+\int\limits _{\varepsilon_{m}^{-}\varepsilon}^{\infty}\left|\mathcal{K}(v)\right|^{2}\left[\int\limits _{\max\left\{ \frac{v}{\eta_{m}^{+}},u^{(0)}\right\} }^{\min\left\{ \varepsilon,u^{(n)}\right\} }\frac{1}{u}\,\mbox{d}u\right]\,\mbox{d}v.
\end{split}
\label{GlpartXPMintegIIIvereinfacht}
\end{equation}
}}{\small \par}

{\small{Now 
\begin{equation}
\begin{split}u^{(0)}\leq\frac{v}{\eta_{m}^{+}}\quad & \Longleftrightarrow\quad-\frac{\eta_{m}^{+}}{2}+\sqrt{\left(\frac{\eta_{m}^{+}}{2}\right)^{2}+v}\leq\frac{v}{\eta_{m}^{+}}\\
\quad & \Longleftrightarrow\quad\sqrt{\left(\frac{\eta_{m}^{+}}{2}\right)^{2}+v}\leq\frac{\eta_{m}^{+}}{2}+\frac{v}{\eta_{m}^{+}}\\
\quad & \Longleftrightarrow\quad\left(\frac{\eta_{m}^{+}}{2}\right)^{2}+v\leq\left(\frac{\eta_{m}^{+}}{2}\right)^{2}+v+\left(\frac{v}{\eta_{m}^{+}}\right)^{2}\quad\Longleftrightarrow\quad0\leq\left(\frac{v}{\eta_{m}^{+}}\right)^{2}
\end{split}
\label{GlrectshapedPSDded3cInt3fX1}
\end{equation}
which is always true. So $\frac{v}{\eta_{m}^{+}}$ is always the lower
limit of the inner integrals. Since 
\begin{equation}
\begin{split}u^{(n)}\leq\frac{v}{\varepsilon_{m}^{-}}\quad & \Longleftrightarrow\quad-\frac{\varepsilon_{m}^{-}}{2}+\sqrt{\left(\frac{\varepsilon_{m}^{-}}{2}\right)^{2}+v}\leq\frac{v}{\varepsilon_{m}^{-}}\\
\quad & \Longleftrightarrow\quad\sqrt{\left(\frac{\varepsilon_{m}^{-}}{2}\right)^{2}+v}\leq\frac{v}{\varepsilon_{m}^{-}}+\frac{\varepsilon_{m}^{-}}{2}\\
\quad & \Longleftrightarrow\quad\left(\frac{\varepsilon_{m}^{-}}{2}\right)^{2}+v\leq\left(\frac{\varepsilon_{m}^{-}}{2}\right)^{2}+v+\left(\frac{v}{\varepsilon_{m}^{-}}\right)^{2}\quad\Longleftrightarrow\quad0\leq\left(\frac{v}{\varepsilon_{m}^{-}}\right)^{2}
\end{split}
\label{GlrectshapedPSDded3cInt3fX2}
\end{equation}
is also always true, $u^{(n)}$ is the upper limit of the first inner
integral. Additionally 
\begin{equation}
\begin{split}\varepsilon\leq u^{(n)}\quad & \Longleftrightarrow\quad\varepsilon\leq-\frac{\varepsilon_{m}^{-}}{2}+\sqrt{\left(\frac{\varepsilon_{m}^{-}}{2}\right)^{2}+v}\\
\quad & \Longleftrightarrow\quad\frac{\varepsilon_{m}^{-}}{2}+\varepsilon\leq\sqrt{\left(\frac{\varepsilon_{m}^{-}}{2}\right)^{2}+v}\\
\quad & \Longleftrightarrow\quad\left(\frac{\varepsilon_{m}^{-}}{2}\right)^{2}+\varepsilon_{m}^{-}\varepsilon+\varepsilon^{2}\leq\left(\frac{\varepsilon_{m}^{-}}{2}\right)^{2}+v\\
\quad & \Longleftrightarrow\quad\varepsilon\left(\varepsilon_{m}^{-}+\varepsilon\right)=m\Delta\varepsilon\leq v.
\end{split}
\label{GlrectshapedPSDded3cInt3fX3}
\end{equation}
}}{\small \par}

{\small{So for $f<\delta$ and $m>0$ we get: 
\begin{equation}
\begin{split}\int\limits _{0}^{\infty} & \left|\mathcal{K}(v)\right|^{2}\left[\int\limits _{0}^{\infty}\frac{1}{u}\cdot G_{0}\left(f-u\right)G_{m}\left(f+\frac{v}{u}\right)G_{m}\left(f-u+\frac{v}{u}\right)\,\mbox{d}u\right]\,\mbox{d}v\\
 & =\int\limits _{0}^{m\Delta\varepsilon}\left|\mathcal{K}(v)\right|^{2}\left[\int\limits _{\frac{v}{\eta_{m}^{+}}}^{u^{(n)}}\frac{1}{u}\,\mbox{d}u\right]\,\mbox{d}v+\int\limits _{m\Delta\varepsilon}^{\infty}\left|\mathcal{K}(v)\right|^{2}\left[\int\limits _{\frac{v}{\eta_{m}^{+}}}^{\varepsilon}\frac{1}{u}\,\mbox{d}u\right]\,\mbox{d}v.
\end{split}
\label{GlpartXPMintegIIIvereinfachtB}
\end{equation}
}}{\small \par}

{\small{Finally we note that 
\begin{equation}
\begin{split}\frac{v}{\eta_{m}^{+}}\leq\varepsilon\quad & \Longleftrightarrow\quad v\leq\eta_{m}^{+}\varepsilon\end{split}
\label{GlGrenzenteilint1xInt3}
\end{equation}
which leads to a corresponding restriction on $v$ for the first integral.
Consequently we arrive at: 
\begin{equation}
\begin{split}\int\limits _{0}^{\infty} & \left|\mathcal{K}(v)\right|^{2}\left[\int\limits _{0}^{\infty}\frac{1}{u}\cdot G_{0}\left(f-u\right)G_{m}\left(f+\frac{v}{u}\right)G_{m}\left(f-u+\frac{v}{u}\right)\,\mbox{d}u\right]\,\mbox{d}v\\
 & =\int\limits _{0}^{m\Delta\varepsilon}\left|\mathcal{K}(v)\right|^{2}\left[\int\limits _{\frac{v}{\eta_{m}^{+}}}^{u^{(n)}}\frac{1}{u}\,\mbox{d}u\right]\,\mbox{d}v+\int\limits _{m\Delta\varepsilon}^{\eta_{m}^{+}\varepsilon}\left|\mathcal{K}(v)\right|^{2}\left[\int\limits _{\frac{v}{\eta_{m}^{+}}}^{\varepsilon}\frac{1}{u}\,\mbox{d}u\right]\,\mbox{d}v\\
 & =\int\limits _{0}^{m\Delta\varepsilon}\left|\mathcal{K}(v)\right|^{2}\mathrm{ln}\left(\frac{-\frac{\varepsilon_{m}^{-}}{2}+\sqrt{\left(\frac{\varepsilon_{m}^{-}}{2}\right)^{2}+v}}{\frac{v}{\eta_{m}^{+}}}\right)\,\mbox{d}v+\int\limits _{m\Delta\varepsilon}^{\eta_{m}^{+}\varepsilon}\left|\mathcal{K}(v)\right|^{2}\mathrm{ln}\left(\frac{\eta_{m}^{+}\varepsilon}{v}.\right)\,\mbox{d}v
\end{split}
\label{GlpartXPMintegIIIvereinfachtC}
\end{equation}
}}{\small \par}

{\small{\rule{0pt}{1ex}}}{\small \par}

{\small{The third part (}}\ref{eqdue-1}{\small{) of $I_{XCI}(f)$
now reads:}}{\small \par}

\begin{equation}
\begin{split}2\left(\frac{P}{2\delta}\right)^{3}\cdot\sum\limits _{m=1}^{N_{c}}\left[\int\limits _{0}^{m\Delta\varepsilon}\left|\mathcal{K}(v)\right|^{2}\mathrm{ln}\left(\frac{-\frac{\varepsilon_{m}^{-}}{2}+\sqrt{\left(\frac{\varepsilon_{m}^{-}}{2}\right)^{2}+v}}{\frac{v}{\eta_{m}^{+}}}\right)\,\mbox{d}v+\int\limits _{m\Delta\varepsilon}^{\eta_{m}^{+}\varepsilon}\left|\mathcal{K}(v)\right|^{2}\mathrm{ln}\left(\frac{\eta_{m}^{+}\varepsilon}{v}.\right)\,\mbox{d}v\right].\end{split}
\label{GlpartXPMintegIIIvereinfacht3end}
\end{equation}
{\small{
}}{\small \par}

\subsection*{Calculation of partial integral (\ref{eq:tre-1}), quadrant III:}

{\small{For the integrand of the inner integral (}}\ref{eq:tre-1}{\small{)
we derive according to (\ref{GlrectshapedPSDded1Int3b}) (note that
we we suppose $f<\delta$): 
\begin{equation}
\begin{split}G_{0}\left(f-u\right)\neq0\quad & \Longleftrightarrow\quad u\leq\varepsilon.\end{split}
\label{GlrectshapedPSDded1Int4}
\end{equation}
}}{\small \par}

{\small{
}}{\small \par}

{\small{For the second factor we have, following (\ref{GlrectshapedPSDded2Int2}):
\begin{equation}
\begin{split}G_{m}\left(f-\frac{v}{u}\right)\neq0\quad & \Longleftrightarrow\quad\frac{v}{u}\geq-\eta_{m}^{+}\quad\text{and}\quad\frac{v}{u}\leq-\varepsilon_{m}^{-}\end{split}
\label{GlrectshapedPSDded2Int4}
\end{equation}
and again we note that the second inequality of (\ref{GlrectshapedPSDded2Int4})
is never fulfilled for $m>0$ and therefore the integral (}}\ref{eq:tre-1}{\small{)
is zero for $m>0$. We thus consider only $m<0$ in the following.
For $m<0$ we get (since we suppose $f<\delta$ and all terms are
positive) 
\begin{equation}
\begin{split}G_{m}\left(f-\frac{v}{u}\right)\neq0\quad & \Longleftrightarrow\quad u\leq\frac{v}{-\eta_{m}^{+}}\quad\text{and}\quad u\geq\frac{v}{-\varepsilon_{m}^{-}}.\end{split}
\label{GlrectshapedPSDded2bInt4}
\end{equation}
}}{\small \par}

{\small{Now (\ref{GlrectshapedPSDded1Int4}) and (\ref{GlrectshapedPSDded2bInt4})
together give: 
\begin{equation}
\begin{split}\frac{v}{-\varepsilon_{m}^{-}}\leq u\leq\min\left\{ \varepsilon,\frac{v}{-\eta_{m}^{+}}\right\} .\end{split}
\label{GlrectshapedPSDded2bR2Int4}
\end{equation}
}}{\small \par}

{\small{Note that this implies: 
\begin{equation}
\begin{split}\quad & \min\left\{ \varepsilon,\frac{v}{-\eta_{m}^{+}}\right\} =\varepsilon\quad\text{iff}\quad v\leq-\eta_{m}^{+}\varepsilon\\
\quad\text{and}\quad\quad & \min\left\{ \varepsilon,\frac{v}{-\eta_{m}^{+}}\right\} =\frac{v}{-\eta_{m}^{+}}\quad\text{iff}\quad v>-\eta_{m}^{+}\varepsilon.
\end{split}
\label{GlrectshapedPSDded2bR3Int4}
\end{equation}
}}{\small \par}

{\small{
}}{\small \par}

{\small{For the third factor we have: 
\begin{equation}
\begin{split}G_{m}\left(f-u-\frac{v}{u}\right)\neq0\quad & \Longleftrightarrow\quad-u-\frac{v}{u}\leq\eta_{m}^{+}\quad\text{and}\quad-u-\frac{v}{u}\geq\varepsilon_{m}^{-}\\
\quad & \Longleftrightarrow\quad-u^{2}-\eta_{m}^{+}u-v\leq0\quad\text{and}\quad-u^{2}-\varepsilon_{m}^{-}u-v\geq0\\
\quad & \Longleftrightarrow\quad u^{2}+\eta_{m}^{+}u+v\geq0\quad\text{and}\quad u^{2}+\varepsilon_{m}^{-}u+v\leq0.
\end{split}
\label{GlrectshapedPSDded3Int4}
\end{equation}
}}{\small \par}

{\small{The first inequality is not a new restriction since by (\ref{GlrectshapedPSDded2Int4})
\begin{equation}
\begin{split}-\frac{v}{u}\leq\eta_{m}^{+}\end{split}
\label{GlrectshapedPSDded2Int4Zusatz}
\end{equation}
and $u\geq0$. We thus get the condition 
\begin{equation}
\begin{split}u^{2}+\varepsilon_{m}^{-}u+v\leq0\quad & \Longleftrightarrow\quad\left(u+\frac{\varepsilon_{m}^{-}}{2}\right)^{2}\leq\left(\frac{\varepsilon_{m}^{-}}{2}\right)^{2}-v.\end{split}
\label{GlrectshapedPSDded3bInt4}
\end{equation}
}}{\small \par}

{\small{The inequality shows that the factor $G_{m}\left(f-u-\frac{v}{u}\right)$
is always $0$ if $\left(\frac{\varepsilon_{m}^{-}}{2}\right)^{2}<v$.
If $\left(\frac{\varepsilon_{m}^{-}}{2}\right)^{2}\geq v$ then (\ref{GlrectshapedPSDded3bInt4})
has solutions and 
\begin{equation}
\begin{split}\left(u+\frac{\varepsilon_{m}^{-}}{2}\right)^{2} & \leq\left(\frac{\varepsilon_{m}^{-}}{2}\right)^{2}-v\\
\quad\Longleftrightarrow\quad u & \leq u^{(1)}:=-\frac{\varepsilon_{m}^{-}}{2}+\sqrt{\left(\frac{\varepsilon_{m}^{-}}{2}\right)^{2}-v}\quad\text{or}\quad u\geq u^{(0)}:=-\frac{\varepsilon_{m}^{-}}{2}-\sqrt{\left(\frac{\varepsilon_{m}^{-}}{2}\right)^{2}-v}.
\end{split}
\label{GlrectshapedPSDded3cInt4}
\end{equation}
}}{\small \par}

{\small{Since (note that $m<0$!) $u^{(1)}\geq-\frac{\varepsilon_{m}^{-}}{2}\geq\varepsilon$,
the first condition is no further restriction if we take (\ref{GlrectshapedPSDded1Int4})
into account. Thus the only remaining restriction is 
\begin{equation}
\begin{split}u\geq u^{(0)}:=-\frac{\varepsilon_{m}^{-}}{2}-\sqrt{\left(\frac{\varepsilon_{m}^{-}}{2}\right)^{2}-v}.\end{split}
\label{GlrectshapedPSDded3cInt4b}
\end{equation}
}}{\small \par}

{\small{Since by (\ref{GlrectshapedPSDded2bR2Int4}) $u^{(0)}$ should
be less than $\varepsilon$ we get 
\begin{equation}
\begin{split}u^{(0)}\leq\varepsilon\quad & \Longleftrightarrow\quad\sqrt{\left(\frac{\varepsilon_{m}^{-}}{2}\right)^{2}-v}\geq-\frac{\varepsilon_{m}^{-}}{2}-\varepsilon\\
\quad & \Longleftrightarrow\quad-v\geq\varepsilon\left(\varepsilon_{m}^{-}+\varepsilon\right)\\
\quad & \Longleftrightarrow\quad v\leq-m\Delta\varepsilon.
\end{split}
\label{GlrectshapedPSDded2bR2Int4F3}
\end{equation}
}}{\small \par}

{\small{So we get a new upper limit for the admissible $v$. Since
for $v\leq-\eta_{m}^{+}\varepsilon$ 
\begin{equation}
\begin{split}u^{(0)}\leq\frac{v}{-\eta_{m}^{+}}\end{split}
\label{GlrectshapedPSDded3cInt4bF1}
\end{equation}
we get for the terms of the partial integral (}}\ref{eq:tre-1}{\small{)
(note $f<\delta$ and $m<0$): 
\begin{equation}
\begin{split}\int\limits _{0}^{\infty} & \left|\mathcal{K}v)\right|^{2}\left[\int\limits _{0}^{\infty}\frac{1}{u}\cdot G_{0}\left(f-u\right)G_{m}\left(f-\frac{v}{u}\right)G_{m}\left(f-u-\frac{v}{u}\right)\,\mbox{d}u\right]\,\mbox{d}v\\
 & =\int\limits _{0}^{-\eta_{m}^{+}\varepsilon}\left|\mathcal{K}(v)\right|^{2}\left[\int\limits _{u^{(0)}}^{\frac{v}{-\eta_{m}^{+}}}\frac{1}{u}\,\mbox{d}u\right]\,\mbox{d}v+\int\limits _{-\eta_{m}^{+}\varepsilon}^{-m\Delta\varepsilon}\left|\mathcal{K}(v)\right|^{2}\left[\int\limits _{u^{(0)}}^{\varepsilon}\frac{1}{u}\,\mbox{d}u\right]\,\mbox{d}v\\
 & =\int\limits _{0}^{-\eta_{m}^{+}\varepsilon}\left|\mathcal{K}(v)\right|^{2}\mathrm{ln}\left(\frac{\frac{v}{-\eta_{m}^{+}}}{-\frac{\varepsilon_{m}^{-}}{2}-\sqrt{\left(\frac{\varepsilon_{m}^{-}}{2}\right)^{2}-v}}\right)\,\mbox{d}v+\int\limits _{-\eta_{m}^{+}\varepsilon}^{-m\Delta\varepsilon}\left|\mathcal{K}(v)\right|^{2}\mathrm{ln}\left(\frac{\varepsilon}{-\frac{\varepsilon_{m}^{-}}{2}-\sqrt{\left(\frac{\varepsilon_{m}^{-}}{2}\right)^{2}-v}}\right)\,\mbox{d}v.
\end{split}
\label{GlpartXPMintegIVvereinfacht}
\end{equation}
}}{\small \par}

{\small{\rule{0pt}{1ex}}}{\small \par}

{\small{The fourth part (}}\ref{eq:tre-1}{\small{) of $I_{XCI}(f)$
now reads:}}
\begin{align}
\notag2\left(\frac{P}{2\delta}\right)^{3}\sum\limits _{m=-N_{c}}^{-1} & \left[\int\limits _{0}^{-\eta_{m}^{+}\varepsilon}\left|\mathcal{K}(v)\right|^{2}\mathrm{ln}\left(\frac{\frac{v}{-\eta_{m}^{+}}}{-\frac{\varepsilon_{m}^{-}}{2}-\sqrt{\left(\frac{\varepsilon_{m}^{-}}{2}\right)^{2}-v}}\right)\,\mbox{d}v\right.\\
 & \left.\qquad\qquad\quad+\int\limits _{-\eta_{m}^{+}\varepsilon}^{-m\Delta\varepsilon}\left|\mathcal{K}(v)\right|^{2}\mathrm{ln}\left(\frac{\varepsilon}{-\frac{\varepsilon_{m}^{-}}{2}-\sqrt{\left(\frac{\varepsilon_{m}^{-}}{2}\right)^{2}-v}}\right)\,\mbox{d}v.\right].
\end{align}
\\

{\small{Now define 
\begin{equation}
\begin{split}\eta_{m}^{-}:=m\Delta-\eta\quad\text{and}\quad\varepsilon_{m}^{+}=m\Delta+(\delta+f).\end{split}
\label{GlrectshapedPSDded2DefdeltasDefszusatz}
\end{equation}
}}{\small \par}

{\small{Then for $m>0$ we get the following correspondences: 
\begin{equation}
\begin{split}\eta_{m}^{-}:=-\eta_{-m}^{+}=-(-m\Delta+\eta)\quad\text{and}\quad\varepsilon_{m}^{+}=-\varepsilon_{-m}^{+}=-(-m\Delta+(\delta+f)).\end{split}
\label{correspondences}
\end{equation}
}}{\small \par}

{\small{This allows to express (\ref{GlpartXPMintegIIvereinfacht})
and (\ref{GlpartXPMintegIVvereinfacht}) in a unified form as a $\sum_{m=1}^{N_{c}}$
instead of a $\sum_{m=-N_{c}}^{-1}$. The second part (}}\ref{eq:tre-1}{\small{)
of $I_{XCI}(f)$ can now be expressed as:}}{\small \par}

\begin{equation}
\begin{split}2\left(\frac{P}{2\delta}\right)^{3}\sum\limits _{m=1}^{N_{c}}\left[\int\limits _{0}^{m\Delta\eta}\left|\mathcal{K}(v)\right|^{2}\mathrm{ln}\left(\frac{-\frac{\eta_{m}^{-}}{2}+\sqrt{\left(\frac{\eta_{m}^{-}}{2}\right)^{2}+v}}{\frac{v}{\varepsilon_{m}^{+}}}\right)\,\mbox{d}v+\int\limits _{m\Delta\eta}^{\varepsilon_{m}^{+}\eta}\left|\mathcal{K}(v)\right|^{2}\mathrm{ln}\left(\frac{\varepsilon_{m}^{+}\eta}{v}\right)\,\mbox{d}v\right].\end{split}
\label{GlpartXPMintegIIvereinfacht3Nwu}
\end{equation}

{\small{The forth part (}}\ref{eq:tre-1}{\small{) of $I_{XCI}(f)$
now reads:}}{\small \par}

\begin{align}
\notag2\left(\frac{P}{2\delta}\right)^{3}\sum\limits _{m=1}^{N_{c}} & \left[\int\limits _{0}^{\eta_{m}^{-}\varepsilon}\left|\mathcal{K}(v)\right|^{2}\mathrm{ln}\left(\frac{\frac{v}{\eta_{m}^{-}}}{\frac{\varepsilon_{m}^{+}}{2}-\sqrt{\left(\frac{\varepsilon_{m}^{+}}{2}\right)^{2}-v}}\right)\,\mbox{d}v\right.\\
 & \left.\qquad\qquad\quad+\int\limits _{\eta_{m}^{-}\varepsilon}^{m\Delta\varepsilon}\left|\mathcal{K}(v)\right|^{2}\mathrm{ln}\left(\frac{\varepsilon}{\frac{\varepsilon_{m}^{+}}{2}-\sqrt{\left(\frac{\varepsilon_{m}^{+}}{2}\right)^{2}-v}}\right)\,\mbox{d}v.\right].
\end{align}
{\small{}}\\
{\small \par}

{\small{For symmetry reasons, the results may be generalized to all
$-\delta<f<\delta$ if we define}}{\small \par}

{\small{
\begin{equation}
\begin{split}\eta:=\delta-|f|\quad\text{and}\quad\varepsilon:=\delta+|f|\end{split}
.\label{GlrectshapedPSDded2DefdeltasDefszusatz2}
\end{equation}
}}{\small \par}

{\small{We are now ready to summarize the formula for expressing $\mathcal{I}_{XCI}(f)\triangleq I_{XCI}(f)/\left(\frac{P}{2\delta}\right)^{3}$
in the case of rectangular shaped input signals for all $-\delta<f<\delta$
and thus finally prove the theorem:}}{\small \par}

{\small{{ 
}}{\footnotesize{
\begin{align}
\notag\mathcal{I}_{XCI}(f) & =\textbf{\ensuremath{2}}\sum\limits _{m=1}^{N_{c}}\left[\int\limits _{0}^{\varepsilon_{m}^{-}\eta}\left|\mathcal{K}(v)\right|^{2}\mathrm{ln}\left(\frac{\frac{v}{\eta_{m}^{+}}}{\frac{\eta_{m}^{+}}{2}-\sqrt{\left(\frac{\eta_{m}^{+}}{2}\right)^{2}-v}}\right)\,\mbox{d}v+\int\limits _{\varepsilon_{m}^{-}\eta}^{m\Delta\eta}\left|\mathcal{K}(v)\right|^{2}\mathrm{ln}\left(\frac{\eta}{\frac{\eta_{m}^{+}}{2}-\sqrt{\left(\frac{\eta_{m}^{+}}{2}\right)^{2}-v}}\right)\,\mbox{d}v\right.\\
\notag & \qquad\qquad\quad\qquad\quad+\int\limits _{0}^{m\Delta\eta}\left|\mathcal{K}(v)\right|^{2}\mathrm{ln}\left(\frac{\frac{\eta_{m}^{+}}{2}+\sqrt{\left(\frac{\eta_{m}^{+}}{2}\right)^{2}+v}}{\frac{v}{\varepsilon_{m}^{+}}}\right)\,\mbox{d}v+\int\limits _{m\Delta\eta}^{\varepsilon_{m}^{+}\eta}\left|\mathcal{K}(v)\right|^{2}\mathrm{ln}\left(\frac{\varepsilon_{m}^{+}\eta}{v}\right)\,\mbox{d}v\\
 & \qquad\qquad\quad\qquad\quad+\int\limits _{0}^{m\Delta\varepsilon}\left|\mathcal{K}(v)\right|^{2}\mathrm{ln}\left(\frac{-\frac{\varepsilon_{m}^{-}}{2}+\sqrt{\left(\frac{\varepsilon_{m}^{-}}{2}\right)^{2}+v}}{\frac{v}{\eta_{m}^{+}}}\right)\,\mbox{d}v+\int\limits _{m\Delta\varepsilon}^{\eta_{m}^{+}\varepsilon}\left|\mathcal{K}(v)\right|^{2}\mathrm{ln}\left(\frac{\eta_{m}^{+}\varepsilon}{v}.\right)\,\mbox{d}v\\
\notag & \left.\qquad\qquad\quad\qquad\quad+\int\limits _{0}^{\eta_{m}^{-}\varepsilon}\left|\mathcal{K}(v)\right|^{2}\mathrm{ln}\left(\frac{\frac{v}{\eta_{m}^{-}}}{\frac{\varepsilon_{m}^{+}}{2}-\sqrt{\left(\frac{\varepsilon_{m}^{+}}{2}\right)^{2}-v}}\right)\,\mbox{d}v+\int\limits _{\eta_{m}^{-}\varepsilon}^{m\Delta\varepsilon}\left|\mathcal{K}(v)\right|^{2}\mathrm{ln}\left(\frac{\varepsilon}{\frac{\varepsilon_{m}^{+}}{2}-\sqrt{\left(\frac{\varepsilon_{m}^{+}}{2}\right)^{2}-v}}\right)\,\mbox{d}v\right].
\end{align}
}}}{\footnotesize \par}

\subsection{Proof for $f>\delta$ (resp.\ $|f|>\delta$)}

{\small{First note that we may restrict the analysis to the case $\delta<f<3\delta$
since the Nonlinearity Double Integral is generally 0 for $f\geq3\delta$.
In section \ref{ProofA} we showed that the partial integrals (\ref{equno-1})
and (\ref{eq:quattro-1}) are zero if $f$ exceeds $\delta$. So the
only contribution to $I_{XCI}(f)$ in the case $\delta<f<3\delta$
is due to partial integrals (\ref{eqdue-1}) and (\ref{eq:tre-1}).
The proof follows the guidelines of that of section \ref{ProofA}.
 In both cases the condition $G_{0}\left(f-u\right)$ leads to 
\begin{equation}
\begin{split}-\eta\leq u\leq\varepsilon.\end{split}
\label{GlrectshapedPSDded1Z}
\end{equation}
}}{\small \par}

{\small{The condition $G_{m}\left(f+\frac{v}{u}\right)$ for the partial
integral (\ref{eqdue-1}) leads to: 
\begin{equation}
\begin{split}\frac{v}{u}\leq\eta_{m}^{+}=m\Delta+\eta\quad\text{and}\quad\frac{v}{u}\geq\varepsilon_{m}^{-},\end{split}
\label{GlrectshapedPSDded2Int3Z}
\end{equation}
}}{\small \par}

{\small{Since $0\geq\eta\geq-2\delta$ the first inequality of (\ref{GlrectshapedPSDded2Int3Z})
is never fulfilled for $m<0$. So the integral (\ref{eqdue-1}) is
zero for $m<0$. For $m>0$ we get since $\eta_{m}^{+}=m\Delta+\eta>\Delta-2\delta>0$:
\begin{equation}
\begin{split}\max\left\{ \frac{v}{\eta_{m}^{+}},-\eta\right\} \leq u.\end{split}
\label{GlrectshapedPSDded2bR2Int3Z2}
\end{equation}
}}{\small \par}

{\small{Now: 
\begin{equation}
\begin{split} & \max\left\{ \frac{v}{\eta_{m}^{+}},-\eta\right\} =-\eta\quad\text{iff}\quad v\leq-\eta\eta_{m}^{+}\\
\quad\text{and}\quad\quad & \max\left\{ \frac{v}{\eta_{m}^{+}},-\eta\right\} =\frac{v}{\eta_{m}^{+}}\quad\text{iff}\quad v\geq-\eta\eta_{m}^{+}.
\end{split}
\label{GlrectshapedPSDded2bR3IntZ3}
\end{equation}
}}{\small \par}

{\small{Taking into account $G_{m}\left(f-u+\frac{v}{u}\right)$ we
deduce the restrictions 
\begin{equation}
\begin{split}\frac{v}{u}\leq\eta_{m}^{+}+u\quad\text{and}\quad\frac{v}{u}\geq\varepsilon_{m}^{-}+u.\end{split}
\label{GlrectshapedPSDded2bR3IntZ4}
\end{equation}
}}{\small \par}

{\small{This implies together with (\ref{GlrectshapedPSDded1Z}) and
(\ref{GlrectshapedPSDded2Int3Z}) the following restrictions: 
\begin{equation}
\begin{split} & \quad v\geq(\varepsilon_{m}^{-}+u)u\geq(\varepsilon_{m}^{-}-\eta)(-\eta)=\eta(\eta-\varepsilon_{m}^{-})\\
\quad\text{and} & \quad\eta_{m}^{+}\geq\varepsilon_{m}^{-}+u\quad\Longleftrightarrow\quad u\leq\eta_{m}^{+}-\varepsilon_{m}^{-}=\delta-f+(\delta+f)=2\delta.
\end{split}
\label{GlrectshapedPSDded2bR3IntZ5}
\end{equation}
}}{\small \par}

{\small{For this maximum $u$ we have using (\ref{GlrectshapedPSDded2Int3Z})
again a further restriction for $v$: 
\begin{equation}
\begin{split}\frac{v}{u}=\frac{v}{2\delta}\leq\eta_{m}^{+}\quad\Longleftrightarrow\quad v\leq2\delta\eta_{m}^{+}.\end{split}
\label{GlrectshapedPSDded2bR3IntZ6}
\end{equation}
}}{\small \par}

{\small{Consequently we have: 
\begin{equation}
\begin{split}\int\limits _{0}^{\infty} & \left|\mathcal{K}(v)\right|^{2}\left[\int\limits _{0}^{\infty}\frac{1}{u}\cdot G_{0}\left(f-u\right)G_{m}\left(f+\frac{v}{u}\right)G_{m}\left(f-u+\frac{v}{u}\right)\,\mbox{d}u\right]\,\mbox{d}v\\
 & =\int\limits _{\eta(\eta-\varepsilon_{m}^{-})}^{-\eta\eta_{m}^{+}}\left|\mathcal{K}(v)\right|^{2}\left[\int\frac{1}{u}\,\mbox{d}u\right]\,\mbox{d}v+\int\limits _{-\eta\eta_{m}^{+}}^{2\delta\eta_{m}^{+}}\left|\mathcal{K}(v)\right|^{2}\left[\int\frac{1}{u}\,\mbox{d}u\right]\,\mbox{d}v
\end{split}
\label{GlpartXPMintegIIIvereinfachtCZ}
\end{equation}
and we have to fill-in the correct integration limits of the inner
integration. In the interval $[\eta(\eta-\varepsilon_{m}^{-}),-\eta\eta_{m}^{+}]$
we derived 
\begin{equation}
\begin{split}u\geq-\eta.\end{split}
\label{GlrectshapedPSDded2bR3IntZ7}
\end{equation}
}}{\small \par}

{\small{In the interval $[-\eta\eta_{m}^{+},2\delta\eta_{m}^{+}]$
we derived 
\begin{equation}
\begin{split}u\geq\frac{v}{\eta_{m}^{+}}.\end{split}
\label{GlrectshapedPSDded2bR3IntZ7}
\end{equation}
}}{\small \par}

{\small{Further we always have: 
\begin{equation}
\begin{split}u\leq\frac{v}{\varepsilon_{m}^{-}+u}.\end{split}
\label{GlrectshapedPSDded2bR3IntZ7b}
\end{equation}
}}{\small \par}

{\small{Since (note that one solution of the quadratic equation doesn't
give a restriction): 
\begin{equation}
\begin{split}u\leq\frac{v}{\varepsilon_{m}^{-}+u}\quad & \Longleftrightarrow\quad\left(u+\frac{\varepsilon_{m}^{-}}{2}\right)^{2}\leq\left(\frac{\varepsilon_{m}^{-}}{2}\right)^{2}+v\\
\quad & \Longleftrightarrow\quad u\leq-\left(\frac{\varepsilon_{m}^{-}}{2}\right)^{2}+\sqrt{\left(\frac{\varepsilon_{m}^{-}}{2}\right)^{2}+v}
\end{split}
\label{GlrectshapedPSDded2bR3IntZ8}
\end{equation}
}}{\small \par}

{\small{Hence we got the integration limits in the inner integral:
\begin{equation}
\begin{split}\int\limits _{0}^{\infty} & \left|\mathcal{K}(v)\right|^{2}\left[\int\limits _{0}^{\infty}\frac{1}{u}\cdot G_{0}\left(f-u\right)G_{m}\left(f+\frac{v}{u}\right)G_{m}\left(f-u+\frac{v}{u}\right)\,\mbox{d}u\right]\,\mbox{d}v\\
 & =\int\limits _{\eta(\eta-\varepsilon_{m}^{-})}^{-\eta\eta_{m}^{+}}\left|\mathcal{K}(v)\right|^{2}\left[\int\limits _{-\eta}^{-\left(\frac{\varepsilon_{m}^{-}}{2}\right)^{2}+\sqrt{\left(\frac{\varepsilon_{m}^{-}}{2}\right)^{2}+v}}\frac{1}{u}\,\mbox{d}u\right]\,\mbox{d}v+\int\limits _{-\eta\eta_{m}^{+}}^{2\delta\eta_{m}^{+}}\left|\mathcal{K}(v)\right|^{2}\left[\int\limits _{\frac{v}{\eta_{m}^{+}}}^{-\left(\frac{\varepsilon_{m}^{-}}{2}\right)^{2}+\sqrt{\left(\frac{\varepsilon_{m}^{-}}{2}\right)^{2}+v}}\frac{1}{u}\,\mbox{d}u\right]\,\mbox{d}v\\
 & =\int\limits _{\eta(\eta-\varepsilon_{m}^{-})}^{-\eta\eta_{m}^{+}}\left|\mathcal{K}(v)\right|^{2}\mathrm{ln}\left(\frac{\frac{\varepsilon_{m}^{-}}{2}-\sqrt{\left(\frac{\varepsilon_{m}^{-}}{2}\right)^{2}+v}}{\eta}\right)\,\mbox{d}v+\int\limits _{-\eta\eta_{m}^{+}}^{2\delta\eta_{m}^{+}}\left|\mathcal{K}(v)\right|^{2}\mathrm{ln}\left(\frac{\frac{-\varepsilon_{m}^{-}}{2}+\sqrt{\left(\frac{\varepsilon_{m}^{-}}{2}\right)^{2}+v}}{\frac{v}{\eta_{m}^{+}}}\right)\,\mbox{d}v.
\end{split}
\label{GlpartXPMintegIIIvereinfachtCZ10}
\end{equation}
}}\\
{\small \par}

{\small{The condition $G_{m}\left(f-\frac{v}{u}\right)$ for the partial
integral (\ref{eq:tre-1}) leads to: 
\begin{equation}
\begin{split}-\frac{v}{u}\leq\eta_{m}^{+}\quad & \text{and}\quad-\frac{v}{u}\geq\varepsilon_{m}^{-}\\
\quad\Longleftrightarrow\quad\frac{v}{u}\geq-\eta_{m}^{+}\quad & \text{and}\quad\frac{v}{u}\leq-\varepsilon_{m}^{-}.
\end{split}
\label{GlrectshapedPSDded2Int3Y}
\end{equation}
}}{\small \par}

{\small{Since $0\geq\eta\geq-2\delta$ the second inequality of (\ref{GlrectshapedPSDded2Int3Y})
is never fulfilled for $m>0$. So the integral (\ref{eq:tre-1}) is
zero for $m>0$. Taking into account $G_{m}\left(f-u-\frac{v}{u}\right)$
we deduce the restrictions 
\begin{equation}
\begin{split}-\frac{v}{u}\leq\eta_{m}^{+}+u\quad\text{and}\quad-\frac{v}{u}\geq\varepsilon_{m}^{-}+u.\end{split}
\label{GlrectshapedPSDded2bR3IntY4}
\end{equation}
we see that the that we get instead of (\ref{GlrectshapedPSDded2Int3Y})
\begin{equation}
\begin{split}-\frac{v}{u}\leq\eta_{m}^{+}\quad & \text{and}\quad-\frac{v}{u}\geq\varepsilon_{m}^{-}+u\end{split}
\label{GlrectshapedPSDded2Int3Yb}
\end{equation}
since $-\eta_{m}^{+}=|m|\Delta-\eta\geq\Delta-2\delta>0$ and $-\varepsilon_{m}^{-}=|m|\Delta+\varepsilon>0$.
Then 
\begin{equation}
\begin{split}u\leq\frac{v}{-\eta_{m}^{+}}\quad & \text{and}\quad u\geq\frac{v}{-\varepsilon_{m}^{-}-u}\end{split}
\label{GlrectshapedPSDded2Int3Y}
\end{equation}
and consequently using (\ref{GlrectshapedPSDded1Z}): 
\begin{equation}
\begin{split}\max\left\{ \frac{v}{-\varepsilon_{m}^{-}-u},-\eta\right\} \leq u.\end{split}
\label{GlrectshapedPSDded2bR2Int3Y4}
\end{equation}
}}{\small \par}

{\small{Now it follows that: 
\begin{equation}
\begin{split} & \max\left\{ \frac{v}{-\varepsilon_{m}^{-}\eta},-\eta\right\} =-\eta\quad\text{iff}\quad v\leq\eta(\varepsilon_{m}^{-}-\eta)\\
\quad\text{and}\quad\quad & \max\left\{ \frac{v}{-\varepsilon_{m}^{-}},-\eta\right\} =\frac{v}{-\varepsilon_{m}^{-}}\quad\text{iff}\quad v\geq\eta(\varepsilon_{m}^{-}-\eta).
\end{split}
\label{GlrectshapedPSDded2bR3IntYY3}
\end{equation}
}}{\small \par}

{\small{Equations (\ref{GlrectshapedPSDded2Int3Y}) and (\ref{GlrectshapedPSDded2Int3Yb})
imply 
\begin{equation}
\begin{split}\eta_{m}^{+}\geq-\frac{v}{u}\geq\varepsilon_{m}^{-}+u\end{split}
\label{GlrectshapedPSDded2bR3IntY6}
\end{equation}
which implies 
\begin{equation}
\begin{split}u\leq2\delta\end{split}
\label{GlrectshapedPSDded2bR3IntY7}
\end{equation}
and 
\begin{equation}
\begin{split}v\leq-2\delta\eta_{m}^{+}.\end{split}
\label{GlrectshapedPSDded2bR3IntY8}
\end{equation}
}}{\small \par}

{\small{On the other hand 
\begin{equation}
\begin{split}v\geq-\eta_{m}^{+}u\quad & \text{and}\quad u\geq-\eta\end{split}
\label{GlrectshapedPSDded2Int3Yb9}
\end{equation}
imply 
\begin{equation}
\begin{split}v\geq\eta_{m}^{+}\eta.\end{split}
\label{GlrectshapedPSDded2Int3Yb10}
\end{equation}
}}{\small \par}

{\small{Thus we get: 
\begin{equation}
\begin{split}\int\limits _{0}^{\infty} & \left|\mathcal{K}(v)\right|^{2}\left[\int\limits _{0}^{\infty}\frac{1}{u}\cdot G_{0}\left(f-u\right)G_{m}\left(f-\frac{v}{u}\right)G_{m}\left(f-u-\frac{v}{u}\right)\,\mbox{d}u\right]\,\mbox{d}v\\
 & =\int\limits _{\eta_{m}^{+}\eta}^{\eta(\varepsilon_{m}^{-}-\eta)}\left|\mathcal{K}(v)\right|^{2}\left[\int\limits _{-\eta}^{\frac{v}{-\eta_{m}^{+}}}\frac{1}{u}\,\mbox{d}u\right]\,\mbox{d}v+\int\limits _{\eta(\varepsilon_{m}^{-}-\eta)}^{-2\delta\eta_{m}^{+}}\left|\mathcal{K}(v)\right|^{2}\left[\int\limits _{}^{\frac{v}{-\eta_{m}^{+}}}\frac{1}{u}\,\mbox{d}u\right]\,\mbox{d}v.
\end{split}
\label{GlpartXPMintegIIIvereinfachtCZneu}
\end{equation}
}}{\small \par}

{\small{In the interval $[\eta(\varepsilon_{m}^{-}-\eta),-2\delta\eta_{m}^{+}]$
we derive from the second inequality (\ref{GlrectshapedPSDded2Int3Y})
like in that (\ref{GlrectshapedPSDded3cInt4}) that 
\begin{equation}
\begin{split}u\geq-\frac{\varepsilon_{m}^{-}}{2}-\sqrt{\left(\frac{\varepsilon_{m}^{-}}{2}\right)^{2}-v}.\end{split}
\label{GlrectshapedPSDded3cInt4}
\end{equation}
}}{\small \par}

{\small{We therefore have 
\begin{equation}
\begin{split}\int\limits _{0}^{\infty} & \left|\mathcal{K}(v)\right|^{2}\left[\int\limits _{0}^{\infty}\frac{1}{u}\cdot G_{0}\left(f-u\right)G_{m}\left(f-\frac{v}{u}\right)G_{m}\left(f-u-\frac{v}{u}\right)\,\mbox{d}u\right]\,\mbox{d}v\\
 & =\int\limits _{\eta_{m}^{+}\eta}^{\eta(\varepsilon_{m}^{-}-\eta)}\left|\mathcal{K}(v)\right|^{2}\left[\int\limits _{-\eta}^{\frac{v}{-\eta_{m}^{+}}}\frac{1}{u}\,\mbox{d}u\right]\,\mbox{d}v+\int\limits _{\eta(\varepsilon_{m}^{-}-\eta)}^{-2\delta\eta_{m}^{+}}\left|\mathcal{K}(v)\right|^{2}\left[\int\limits _{-\frac{\varepsilon_{m}^{-}}{2}-\sqrt{\left(\frac{\varepsilon_{m}^{-}}{2}\right)^{2}-v}}^{\frac{v}{-\eta_{m}^{+}}}\frac{1}{u}\,\mbox{d}u\right]\,\mbox{d}v.
\end{split}
\label{GlpartXPMintegIIIvereinfachtCZneu2}
\end{equation}
}}{\small \par}

{\small{Using the correspondences (\ref{correspondences}) we finally
arrive at 
\begin{equation}
\begin{split}\int\limits _{0}^{\infty} & \left|\mathcal{K}(v)\right|^{2}\left[\int\limits _{0}^{\infty}\frac{1}{u}\cdot G_{0}\left(f-u\right)G_{m}\left(f-\frac{v}{u}\right)G_{m}\left(f-u-\frac{v}{u}\right)\,\mbox{d}u\right]\,\mbox{d}v\\
 & =\int\limits _{-\eta\eta_{m}^{-}}^{-\eta(\varepsilon_{m}^{+}+\eta)}\left|\mathcal{K}(v)\right|^{2}\mathrm{ln}\left(-\frac{v}{\eta\eta_{m}^{-}}\right)\,\mbox{d}v+\int\limits _{-\eta(\varepsilon_{m}^{+}+\eta)}^{2\delta\eta_{m}^{-}}\left|\mathcal{K}(v)\right|^{2}\mathrm{ln}\left(\frac{\frac{v}{\eta_{m}^{-}}}{\frac{\varepsilon_{m}^{+}}{2}-\sqrt{\left(\frac{\varepsilon_{m}^{+}}{2}\right)^{2}-v}}\right)\,\mbox{d}v.
\end{split}
\label{GlpartXPMintegIIIvereinfachtCZneu3}
\end{equation}
}}{\small \par}

{\small{which completes the proof.$\,\,\Box$}}
\end{document}